\documentclass[a4paper,11pt]{article}
\usepackage{jheppub}
\usepackage{physics}
\usepackage{tikz}
\usepackage{float}
\usepackage{multirow}
\usepackage{rotating}

\usepackage{upgreek}

\usepackage[T1]{fontenc} 

\usepackage{etoolbox}

\usepackage{extarrows} 
\usepackage{amsmath}
\usepackage{graphicx}
\usepackage{dcolumn}

\usepackage{bm}
\usepackage{wasysym}
\usepackage{verbatim}
\usepackage{color}
\usepackage{mathtools,slashed}

\newcommand{\mpl}{M_{\mbox{\tiny{Pl}}}}

\newcommand{\gBL}{g_{_{\mB\mL}}}
\newcommand{\gY}{g_{_{Y}}}
\newcommand{\gR}{g_{_{R}}}
\newcommand{\gL}{g_{_{L}}}
\newcommand{\mN}{\mathcal{N}}
\newcommand{\mB}{\mathsf{B}}
\newcommand{\mL}{\mathsf{L}}
\newcommand{\bF}{\boldsymbol{W}}

\newcommand{\bW}{\boldsymbol{W}}
\newcommand{\bff}{\boldsymbol{f}}

\newcommand{\p}{\partial}

\newcommand{\bx}{{\bf{x}}}
\newcommand{\bT}{{\bf{T}}}

\newcommand{\bPhi}{\boldsymbol{\Phi}}
\newcommand{\bDelta}{\boldsymbol{\Delta}}

\newcommand{\mH}{\mathcal{H}}

\newcommand{\an}{\quad \textmd{and} \quad }
\newcommand{\bs}{\boldsymbol{\sigma}}
\newcommand{\bt}{\boldsymbol{\tau}}

\newcommand{\I}{{\bf{I}}_2}

\newcommand{\bea}{\begin{eqnarray}}
\newcommand{\eea}{\end{eqnarray}}

\newcommand{\bN}{{\bf{N}}}
\newcommand{\bn}{{\boldsymbol{\nu}}}
\newcommand{\treh }{T_{\rm reh}}

\newcommand{\where}{\quad \textmd{where} \quad}
\newcommand{\x}{\tilde{\tau}}

\newcommand{\bk}{{\bf{k}}}
\newcommand{\bfe}{{\boldsymbol{e}}}
\newcommand{\uL}{\underline{\mL}}
\newcommand{\bD}{{\boldsymbol{\Delta}}}

\begin{document}

\preprint{CERN-TH-2021-034}


\title{\boldmath Chiral Anomaly in SU(2)${}_R$-Axion Inflation and the New Prediction for Particle Cosmology} \date{\today}
\author{Azadeh Maleknejad}
\affiliation{Theoretical Physics Department, CERN, 1211 Geneva 23, Switzerland
}
\emailAdd{azadeh.maleknejad@cern.ch}
\abstract{Upon embedding the axion-inflation in the minimal left-right symmetric gauge extension of the SM with gauge group $SU(2)_L\times SU(2)_R \times U(1)_{\mB-\mL}$, [arXiv:2012.11516] proposed a new particle physics model for inflation. In this work, we present a more detailed analysis. As a compelling consequence, this setup provides a new mechanism for simultaneous baryogenesis and right-handed neutrino creation by the chiral anomaly of $W_R$ in inflation. The lightest right-handed neutrino is the dark matter candidate. This setup has two unknown fundamental
scales, i.e., the scale of inflation and left-right symmetry breaking $SU(2)_R\times U(1)_{\mB-\mL}\rightarrow U(1)_{Y}$. Sufficient matter creation demands the left-right symmetry breaking scale happens shortly after the end of inflation. Interestingly, it prefers left-right symmetry breaking scales above $10^{10}~GeV$, which is in the range  suggested by the non-supersymmetric SO(10) Grand Unified Theory with an intermediate left-right symmetry scale. Although $W_R$ gauge field generates equal amounts of right-handed baryons and leptons in inflation, i.e. $\mB-\mL=0$, in the Standard Model sub-sector $\mB-\mL_{\rm SM}\neq 0$. A key aspect of this setup is that $SU(2)_R$ sphalerons are never in equilibrium, and the primordial $\mB-\mL_{\rm SM}$ is conserved by the Standard Model interactions.  This setup yields a deep connection between CP violation in physics of inflation and matter creation (visible and dark); hence it can naturally explain the observed coincidences among cosmological parameters, i.e., $\eta_{\mB}\simeq 0.3 P_{\zeta}$ and $\Omega_{\rm DM}\simeq 5\Omega_{\mB}$. The new mechanism
does not rely on the largeness of the unconstrained CP-violating phases in the neutrino sector
nor fine-tuned masses for the heaviest right-handed neutrinos.
 The $SU(2)_R$-axion inflation comes with a cosmological smoking gun; chiral, non-Gaussian, and blue-tilted gravitational wave background, which can be probed by future CMB missions and laser interferometer detectors.  }

\maketitle
\flushbottom

\section{Introduction}

 Modern cosmology has been remarkably successful in describing the Universe from a second after the Big Bang until today. However,
its physics before that time is still much less certain. It
profoundly involves particle theory beyond the Standard Model (BSM) to explain
long-standing puzzles: I) the origin of the observed matter
asymmetry, II) neutrino mass, III) nature of dark matter, and IV) cosmic inflation. Apart from the above problems, the standard model of particle physics (SM) faces some conceptual issues: i) ad hoc parity violation, ii) accidental $\mB-\mL$ global symmetry, iii) vacuum instability, and iv) strong CP problem.

  \begin{figure}[h]
  \centering
  \includegraphics[height=0.33\textheight]{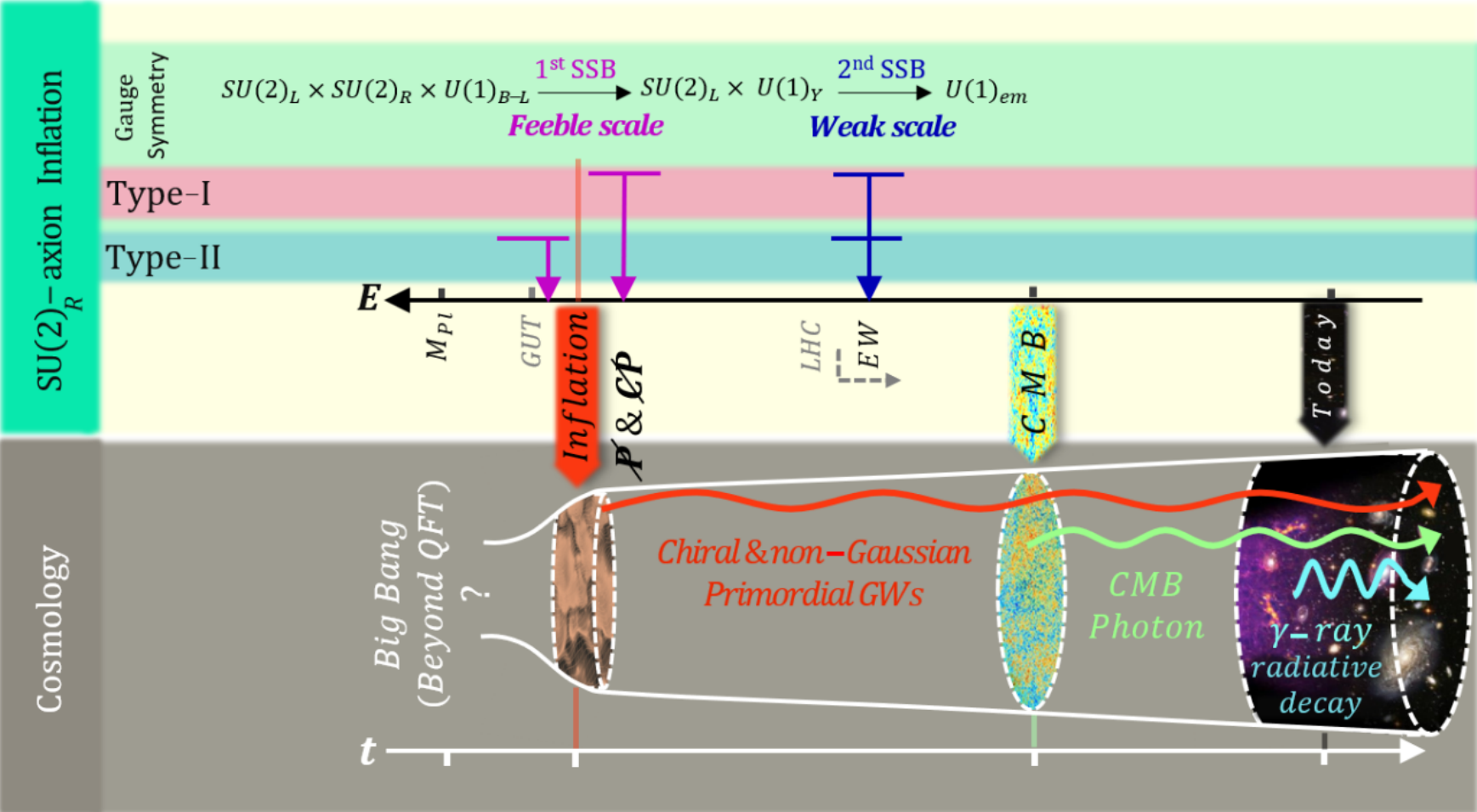} \\
   \caption{The $SU(2)_R$-axion inflation throughout cosmic history. This setup has two new fundamental scales, scale of inflation $\Lambda_{\rm inf}=\sqrt{H\mpl}$, and scale of first SSB, $\Lambda_F$. Scenarios with $\Lambda_{\rm inf}>\Lambda_{F}$ ($\Lambda_{\rm inf}<\Lambda_{F}$) are called type-{\bf{I}} (type-{\bf{II}}).}
  \label{cosmic-history} 
\end{figure}

Recently, \cite{Maleknejad:2020yys} proposed a new setup for physics of inflation by embedding axion inflation in $SU(2)_R$ gauge extensions of the SM. For concreteness, as the most minimal realization of this idea \footnote{This is the minimal realization that can produce a non-zero $\mB-\mL$ in the SM sector, i.e. $\mB_{\rm SM}-\mL_{\rm SM}\neq 0$.}, the axion inflaton is coupled to $SU(2)_R$ gauge field in the minimal left-right symmetric model (LRSM) \cite{Pati:1974y, Mohapatra:1974gc, Senjanovic:1975rk,Davidson:1978pm}. This new particle physics model for inflation gives rise to a new mechanism for simultaneous baryogenesis and Right-Handed Neutrino (RHN) creation through the chiral anomaly of $SU(2)_R$, which provides the source of CP violation in inflation. This new mechanism
does not rely on the largeness of the unconstrained CP-violating phases in the neutrino sector
nor fine-tuned masses for the heaviest right-handed neutrinos. On the other hand, it makes a deep connection between inflation, baryon asymmetry, and DM relic density. Therefore, it can naturally explain the observed coincidences among cosmological parameters, i.e., $\eta_{\mB}\simeq 0.3 P_{\zeta}$ and $\Omega_{DM}\simeq 5\Omega_{\mB}$. If the primordial relic density of the lightest RHN makes all the DM today, $\Omega_{\rm DM}$ specifies its mass as $1.7~GeV$. Therefor its radiative decay may produce active neutrinos and gamma-rays with energy $E_{\gamma} =m_{\rm{N}_1}/2$ in the highly-dense DM regions. As a compelling consequence, this setup can simultaneously provide plausible explanations for the phenomena (I-IV) named earlier. In this work, we present a more detailed analytical and numerical analysis.

Originally proposed to explain P violation in low energy processes \cite{Pati:1974y}, LRSM predicted massive neutrinos years before experiment. Among its additional compelling consequences are: natural $\mB-\mL$ symmetry \cite{Mohapatra:1980qe}, natural entailed seesaw mechanisms \cite{Mohapatra:1980yp}, solution to vacuum stability problem \cite{Maiezza:2016ybz}, and strong CP problem without an axion. The LRSM can solve the conceptual issues (i-iv) named earlier.
In the minimal LRSM, the Electro-Weak (EW) gauge symmetry is extended to $SU(2)_L\times SU(2)_R \times U(1)_{\mB-\mL}$ \cite{Pati:1974y, Mohapatra:1974gc, Senjanovic:1975rk,Davidson:1978pm}.
As a result, it introduces a new fundamental scale, $\Lambda_F$, during which the extended gauge symmetry is broken as $SU(2)_R \times U(1)_{\mB-\mL}\rightarrow U(1)_Y$. Upon embedding axion-inflation in LRSM, we have two unknown energy scales, the scale of inflation and $\Lambda_F$. Based on that we can classify the setup into two types; type-{\bf{I}} with $\sqrt{H\mpl}>\Lambda_F$ and type-{\bf{II}} with $\sqrt{H\mpl}<\Lambda_F$.  (See Fig. \ref{cosmic-history})

Axion fields are abundant in string theory, and therefore very well-motivated candidates
for the inflaton field \cite{Freese:1990rb, Pajer:2013fsa, McAllister:2014mpa}. Thanks to  their natural shift symmetry, their effective potential is protected from
dangerous quantum corrections, which guarantees the flatness of the potential. Besides their appealing theoretical stability, models of axion inflation are attractive
phenomenologically and are naturally coupled to gauge fields.
Non-Abelian gauge fields may contribute to the physics of inflation while respecting the cosmological symmetries \cite{Maleknejad:2011sq,Maleknejad:2011jw}. The first inflationary model based on non-Abelian gauge fields has been introduced as Gauge-flation \cite{Maleknejad:2011sq}, which is an EFT of a larger model, i.e., Chromo-natural \cite{Adshead:2012kp}, after integrating out the axion \cite{SheikhJabbari:2012qf}. \footnote{In Chromo-natural, the form of the axion potential is assumed to be cosine, which requires a large value of $\lambda$ to support slow-roll inflation \cite{Adshead:2012kp}. However, the large coupling is hard to achieve in a controlled string compactification \cite{Baumann:2014nda}. Therefore, we are interested in flat axion potentials and small values of $\lambda$, e.g. $f \lesssim 0.01$ and $\lambda \lesssim 0.1$ \cite{Maleknejad:2016qjz}.} Inspired by the original models, several more inflationary
models with the $SU(2)$ fields have been proposed and studied, which share the same key
features. In this work, we consider the minimal realization of $SU(2)$-axion inflation introduced and studied in \cite{Maleknejad:2016qjz}. For review see \cite{Maleknejad:2012fw}, Sec. 2 of \cite{Maleknejad:2018nxz} and references therein. Including $SU(2)$ gauge fields in physics of inflation give rise to a rich phenomenology which we summarize in the following. 
 The $SU(2)$-axion inflation \cite{Maleknejad:2016qjz} is a natural setting for warm inflation \cite{Bastero-Gil:2016qru,Kamali:2019ppi,Berghaus:2019whh}. The Chern-Simons interaction drains kinetic energy from the axion and injects it into the radiation.  This gauge field produces particles in inflation; charged Higgs via the Schwinger effect \cite{Lozanov:2018kpk} and charged fermions by both Schwinger effect and chiral anomaly \cite{Maleknejad:2020yys, Maleknejad:2019hdr, Mirzagholi:2019jeb}. Another consequence of this Schwinger effect is sourced primordial gravitational waves \cite{Maleknejad:2018nxz}. All the Sakharov conditions \cite{Sakharov:1967dj} are satisfied in inflation \cite{Maleknejad:2014wsa, Maleknejad:2016dci}, hence it provides a natural setting to explain the matter asymmetry (see Sec. \ref{B-L-N}). \footnote{Within the SM and through (global) gravitational anomaly, this setting naturally accompanies an inflationary leptogenesis \cite{Maleknejad:2016dci, Caldwell:2017chz, Alexander:2018fjp}.}  As a cosmological smoking gun, it predicts chiral \cite{Maleknejad:2012fw, Adshead:2013qp, Dimastrogiovanni:2012ew} and non-Gaussian \cite{Agrawal:2018mrg} Gravitational Wave Background (GWB) which leads to parity odd CMB cross-spectra (see \cite{Campeti:2020xwn} and Sec. \ref{Obser-sec})  \cite{Thorne:2017jft}.

\begin{figure}[h]
  \centering\includegraphics[height=0.33\textheight]{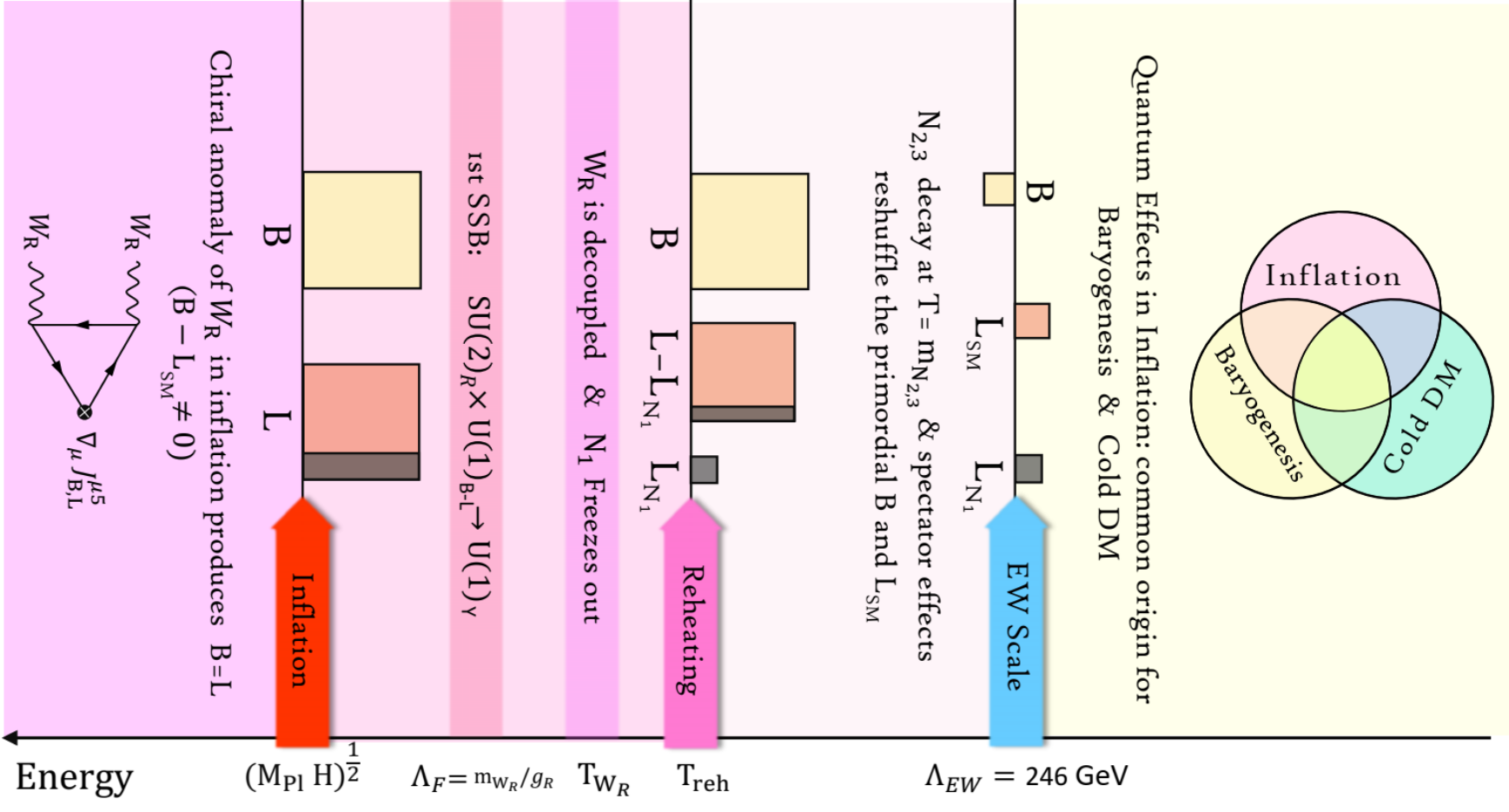} \\
   \caption{Summary of the mechanism: Illustration shows the evolution of baryons $\mB$ (yellow box), SM leptons $\mL_{_{\rm SM}}$ (pink box) and RHNs $\mL_{\rm N}=\sum_{i=1}^3 \mL_{{\rm N}_i}$ (gray box) during cosmic evolution. Here $\mB \equiv \mB_{\rm SM}$ and $\mL\equiv \mL_{\rm SM}+\mL_{\rm N}$. 
   The CP violation by the chiral anomaly of $W_R$ in inflation simultaneously produces baryons, SM leptons, and RHNs in inflation. The lightest RHN (our DM candidate $\rm{N}_1$) freezes out at $T=T_{W_R}$. Between reheating and electoweak scale, the spectator effects reshuffle the primordial densities while $\rm{N}_{2,3}$ decay at $T=m_{\rm N_{2,3}}$.
   The net baryon density and dark matter today are the remnant of that quantum effect in inflation.    Notice that $T_{\rm reh}<m_{W_R}$ condition is essential to keep $W_R$ sphalerons out of equilibrium (see App \ref{Sphaleron-App}). The $W_L$ sphalerons, however, contribute to the spectator effects and washout $\mB+\mL_{\rm SM}$ but conserve $\mB-\mL_{\rm SM}$.   }
  \label{evolution} 
\end{figure}

The new setup proposed in \cite{Maleknejad:2020yys} extended the field content of the minimal LRSM with an axion field which drives the cosmic inflation. It is assumed that the axion and $SU(2)_R$ gauge field are coupled by a Chern-Simons interaction, i.e., $SU(2)_R$-axion inflation. \footnote{In principle we can couple the axion to both $SU(2)_R$ and $SU(2)_L$ gauge fields. However, any primordial left-handed baryons and leptons produced by $SU(2)_L$ (i.e. $\mB_{\rm SM}=\mL_{\rm SM}$) will be completely washed out by the $SU(2)_L$ sphaleon effects which are in thermal equilibrium between $T_{\rm reh}$ and $m_{W_L}$. Therefore in the minimal realization of this idea we neglect this interaction.} 
Here both Parity and CP are spontaneously violated by the axion and its interaction with $\bW_R$ in the physics of inflation. Within this setup, $SU(2)_R$ gauge field is generated in inflation and creates right-handed chiral fermions coupled to it, i.e., SM baryons $\mB$, SM leptons $\mL_{\rm SM}$, and three Right-Handed Neutrinos (RHN) $\mL_{\rm N}$. In type-{\bf{I}} scenario in which the first SSB happens after inflation, equal baryon and lepton numbers are created in inflation, i.e. $\mB=\mL$ where $\mL\equiv \mL_{\rm SM}+\mL_{\rm N}$, yet $\mB-\mL_{\rm SM}\neq 0$. Shortly after inflation, the first SSB happens at $\Lambda_F$, and eventually, the $SU(2)_R$ interactions freeze out at temperature $T_{W_R}$. The lightest of RHNs with feeble Yukawa couplings (our DM candidate) is decoupled at this point, while $\rm{N}_{2,3}$ decay at $T=m_{\rm N_{2,3}}$. Between reheating and EW scale, the spectator effects reshuffle the primordial densities. The summary of this new mechanism for simultaneous baryogenesis and RHN production is presented in Fig. \ref{evolution}. 
Ref. \cite{Maleknejad:2020yys} was the first step to further, more involved analysis on the rich and multifaceted phenomenology of the gauge extensions of the SM in inflation physics. In the current work, we present a more detailed analytical and numerical analysis of the setup.

The paper is structured as follows. In Sec. \ref{Gauge-axion-inflation} we review the setup of $SU(2)_R$-axion inflation embedded in LRSM. In Sec. \ref{inf-PP}, we
work out the inflationary particle production. In Sec. \ref{Post-flation}, we study the post reheating evolution of the system. Next in
Sec. \ref{today}, we work out the final baryon asymmetry and dark matter in the modern era.
Sec. \ref{Obser-sec} presents a quick discussion on the setup's observational constraints and signatures. We finally conclude in
Sec. \ref{Sec/Sec-Concl}. Technical details of the computations and the underlying mathematical tools are provided in App.s \ref{LRSM-overview}-\ref{spectator-App}.

\textbf{Notations and conventions:~} In this work, we deal with 4 and 2-component spinors, which are acted upon by $4\times 4$ and $2\times 2$  matrices, respectively. The 4-spinors and $4\times 4$ matrices are remained unchanged, while the 2-spinors and $2\times 2$ matrices are written in boldface. The $2\times 2$ identity matrix is represented as $\I$. $ L$ and $R$ subscripts denote the left- and right-handed fermions. The lepton and baryon numbers are presented by $\mathsf{L}$ and $\mathsf{B}$. The Hubble parameter in inflation is denoted by $H$ and $\mpl=(8\pi G)^{-1}$ is set to one, unless otherwise specified. We use the Einstein summation notation, i.e., repeated indices (one upper and one lower) are summed. The beginning of the Latin alphabets, i.e. $a,b,c$ denote the $SU(2)$ group indices. Greek letters starting from the middle of the alphabet, i.e., $\mu, \nu, \dots$ are used for the space-time indices, whereas the starting ones, i.e., $\alpha, \beta, \dots$, present the indices of the tangent space (non-coordinate) bases.

\section{Framework}
\label{Gauge-axion-inflation}

The aim of this work is to embed the axion inflation in the gauge extensions of the SM when the inflaton is directly coupled to the BSM fields. Here we consider the model proposed in \cite{Maleknejad:2020yys}, i.e. SU(2)$_{R}$-axion inflation, in which the axion-inflation is embedded in the minimal Left-Right Symmetric Model (LRSM). The minimal gauge group that implements the hypothesis of left-right symmetry is \cite{Pati:1974y, Mohapatra:1974gc, Senjanovic:1975rk,Davidson:1978pm}
\bea
\mathcal{G} \equiv SU(2)_L\times SU(2)_R \times U(1)_{\mB-\mL},
\eea
where the color ($SU(3)$ group) is suppressed. The subscripts $L$ and $R$ denotes left- and right-handed fields, while $\mB$ and $\mL$ represent baryon and lepton numbers respectively. The LRSM includes three gauge fields; $\bF_{L,R}$ are associated with $SU(2)_{L,R}$ and $B_{\mu}$ corresponds to
 $U(1)_{\mB-\mL}$. The fermionic content is consists of the SM quarks and leptons extended by three RHNs as
\bea
q_{iL,R} = ~\begin{pmatrix}
{\boldsymbol{u}}_{i} \\ {\boldsymbol{d}}_{i}
\end{pmatrix}_{\!L,R} \an 
l_{iL,R} = ~\begin{pmatrix}
{\bn}_{i} \\ {\boldsymbol{l}}_{i}
\end{pmatrix}_{\!L,R},
\eea
where $\bn_{iR}$ are three RHNs. The right-handed fermions interact with $\bW_R$ gauge field which is  $SU(2)_R$-valued, and is given as
\bea
\bF_{R} = W^a_{R} \bT_a \an [\bT_a,\bT_b] = i \epsilon_{abc} \bT_c.
\eea
The extended Higgs sector of the model consists of a Higgs bi-doublet $\bPhi$, and $SU(2)_{L,R}$ triplets $\bD_{L,R}$. The Spontaneous Symmetry Breaking (SSB) structure is
\bea
SU(2)_L\times SU(2)_R \times U(1)_{\mB-\mL} \xrightarrow[\text{1st SSB}]{T<\Lambda_F}  SU(2)_L \times U(1)_{Y} \xrightarrow[\text{2nd SSS}]{T<\Lambda_{W}} U(1)_{em}.\nonumber
\eea
The first SSB occurs at $T=\Lambda_F$ which breaks the left-right symmetry and gives a VEV to the $SU(2)_R$ triplet, i.e. $\langle \bD_R \rangle \neq 0$. That gives mass to $W^{\pm}_R$, $Z_R$, and provides Majorana masses for $\bN_i\equiv \bn_i+\bn_i^c$. Next, when the temperature gets below EW scale, $T<\Lambda_{W}$, the second SSB happens, and the Higgs bi-doublet acquires a VEV, i.e., $\langle \bPhi \rangle \neq 0$. It gives Dirac masses to the SM particles, active neutrinos included. 
In the minimal LRSM, the origin of the mass for the SM neutrino is a hybrid (I+II) seesaw mechanism \cite{Mohapatra:1980yp}.
For an overview on LRSM, see Appendix \ref{LRSM-overview} and the references therein.

\subsection{SU(2)$_R$-Axion Inflation}

Cosmic inflation is given by 
 Friedmann-Lemaitre-Robertson-Walker (FRLW) metric
\bea
ds^2=-dt^2+a^2(t) \delta_{ij} dx^idx^j,
\eea
in which the Hubble parameter is almost constant $H(t)\simeq H$, and the scale factor is  $a(t) \simeq e^{Ht}$. As for the inflaton field we consider an axion field $\varphi$ which is coupled to the $\bW_R$ gauge field in the LRSM as \cite{Maleknejad:2020yys} 
\bea\label{setup}
\mathcal{L}_{Inf} = -\frac12 \p_{\mu} \varphi^2 - V(\varphi) - \frac12 {\rm{Tr}}[\bF_{R\mu\nu} \bF_R^{\mu\nu}] - \frac{\lambda \varphi}{f} {\rm{Tr}}[\bF_{R\mu\nu} \tilde\bF_R^{\mu\nu}],
\eea
where $\lambda\lesssim 1$ is a dimensionless parameter, $f\lesssim 10^{-1} \mpl$ is the axion decay constant, $\bF^{\mu\nu}_R$ is the strength tensor of $\bW_R^\mu$ as
\bea
\bW_{R\mu\nu} \equiv \p_{\mu} \bW_{R\nu} - \p_{\nu} \bW_{R\mu} - i g_R [ \bW_{R\mu}, \bW_{R\nu}],
\eea
and $\tilde\bF^{\mu\nu}_R\equiv\frac12\frac{\epsilon^{\mu\nu\lambda\sigma}}{\sqrt{-g}} \bF_{R\lambda\sigma}$. For the sake of generality, we assume $V (\varphi)$ is an arbitrary axion potential, flat enough to support the slow-roll inflation. This SU(2)-axion inflation model and its cosmic perturbations, for a generic dark $SU(2)$ field, has been introduced and studied in \cite{Maleknejad:2016qjz} (See also \cite{Maleknejad:2016dci, Caldwell:2017chz}). One of
the most popular and well-motivated axion models of inflation to provide the flat potential is the axion monodromy. While the underlying periodicity of the theory continues to protect the inflaton potential from corrections, here the periodic field space of the axion is effectively unfolded due to the monodromy \cite{Silverstein:2008sg,Flauger:2009ab, McAllister:2014mpa}.

In this setup, we have two unknown high energy scales, i.e., the scale of inflation $\Lambda_{inf}=\sqrt{\mpl H}$, and LR symmetry breaking scale $\Lambda_F$. Besides, the $SU(2)_R$ may or may not acquire a VEV. Therefore, we can distinguish four different types of scenarios, which are classified in Table \ref{scenario}. Scenario $\rm{\bf{I}}$ and $\rm{\bf{I}}_{v}$ describe the case $\Lambda_{inf}>\Lambda_{F}$, while $\rm{\bf{II}}$ and $\rm{\bf{II}}_{v}$ otherwise, i.e. $\Lambda_{ inf}<\Lambda_{F}$. Moreover, the $\rm{v}$ subscript denotes systems in which the $SU(2)_R$ acquires a VEV in inflation. In scenarios $\rm{\bf{II}}$ and $\rm{\bf{II}}_{v}$, the $\bW_{R}$ is massive in inflation. \footnote{For different but related models based on massive $SU(2)$ fields coupled to an $SU(2)$-doublet see \cite{Adshead:2016omu,Adshead:2017hnc}.} In this work we solely focus on types ${\bf{I}}$ and ${\bf{II}}$ and leave ${\bf{I_v}}$ and ${\bf{II_v}}$ for future work.

\begin{table}[H]
\begin{center}
\begin{tabular}
{|c|c|c|}
\hline
\rule{0pt}{15pt}
 & $\Lambda_{inf} > \Lambda_{F}$  & $\Lambda_{inf} < \Lambda_{F}$ \\ [1.8ex]
 \hline
 \rule{0pt}{15pt}
$\langle \bF_{R} \rangle =0$ & {\bf{I}} & {\bf{II}} \\[1.8ex]
\hline
\rule{0pt}{15pt}
$\langle \bF_{R} \rangle \neq0$ & ${\bf{I_{v}}}$ & ${\bf{II}_{v}}$ \\[1.8ex]
\hline
\end{tabular}
\end{center}
\caption{Different scenarios of SU(2)$_{R}$-axion inflation. Based on the scale of inflation $\Lambda_{inf}=\sqrt{\mpl H}$, scale of left-right symmetry breaking $\Lambda_{F}$, and the (possible) $SU(2)_R$ field's VEV in inflation, one can separate four different types of scenarios.}
\label{scenario}
\end{table}

\subsubsection*{Right-handed fermions in SU(2)$_{R}$-axion inflation:}

The $U(1)_{\mB-\mL}$ and $SU(2)_L$ gauge fields and left-handed fermions in inflation have conformal symmetry, hence are negligible in physics of inflation. The $\bF_R$ gauge field, however, is coupled to the axion which breaks its conformal symmetry and sources it in inflation. The right-handed fermions are coupled to $\bW_R$ as (see Eq. \eqref{L-Psi}) \footnote{Notice that  $U(1)_{\mB-\mL}$ is decaying and unimportant in inflation, and it is neglected here.}
\bea
\mathcal{L} \supset \sum_{i} \bar{q}_{iR}\big(i \bs^{\mu} D_{\mu} - i\gR \bs^{\mu} \bF_{R\mu} \big) q_{iR} + \bar{l}_{iR}\big(i \bs^{\mu} D_{\mu} - i\gR \bs^{\mu} \bF_{R\mu} \big) l_{iR},
\eea
where $D_{\mu}$ is the spinor covariant  derivative. 
The generated $\bW_R$ gauge field, hence, produces right-handed quarks and leptons in inflation.
Finally, the right-handed fermions can have an effective interaction with the axions as
\bea
\mathcal{L}_{5} = \frac{\tilde\lambda\varphi}{f} \nabla_{\mu} J^{\mu}_R,
\eea
where $\tilde\lambda$ is a constant of the order of $\lambda$, and the right-handed current is
\bea\label{JR-JL}
J^{\mu}_{R}= \sum_i \bar{q}_{iR} \bs^{\mu} q_{iR} + \bar{l}_{iR} \bs^{\mu} l_{iR}.
\eea
 There are two source terms for the fermions, i.e. the $SU(2)_R$ gauge field and its axion. However, the axion cannot generate Weyl fermions. The reason is that a Peccei-Quinn type $U_{PQ}(1)$ rotation of fermions as \cite{Peccei:1977hh} 
\bea
\Psi_R \rightarrow e^{-\frac{i\tilde\lambda}{f}\varphi} \Psi_R,
\eea
removes the axion interaction and transforms the fermion mass matrix as \cite{Weinberg:1996kr}
\bea
\mathcal{M} \rightarrow e^{\frac{2i\tilde\lambda}{f}\varphi} \mathcal{M}.
\eea
Therefore, the axion only contributes to the generation of massive fermions in inflation.

\section{Inflationary Particle Production}
\label{inf-PP}

In this inflation model, $\bF_{R}$ gauge field is generated by the axion. The field equation of $\bF_{R\mu}$ in the massless case (type-{\bf{I}} scenarios) is
\bea
(\nabla_{\mu} -i\gR \bF_{R\mu}) [\bF^{\mu\nu}_R + \frac{\lambda\varphi}{f} \tilde{\bF}^{\mu\nu}_R ] = 0,
\eea
and in the massive case (type-{\bf{I}} scenarios)  $W_{R}^{\pm}$ and $Z^0_R$ acquire $m_{W_R}$ and $m_{Z_R}$ respectively. Moreover, apart from the exponential expansion of the Universe, both axion and $\bF_{R}$ gauge field are active in inflation and produce right-handed quarks and leptons. The $P$ and $C$ are maximally broken by the chiral nature of the $SU(2)_R$ interaction, and $CP$ is violated by the Chern-Simons interaction. Both right-handed baryon and lepton numbers are violated by the non-perturbative effects of the $\bW_{R}$, i.e. chiral (Adler-Bell-Jackiw) anomaly \cite{Adler:1969gk,Bell:1969ts}.  The sterile neutrinos are massless (massive with mass $m_{\bN_i}$) in scenario type-{\bf{I}} (type-{\bf{II}}). That gives the right-handed baryons and leptons the following anomalies
\bea
\nabla_{\mu} J^{\mu R}_{\mB} &=&  - \frac{\gR^2 \mN_{R}}{16\pi^2} {\rm{Tr}}[\bF^{\mu\nu}_R\tilde{\bF}_{R\mu\nu}],\\
\nabla_{\mu} J^{\mu R}_{\mL} &=&  - \frac{\gR^2 \mN_{R}}{16\pi^2} {\rm{Tr}}[\bF^{\mu\nu}_R\tilde{\bF}_{R\mu\nu}] + 2im_{\rm{N}_i} \bar{\bn}_{iR} \bn_{iR},
\eea
where $J^{\mu R}_{\mB,\mL}$ is the baryon and lepton number densities, and $\mN_R$ is the number of right-handed fermion generations. Note that the $\mB$ and $\mL$ violating interactions of the left-handed fermions remains negligible in inflation. The Chern-Simons term can be written as a total derivative
\bea
\sqrt{-g}{\rm{Tr}}[\bF^{\mu\nu}_R\tilde{\bF}_{R\mu\nu}] = 2\p_{\mu}\big( \sqrt{-g} K^{\mu}\big),
\eea
where $K_{\mu}$ is the Chern-Simons current, i.e.
\bea\label{CS-current}
K^{\mu} = \epsilon^{\mu\nu\lambda\sigma} {\rm{Tr}}[\bF_{R\nu}\p_{\lambda}\bF_{R\sigma} - \frac{ 2i\gR}{3}  \bF_{R\nu}\bF_{R\lambda}\bF_{R\sigma}].
\eea
In our setup $\mN_{L}=\mN_{R}=3$, hence we neglect the effect of global gravitational anomaly. \footnote{In the context of Einstein gravity and with SM fermions, i.e.  $\mN_{L}-\mN_{R}=3$, the effect of global gravitational leptogenesis in the minimal $SU(2)$-axion model \cite{Maleknejad:2016qjz} is studied in \cite{Maleknejad:2016dci, Caldwell:2017chz}.} The total baryon and lepton numbers are related to their corresponding quantities in SM as
\bea
n_{\mB}=n_{\mB_{\rm SM}} \an n_{\mL}=n_{\mL_{\rm SM}} + \sum_{i} n_{\rm{N}_{i}},
\eea
in which $n_{\mL_{\rm SM}}$ and $n_{\rm{N}_{i}}$ are the contributions of the SM leptons and the $i$th RHN in the total lepton number respectively.
In this section, $g_{_{L,R}}$ are the gauge couplings at the scale of inflation which are computed in App. \ref{RG}. For a high scale inflation, e.g. around $H\sim 10^{14}$ GeV, we have 
\bea
\gL(H) \simeq 0.56 \an   0.3 \leq  \gL(H) \leq 0.56.
\eea
The details of the discussion depend on whether the first SSB happens before or after inflation. Therefore, these two cases will be treated separately. Before going any further, let us fix the notations that will be used in this section. 
For later convenience, we define $\mH$ as
\bea
\mH \equiv aH,
\eea
which in terms of conformal time, i.e. $dt=ad\tau$, and during slow-roll is $
\mH \simeq - \frac{1}{\tau}$. The rescaled physical momentum is defined as
\bea
\x \equiv \frac{k}{aH} \simeq -k\tau.
\eea
Finally, we define dimensionless parameters $\xi$ and $\tilde\xi$ as
\bea
\xi\equiv \frac{\lambda\dot\varphi}{2fH} \an \tilde\xi\equiv \frac{\tilde\lambda}{\lambda}\xi.
\eea
During slow-roll $\xi$ has slow-roll dynamics, and is related to the slow-roll parameter $\epsilon\equiv -\frac{\dot H}{H^2}$ as
\bea
\xi(t) \sim \frac{1}{\sqrt{2}}\frac{\lambda\mpl}{f H} \sqrt{\epsilon(t)}.
\eea
As a result, $\xi$ gradually increases with time.

\subsection{Scenario Type-I}\label{Sec-T1}

In scenario type-{\bf{I}}, the $\bF_R$ and right-handed fermions are all massless in inflation. Here we first study the gauge field production by the axion. Next, we turn to the fermion production by the gauge field.

\subsubsection*{Massless SU(2)$_R$ gauge bosons:}

\begin{figure}[htb]
  \centering\includegraphics[height=0.2\textheight]{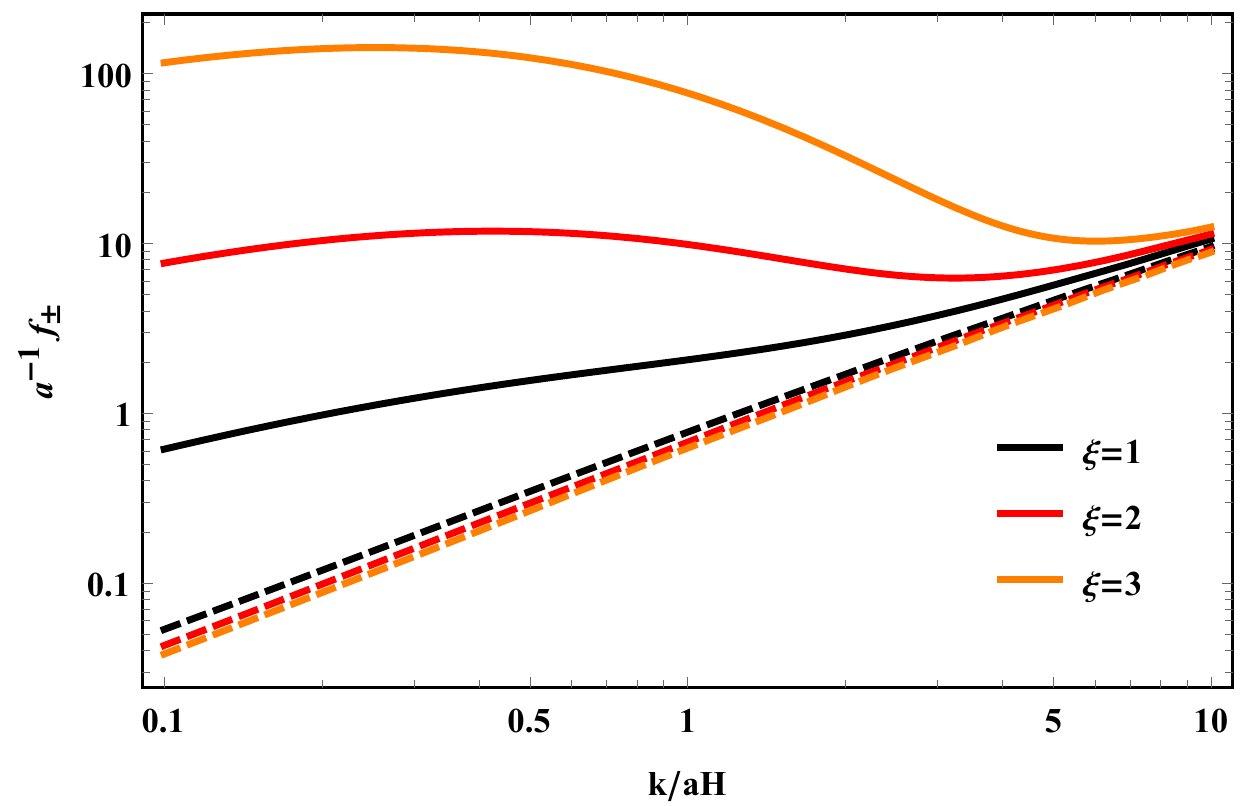} \includegraphics[height=0.2\textheight]{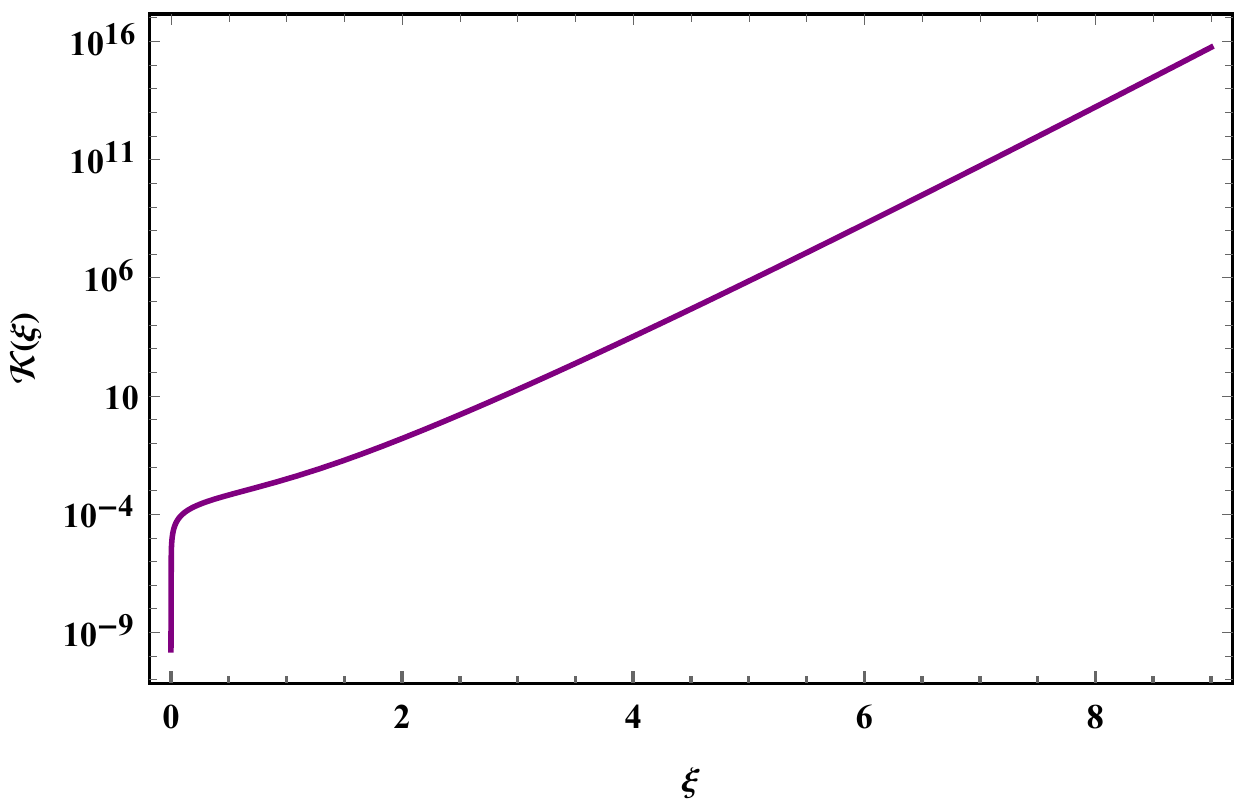}\\
   \caption{The $SU(2)_R$ gauge boson and right-handed fermion production in type-{\bf{I}} scenarios. Left panel: The polarization states of the $SU(2)_R$ field vs $\frac{k}{aH}$ for different values of $\xi$. The plus and minus helecity states are presented with solid and dashed lines respectively. Right panel: the $\mathcal{K}(\xi)$ parameter which quantify the value of fermion production in inflation.}
  \label{fig:gauge-field-massless} 
\end{figure}

The linearized field equation of $SU(2)_R$ is
\bea
\p_{\tau}^2 \bF_{Ri} -  \p_j^2\bF_{Ri} + a^2 \p_{i}( \nabla_{\mu} \bF^{\mu}_R) + 2a \mH\p_i \bF_{R0} - \frac{\lambda\dot\varphi}{fH} \epsilon^{ijk} \mH  \p_j\bF_{Rk}  \simeq 0,\nonumber\\
\eea
and a constraint equation $\nabla_{\mu} \bF^{\mu}_R =0$.  The (charged) gauge field has two degrees of freedoms and can be decomposed in terms of its two transverse modes as
\bea
\bF_{Ri}(\tau, \bx) = \sum_{\sigma=\pm 1} \int d^3k \bigg( a_{\bk,\sigma} \bff_{\sigma}(\tau,\bk) e_{\sigma i} (\bk) e^{i\bk.\bx} + b^{\dag}_{\bk,\sigma} \bff^{*}_{\sigma}(\tau,\bk) e^{*}_{\sigma i}(\bk)  e^{-i\bk.\bx}\bigg),
\eea
where $a_{\bk,\sigma}$ ($b_{\bk,\sigma}$) is the annihilation operator of the particle (anti-particle), and $\bfe_{\pm}(\bk)$ are $\pm 1$ helicity  polarization vectors, defined as
\bea\label{epm}
\bfe_{\pm}(\bk) \equiv  \frac{1}{\sqrt{2}}({\boldsymbol{\hat\theta}} \mp i {\boldsymbol{\hat{\phi}}}),
\eea
where $\boldsymbol{\hat{r}}=-\hat{\bk}$, $\boldsymbol{\hat{\theta}}$ and $\boldsymbol{\hat{\phi}}$ are the local orthogonal unit vectors in the directions of increasing $r$, $\theta$, and $\phi$. The polarization vectors satisfy in the following equations
\bea
\bk.\bfe_{\pm}(\bk)=0 \an \bk \times \bfe_{\pm}(\bk) = \mp i k~\bfe_{\pm}(\bk).
\eea
The function $\bff_{\sigma}(\bk)$ can be expanded in terms of the mode functions as
\bea
\bff_{\sigma}(\tau, \bk)=f^a_{~\sigma}(\tau,\bk)\bT_a,
\eea
which are governed by the field equations below
\bea\label{FEF}
\p_{\tau}^2 f^a_{~\pm} +(k^2 \mp 2 k\mH \xi  ) f^a_{~\pm} \simeq 0.
\eea
The field equation can be written as a Whittaker equation with parameters
\bea
\kappa_{\pm} = \mp i\xi \an \mu^2 = \frac14.
\eea
 Imposing the Bunch-Davies vacuum condition in the asymptotic past, we have the mode functions as  \cite{Maleknejad:2016qjz}
 \bea\label{mode-gf-massless}
 f^a_{~\pm}(\tau,\bk) = \frac{e^{i\kappa_{\pm}\pi/2}}{(2\pi)^{\frac32}\sqrt{2k}} W_{\kappa_{\pm},\mu}(2ik\tau).
 \eea
Therefore one helicity state of the gauge field (plus/minus for positive/negative $\xi$) has a short period of tachyonic growth (See the left panel of Fig. \ref{fig:gauge-field-massless}). \footnote{
It leads to particle production and backreaction to the background \cite{Maleknejad:2018nxz}.}

\subsubsection*{Massless right-handed fermions:}

In scenario type-${\bf{I}}$, all the fermions are massless in inflation. Therefore, the right-handed fermions are generated by the $SU(2)_R$ gauge field as
\bea
\nabla_{\mu} J^{\mu R}_{\mB,\mL} =  - \frac{\gR^2 \mN_{R}}{16\pi^2} {\rm{Tr}}[\bF^{\mu\nu}_R\tilde{\bF}_{R\mu\nu}],
\eea
which implies that the produced right-handed fermions are given by the Chern-Simons current \eqref{CS-current}. After linearization, the Chern-Simons charge is given as
\bea\label{ncS}
n_{\rm{CS}} \equiv \int d^3k K^0 \simeq \frac{1}{a^3} \int d^3k ~\epsilon^{ijk}~\langle {\rm{Tr}}\big[ \bF_i \p_j \bF_k \big]_R\rangle.
\eea
For later convenience, we define $\mathcal{K}(\xi)$ as
\bea
\mathcal{K}(\xi) \equiv  \frac{9}{4(2\pi)^4} \sum_{\sigma=\pm}  \sigma ~ e^{i\kappa_{\sigma}\pi} \int \x^3 d\ln\x  W^{*}_{\kappa_{\sigma},\mu}(-2i\x)W_{\kappa_{\sigma},\mu}(-2i\x),
\eea
with $\kappa_{\pm}= \mp i\xi$ and $\mu=\frac12$. 
Using solution \eqref{mode-gf-massless}, we can write $n_{\rm{CS}}$ as
\bea
n_{\rm{CS}} \simeq \frac{8\pi^2}{3}  H^3 \mathcal{K}(\xi),
\eea
That gives the lepton and baryon number densities as
\bea\label{B-L-massless}   
n_{\mB}=n_{\mL} \simeq -\gR^2 H^3 \mathcal{K}(\xi).
\eea
The number density of $\rm{N}_i$ is 
\bea\label{nNi}
n_{\rm{N}_i}= - \frac16 \gR^2 H^3 \mathcal{K}(\xi).
\eea
The right panel of Fig \ref{fig:gauge-field-massless} shows $\mathcal{K}(\xi)$ vs $\xi$. It increases exponentially with the increase of $\xi$ as
\bea
\mathcal{K}(\xi) \propto \frac{1}{(2\pi)^4} e^{2\xi\pi}.
\eea

\subsection{Scenario Type-II}\label{Sec-T2}

In scenario type-{\bf{II}}, the gauge symmetry $SU(2)_R \times U(1)_{\mB-\mL}$ breaks to $U(1)_Y$. It gives masses to the gauge boson, charged and neutral, i.e.
\bea
W^{\pm}_R = \frac{1}{\sqrt{2}} ( W^1_R \mp i W^2_R) \an Z^0_R = (\gR W^3_R - \gBL B)/\sqrt{\gR^2+\gBL^2},
\eea
as well as at least two of the right-handed fermions. Here we first study the gauge field production by the axion. Next, we turn to the fermion production by the gauge field and the axion.

\subsubsection*{Massive SU(2)$_R$ Gauge Bosons:}

The linearized field equation of $W^{\pm}_R$ is 
\bea
&&\p_{\tau}^2 W^{\pm}_{Ri} -  \p_j^2 W^{\pm}_{Ri} + a^2 \p_{i}( \nabla_{\mu} W^{\pm\mu}_R) + 2a \mH\p_i W^{\pm}_{R0} - \frac{\lambda\dot\varphi}{fH} \epsilon^{ijk} \mH  \p_j W^{\pm}_{Rk} +  \frac{m^2_{_{W_R}}}{H^2} \mH^2 W^{\pm}_{Ri} \simeq 0,\nonumber\\
\eea
with a constraint equation 
\bea\label{constraint}
\nabla_{\mu} W^{\pm\mu}_R =0.
\eea
The neutral component $Z^0_R$, satisfies the same equations with $m_{_{W_R}}$ replaced by $m_{_{Z_R}}$. 
Since the gauge field is massive, in addition to the two transverse modes with polarization vectors $\bfe^{\pm}(\bk)$, there is another dynamical degree of freedom associated with $k_i\bF_R^i(\bk)$. For ease of notation, we define 
\bea
W^{\alpha}_R \equiv (W^{+}_R, W^{-}_R, Z^0_R).
\eea
The gauge field in the massive case is given as
\bea
W^{\alpha}_{Ri}(\tau, \bx) = \sum_{\sigma=1}^{3} \int d^3k \bigg( a_{\bk,\sigma} f^{\alpha}_{\sigma}(\tau,\bk) e_{\sigma i}(\bk) e^{i\bk.\bx} + b^{\dag}_{\bk,\sigma} f^{*\alpha}_{\sigma}(\tau,\bk) e^{*}_{\sigma i}(\bk)  e^{-i\bk.\bx}\bigg),
\eea
where the polarization states are defined as
\bea
\bfe_{1,2}(\bk) \equiv \bfe_{\pm}(\bk) \an \bfe_{3}(\bk) \equiv \hat{\bk}.
\eea
Note that superscript $\pm$ denotes the charged of the field and subscript $\pm$ represents its helicity state.  
In addition to these dynamical fields, massive gauge field has $W_{R0}^{\alpha}$ which is non-dynamical, specified by the constraint Eq. \eqref{constraint}
\bea\label{constriant-1}
\p_{\tau} W^{\alpha}_{R0}  + 3\mH W^{\alpha}_{R0} - \frac{1}{a} \p_i W^{\alpha}_{Ri} =0.
\eea
Since it is only coupled to the longitudinal mode, it can be expanded as
\bea
W^{\alpha}_{R0}(\tau, \bx) =  \frac1a \int d^3k \bigg( a_{\bk,3} f^{\alpha}_{0}(\tau,\bk)  e^{i\bk.\bx} + b^{\dag}_{\bk,3} f^{*\alpha}_{0}(\tau,\bk)  e^{-i\bk.\bx}\bigg).
\eea
The field equation of the transverse modes with $\sigma=1,2 $ (plus and minus helicity states) are given as
\bea\label{FEF-m}
\p_{\tau}^2 f^{\alpha}_{~\pm} +(k^2 \mp 2  k\mH \xi + \frac{m_{_{W_R}}^2}{H^2} \mH^2 ) f^{\alpha}_{~\pm} \simeq 0.
\eea
The field equation of the longitudinal mode with $\sigma=3$ is given as
\bea\label{FEF-L}
\p_{\tau}^2 f^{\alpha}_{~3} + (k^2 + \frac{m_{_{W_R}}^2}{H^2} \mH^2 ) f^{\alpha}_{~3} + 2ik\mH f^{\alpha}_0 \simeq 0,
\eea
which is coupled to the $W_{R0}^{\alpha}$ is given by the constraint Eq \eqref{constriant-1} as
\bea
\p_{\tau}f^{\alpha}_0  + 2\mH f^{\alpha}_0 - ik f^{\alpha}_3 =0.
\eea

\begin{figure}[htb]
  \centering\includegraphics[height=0.2\textheight]{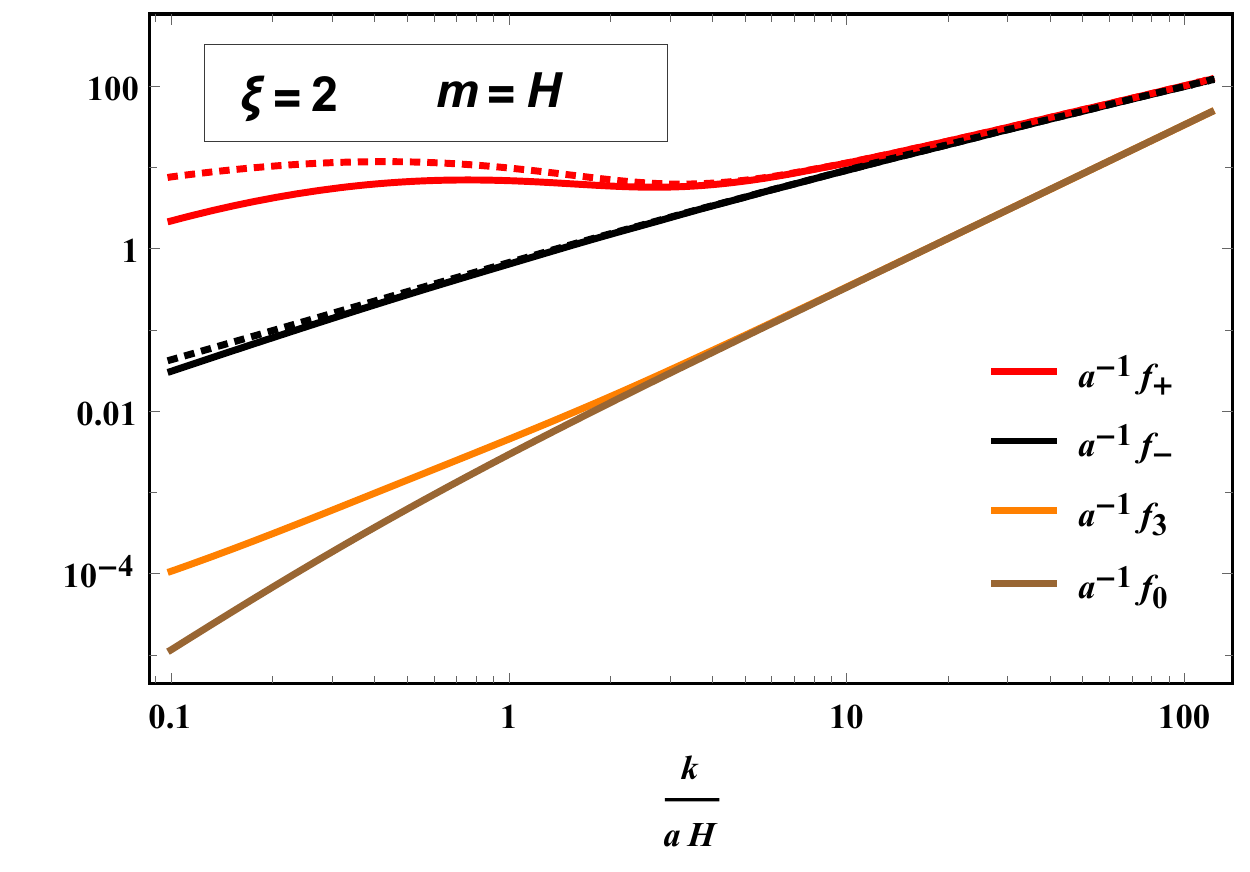} \includegraphics[height=0.2\textheight]{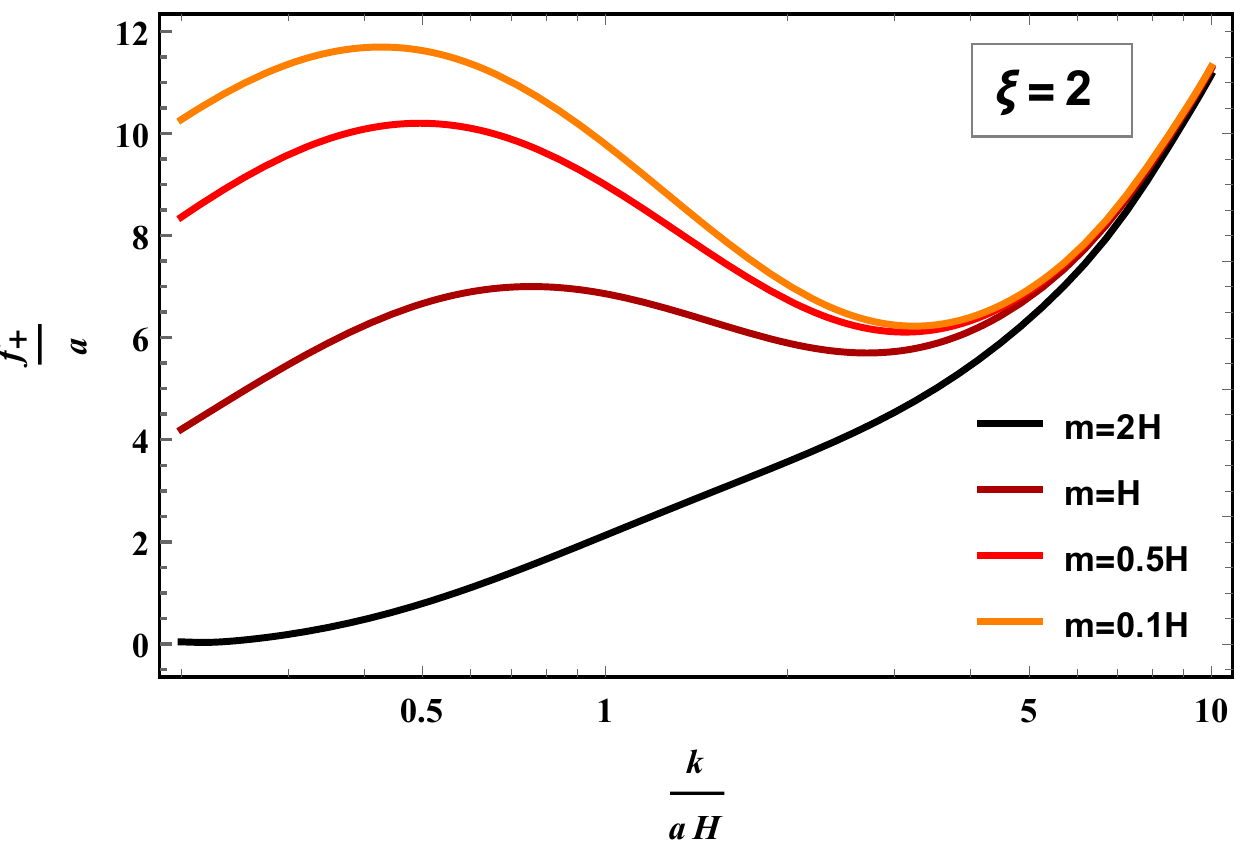}\\
   \caption{The polarization states of massive gauge field vs $\frac{k}{aH}$ for $\xi=2$. Left panel shows the four components of the massive field ($\frac{f_{+}}{a}$, $\frac{f_{-}}{a}$,$\frac{f_{3}}{a}$,$\frac{f_{0}}{a}$) with $m=H$. The dotted lines shows the corresponding modes in the massless case. Right panel shows the enhanced mode $f_{+}$ vs $\frac{k}{a}$ for different values of mass.}
  \label{fig:gauge-field-massive} 
\end{figure}

Like the massless case, the field equation of the transverse modes can be written as a Whitaker equation with parameters 
\bea
\kappa_{\pm} = \mp i\xi \an \mu^2_{\alpha} = \frac14 - \frac{m_{\alpha}^2}{H^2}.
\eea
Imposing the Bunch-Davies vacuum condition in the asymptotic past, we have
 \bea\label{fpm-a}
 f^{\alpha}_{~\pm}(\bk,\tau) = \frac{e^{i\kappa_{\pm}\pi/2}}{(2\pi)^{\frac32}\sqrt{2k}} W_{\kappa_{\pm},\mu_{\alpha}}(2ik\tau).
 \eea
Since the longitudinal mode $f_3^{\alpha}$ and hence $f^{\alpha}_0$ are not coupled to the axion, they are strictly decaying and unimportant in inflation (see left panel of Fig. \ref{fig:gauge-field-massive}). Therefore, similar to the type-{\bf{I}} case, the cosmological relevant modes in type-{\bf{II}} scenarios are the transverse modes as well. Again $f_{+}$ polarization mode is generated by the axion which is shown in the right panel of Fig. \ref{fig:gauge-field-massive}.

\subsubsection*{Massive right-handed neutrinos:}

\begin{figure}[h]
\includegraphics[height=0.2\textheight]{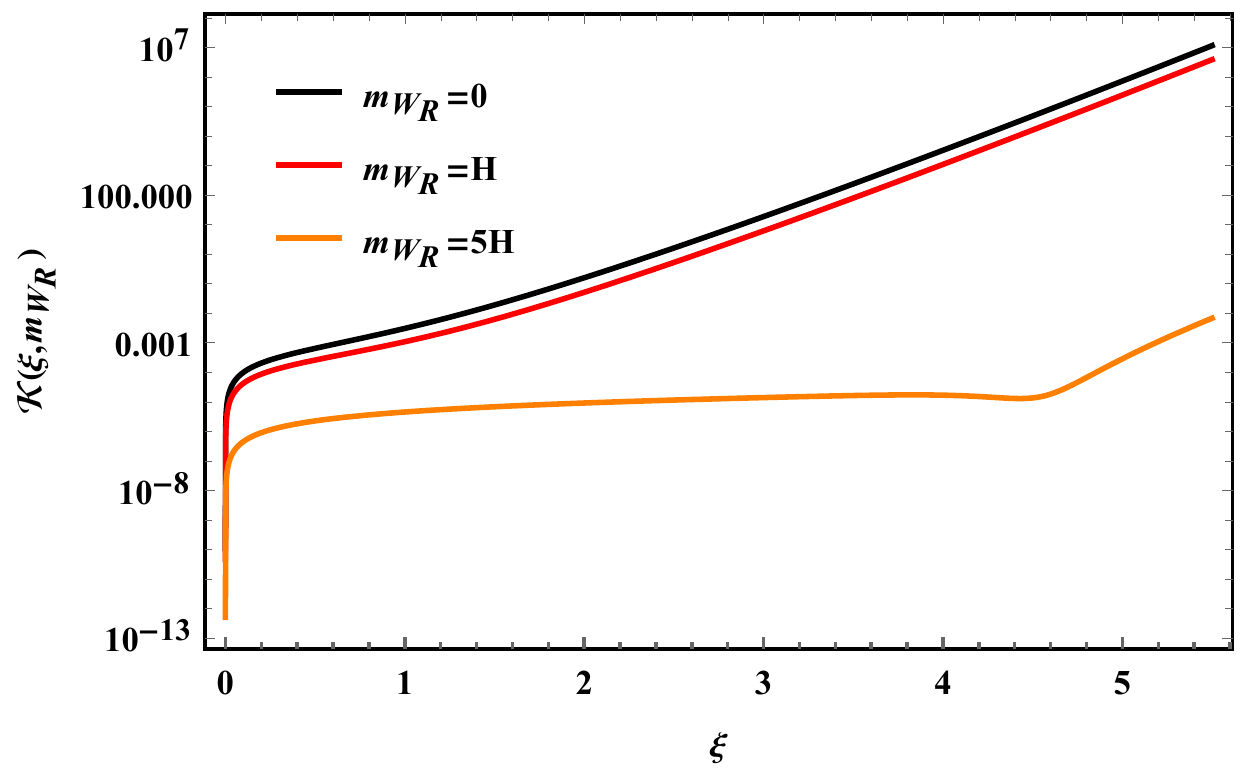} \includegraphics[height=0.2\textheight]{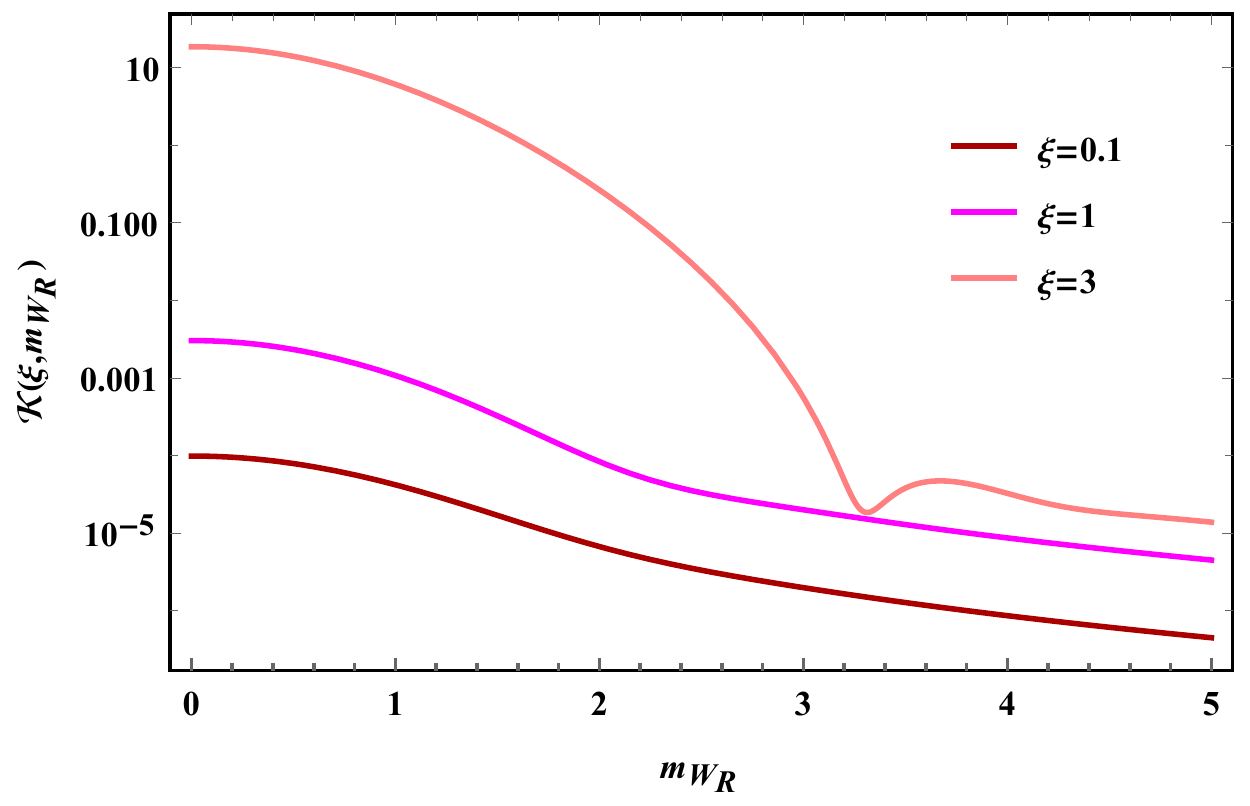}
\caption{The factor $\mathcal{K}(\xi,m_{_{W_R}})$ vs $\xi$ (left panel) and vs $m_{_{W_R}}$ (right panel).  }
\label{fig:cK} 
\end{figure}

In scenario type-${\bf{II}}$, the SM fermions are massless in inflation while at least two of the sterile neutrinos are massive. Therefore, we have
\bea
\nabla_{\mu} J^{\mu}_{\mB}  &=&  - \frac{3\gR^2}{16\pi^2} {\rm{Tr}}\big[\bF^{\mu\nu}\tilde{\bF}_{\mu\nu}\big]_R,\\
\nabla_{\mu} J^{\mu}_{\mL} &=& 2im_{\rm{N}_i} \bar{\bn}_{iR} \bn_{iR} - \frac{3\gR^2}{16\pi^2} {\rm{Tr}}\big[\bF^{\mu\nu}\tilde{\bF}_{\mu\nu}\big]_R.
\eea
The Chern-Simons charge given in Eq. \eqref{ncS} can be written as 
\bea
n_{\rm{CS}} \simeq \frac{8\pi^2}{9}  H^3 \big[ 2\mathcal{K}(\xi,m_{_{W_R}}) + \mathcal{K}(\xi,m_{_{Z_R}})  \big],
\eea
where $\mathcal{K}(\xi,m_{\alpha})$ is
\bea
\mathcal{K}(\xi,m_{\alpha}) \equiv  \frac{9}{4(2\pi)^4} \sum_{\sigma=\pm} \sigma e^{i\kappa_{\sigma}\pi} \int \x^3 d\ln\x W^{*}_{\kappa_{\sigma},\mu_{\alpha}}(-2i\x)W_{\kappa_{\sigma},\mu_{\alpha}}(-2i\x),
\eea
with $\kappa_{\pm}= \mp i\xi$ and $\mu^2_{\alpha}=\frac14-\frac{m_{\alpha}^2}{H^2}$. 
Therefore the baryon number density is given as
\bea\label{B-massive}
n_{\mB} \simeq - \frac13 \gR^2 H^3 \big[ 2\mathcal{K}(\xi,m_{_{W_R}}) + \mathcal{K}(\xi,m_{_{Z_R}})\big].
\eea
The prefactor $\mathcal{K}(\xi,m_{\alpha})$ with respect to $\xi$ and $m_{W_R}$ is shown in the left and right panels of Fig. \ref{fig:cK}. It increases (decreases) with the increase of $\xi$ ($m_{\alpha}$) and for $\xi>m_{
\alpha}$ is
\bea
\mathcal{K}(\xi,m_{\alpha}) \propto \frac{1}{(2\pi)^4} ~e^{2 \xi\pi }.
\eea

\begin{figure}[htb]
 \includegraphics[height=0.2\textheight]{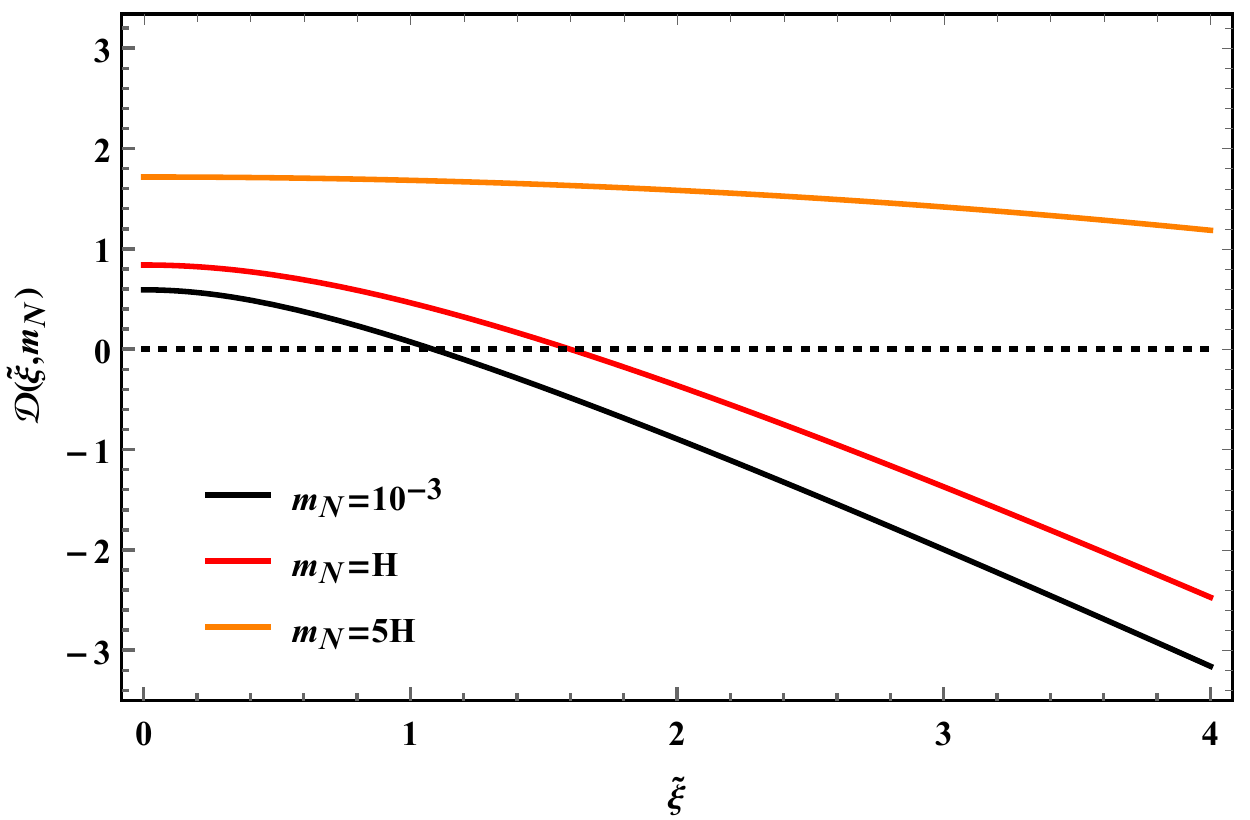}
 \includegraphics[height=0.2\textheight]{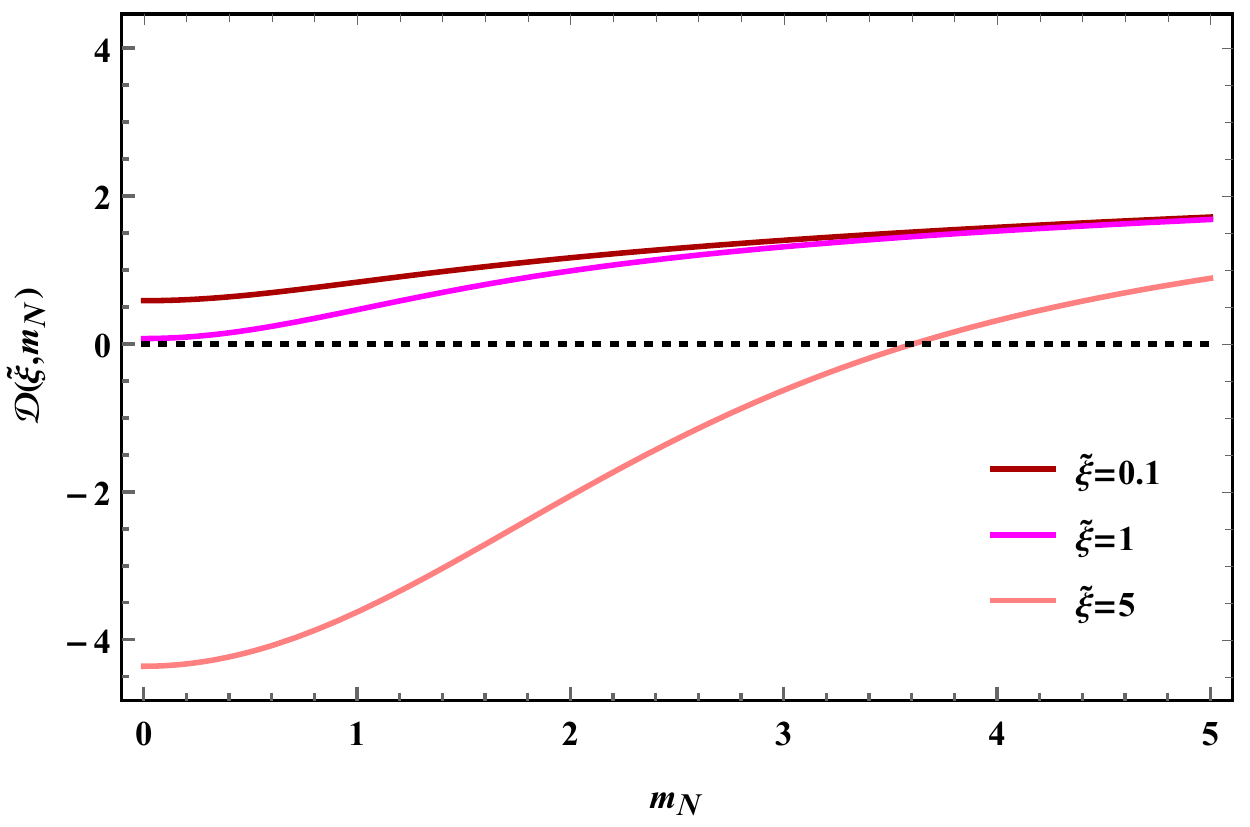}
\caption{The prefactor $\mathcal{D}(\tilde\xi,m_{\rm{N}_i})$ vs $\xi$ (left panel) and vs $m_{\rm{N}_i}$ (right panel). }
\label{fig:cD} 
\end{figure}

Due to the mass of the sterile neutrinos, the calculation of the lepton number is more involved and requires the mode functions. The leptonic field equations are
\bea \label{EOM-Lepton} 
 (i\bs^{\mu}\p_{\mu} + \frac{3i}{2} H+ \gR \bs^{\mu}\bF_{R\mu} -  \frac{\tilde\lambda \dot\varphi}{f}) l_{iR} - m_{\rm{N}_i} ~ \bn^c_{iR}=0,
\eea
\bea\label{Int-n-m}
\bar{n}_{\rm{N}_i} \equiv \int d^3k \langle \bn_{iR}^{\dag} \bn_{iR}\rangle = - H^3  \sum_{i} ~\frac{\tilde\xi}{\pi} \bigg(\frac{m_{\rm{N}_i}}{H}\bigg)^2 ~\mathcal{D}(\tilde\xi, m_{\rm{N}_i}), 
\eea
where the bar emphasises that, unlike chiral anomaly, it is a classical effect. Using the point-splitting regularization, we computed $\bn_{i,R}$ and $\mathcal{D}(\tilde\xi,m_{\rm{N}_i})$ analytically in App. \ref{massive-N}.
The exact form of $\mathcal{D}(\tilde\xi,m_{\rm{N}_i})$ is presented in Eq. \eqref{fin-J+---} and we show it in Fig. \ref{fig:cD}.
Here we discuss its qualitative behavior which in the large and small mass limits is
\bea
\mathcal{D}(\tilde\xi,m_{\rm{N}_i}) \simeq \begin{cases}
\frac{2}{\pi}~  \bigg[  \ln\big(\frac{m_{\rm{N}_i}}{H}\big) - \psi^{(0)}(1)  + \frac12 \bigg]  & \quad \textmd{for} \quad  \frac{m_{\rm{N}_i}}{H}\gg 1 ,\\
  -\frac{4}{3}  \lvert \tilde\xi \rvert  & \quad \textmd{for} \quad  \frac{m_{\rm{N}_i}}{H}\ll 1. 
\end{cases}
\eea
The $\bar{n}_{\rm{N}_i}$ is directly proportional to (and an odd function of) $\tilde\xi$ which is the (classical) source of particle production. Moreover, it increases with the mass of the sterile neutrinos, $m_{\rm{N}_i}$, which is the cause of their chiral symmetry breaking in inflation.   
Therefore, the total number density of RHNs are given as
\bea\label{nNi-massive}
n_{\rm{N}_{i}} = \bar{n}_{\rm{N}_{i}} - \frac{1}{6}~H^3 \gR^2 \big[ \frac23\mathcal{K}(\xi,m_{_{W_R}}) + \frac13\mathcal{K}(\xi,m_{_{Z_R}})\big].
\eea
The final total lepton number is
\bea\label{LN-massive-}
n_{\mL} \simeq - ~\bigg[ \frac{\gR^2}{3} \big[ 2\mathcal{K}(\xi,m_{_{W_R}}) +  \mathcal{K}(\xi,m_{_{Z_R}})\big] + \sum_{i} \frac{\tilde{\xi}}{\pi} \bigg(\frac{m_{\rm{N}_i}}{H}\bigg)^2 \mathcal{D}(\tilde\xi,m_{\rm{N}_i})\bigg] H^3.
\eea

\subsection{Baryon and Lepton Numbers}\label{B-L-N}

Here we summarize the main features of inflationary baryon and lepton generation.

\begin{figure}[htb]
  \centering\includegraphics[height=0.2\textheight]{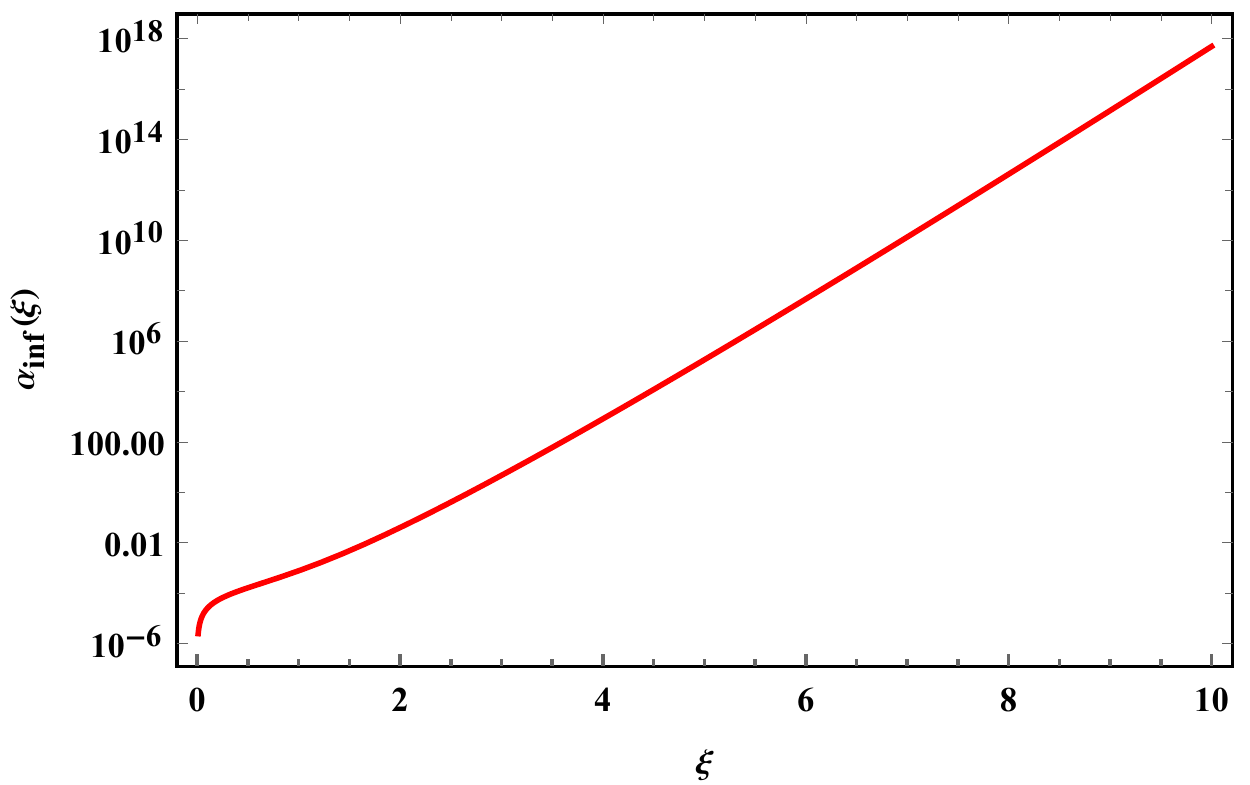}\\
   \caption{The $\alpha_{\rm inf}(\xi)$ vs $\xi$ in Eq. \eqref{alpha-inf-}.}
  \label{fig:alpha-inf-} 
\end{figure}

\begin{itemize}
\item The transverse modes of $\bW_R$ are generated by the axion that subsequently sources right-handed baryons and leptons.
\item All the Sakharov conditions required for BAU are satisfied in inflation: i) Out of thermal equilibrium condition holds during inflation, ii) $C$ is violated by the chiral nature of the $SU(2)_R$ interaction, iii) $\mB$, $\mL$, and $CP$ are violated by the non-perturbative effects of $\bW_R$.
\item $n_\mB$ and $n_\mL$ are the total baryon and lepton number densities respectively. $n_\mB$ and $n_{\mL_{\rm SM}}$ are the contributions of the SM fermions. The RHNs number density is $n_{\rm{N}}= n_\mL - n_{\mL_{\rm SM}}$.
\item The $\mB-\mL$ is conserved (violated) in scenario type-{\bf{I}} (type-{\bf{II}} ). However, $\mB-\mL_{\rm SM}= \mL_{\rm{N}}$ is violated in both scenarios. 
\item In scenario type-{\bf{I}}, the baryon and lepton numbers are both generated by the chiral anomaly of $\bF_{R}$ in inflation, i.e. $n_{\mB}=n_{\mL} \simeq  - \frac{3\gR^2}{8\pi^2} n_{\rm CS}$ (Eq. \eqref{B-L-massless}). It can be written as
\bea
n_{\rm{B}} = \alpha_{\rm inf}(\xi) H^3,
\eea
where $ \alpha_{\rm inf}(\xi)$ is given as (see Fig. \ref{fig:alpha-inf-})
\bea\label{alpha-inf-}
\alpha_{\rm inf}(\xi) \simeq -\gR^2 \mathcal{K}(\xi).
\eea
\item In scenario type-{\bf{II}}, the baryon number is specified entirely by the chiral anomaly of $\bW_R$, i.e. $n_{\mB}\simeq  - \frac{3\gR^2}{8\pi^2} n_{\rm CS}$ (Eq. \eqref{B-massive}). In the leptonic sector, however, the massive RHNs are also generated by the axion. Therefore, the total lepton number is $ n_\mL\simeq n_\mB + \bar{n}_{\rm{N}} $ where $\bar{n}_{\rm{N}}$ is the RHNs produced by the axion (Eq. \eqref{Int-n-m}).
\item The number density of the RHNs generated in inflation is
\bea
n_{\rm{N}_i} = \frac13 \tilde\alpha_{\rm inf}(\xi, m_{\rm{N}_i}) H^3,
\eea
where $\alpha_{\rm inf}$ for scenarios type-{\bf{I}} (Eq. \eqref{nNi}) and {\bf{II}} (Eq. \eqref{nNi-massive}) are given as 
\bea\label{alpha-inf}
\tilde\alpha_{\rm inf}(\xi, m_{\rm{N}_i}) \simeq \bigg\{%
\begin{array}{ll}
- \frac12 \gR^2 \mathcal{K}(\xi) & ~~\textrm{Type-\bf{I}},\\
- \bigg( \frac12 \gR^2 \big[\frac23 \mathcal{K}(\xi, m_{_{W_R}}) + \frac13 \mathcal{K}(\xi, m_{W_Z})\big] + \frac{3\tilde\xi}{\pi} \big(\frac{m_{\rm{N}_i}}{H}\big)^2 \mathcal{D}(\tilde\xi,m_{\rm{N}_i}) \bigg) & ~~\textrm{Type-\bf{II}}.
\end{array} \nonumber\\
\eea
\end{itemize}

\section{Post Reheating Evolution}
\label{Post-flation}
\begin{figure}[h]  
  \centering\includegraphics[height=0.085\textheight]{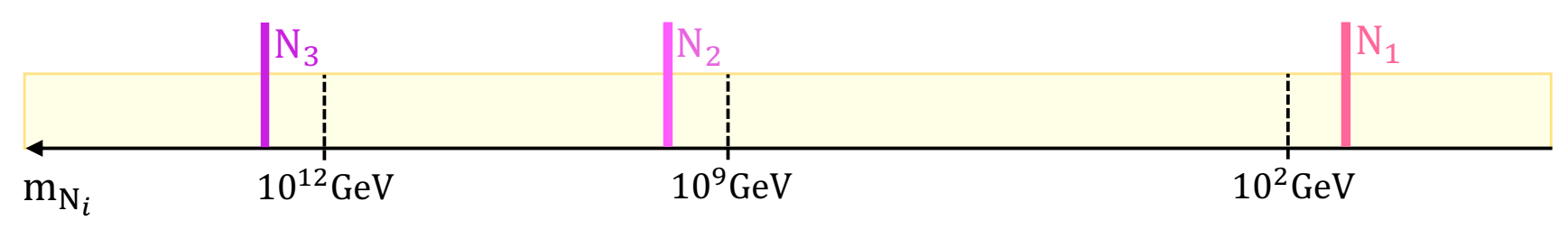}\\
   \caption{ For concreteness we assume this hierarchical mass spectrum for right-handed neutrinos.  The mass of $\rm N_1$ is specified by the relic density of DM in Sec. \ref{NCDM}.}  \label{fig:mNi} 
\end{figure}

To study the post-inflationary evolution, we need to specify our parameter space. For the sake of concreteness, we restrict the current analysis by assuming the following conditions considered by the author in \cite{Maleknejad:2020yys}: \textbf{Condition C1)} A hierarchical mass spectrum for the RH neutrinos (as implied by the neutrino oscillations) as
\bea\label{mass-h-RHN}     
m_{\rm{N}_3}\gtrsim 10^{12}~GeV \gg m_{\rm{N}_2} \gtrsim 10^{9}~GeV \gg m_{\rm{N}_1},
\eea
where $\bN_1$ is much lighter than the EW scale with feeble Yukawa interactions and hence a DM candidate. (See Fig. \ref{fig:mNi}) \textbf{Condition C2)} The $\bF_R$ field is never in thermal equilibrium with the thermal bath, i.e. $T_{W_R}>T_{\rm reh}$. \textbf{Condition C3)} The CP-violating phases in the neutrino sector, unconstrained by the current data, are not large enough to create the observed BAU. Reference \cite{Maleknejad:2020yys} introduces a stronger version of the above conditions by imposing a more restrictive version of C2. 
\textbf{Restricted condition C2)} The post-inflationary generation of RHNs via $\bF_R$ interactions is negligible compared to their pre-existing relics. We discuss and quantify conditions C2 \& restricted C2 in Sec. \ref{ThE} and condition C3 in Sec. \ref{spectator}.

\subsection{Thermal Evolution}\label{ThE}

Reheating starts at some point after the end of inflation and ends at the formation of a dominant thermal bath with temperature $\treh$. Here, we consider the phenomenological reheating model below
\bea\label{reh-inf}
\rho_{\rm reh}=\delta_{\rm reh} \bigg(\frac{a_{\rm inf}}{a_{\rm reh}}\bigg)^{\!4}\rho_{\rm inf}   ,
\eea
in which $\rho_{\rm inf}$ and $\rho_{\rm reh}$ are the energy density at the end of inflation and reheating respectively. Moreover, $\delta_{\rm reh}$ is the efficiency of the reheating process given as (See App. \ref{reheat})
\bea\label{phe-reheat}
\delta_{\rm reh} \approx \exp[-(3w_X-1)\Delta N], \quad  \Delta N\equiv \ln\big(\frac{a_{\rm reh}}{a_{\rm inf}}\big),
\eea
where $w_X$ is the effective equation of state in the intermediate period between the end of inflation and the formation of the thermal bath. The radiation energy density is given by
\bea\label{reheat}
\rho_{\rm rad}(T)=\frac{\pi^2}{30}g_{\rm eff} T^4\,,
\eea
where $g_{\rm eff}$ is the effective number of relativistic degrees of freedom. For SM particles at the time of reheating we have $g_{\rm eff}=427/4$. 
The reheating temperature is 
\bea\label{Treh}
\frac{T_{\rm reh}}{\mpl} \approx \bigg(\frac{90}{g_{\rm eff}}\bigg)^{\frac14}  \bigg(\frac{1}{\pi}\frac{H}{\mpl}\bigg)^{\frac12}~\exp[-\frac{3(w_X+1)}{4}\Delta N].
\eea

The photon number density at the time of reheating is
\bea\label{photon}
n_{\gamma,\rm reh}=\frac{2\zeta(3)}{\pi^2}T^3_{\rm reh},
\eea
where $\zeta(x)$ is the Riemann zeta function and $\zeta(3)\simeq 1.2$.
The photon number density today is related to $n_{\gamma,\rm reh}$ as
\bea\label{photon}
n_{\gamma,0} = S~ n_{\gamma,\rm reh}~\bigg(\frac{a_{\rm reh}}{a_{0}}\bigg)^3
\eea
where $S$ is the entropy injection factor. It captures the increase of entropy by the out of thermal  equilibrium decay of heavy RHNs, and is worked out in App. \ref{entropy-sub}. We found that the entropy injection is negligible in our setup, i.e. \footnote{Contrary to our setup, the late decay of long-lived $\bN_{2,3}$ (with lifetime up to a second) which are produced via {\it{freeze-out}} mechanism can generate a sizable amount of entropy in the LRSM \cite{Nemevsek:2012cd}. That requires a reheat temperature as low as a few $MeV$ and $\bN_{2,3}$ masses in the $100~MeV$ range. }
\bea
S\simeq 1.
\eea

\begin{figure}[h]
  \centering\includegraphics[height=0.3\textheight]{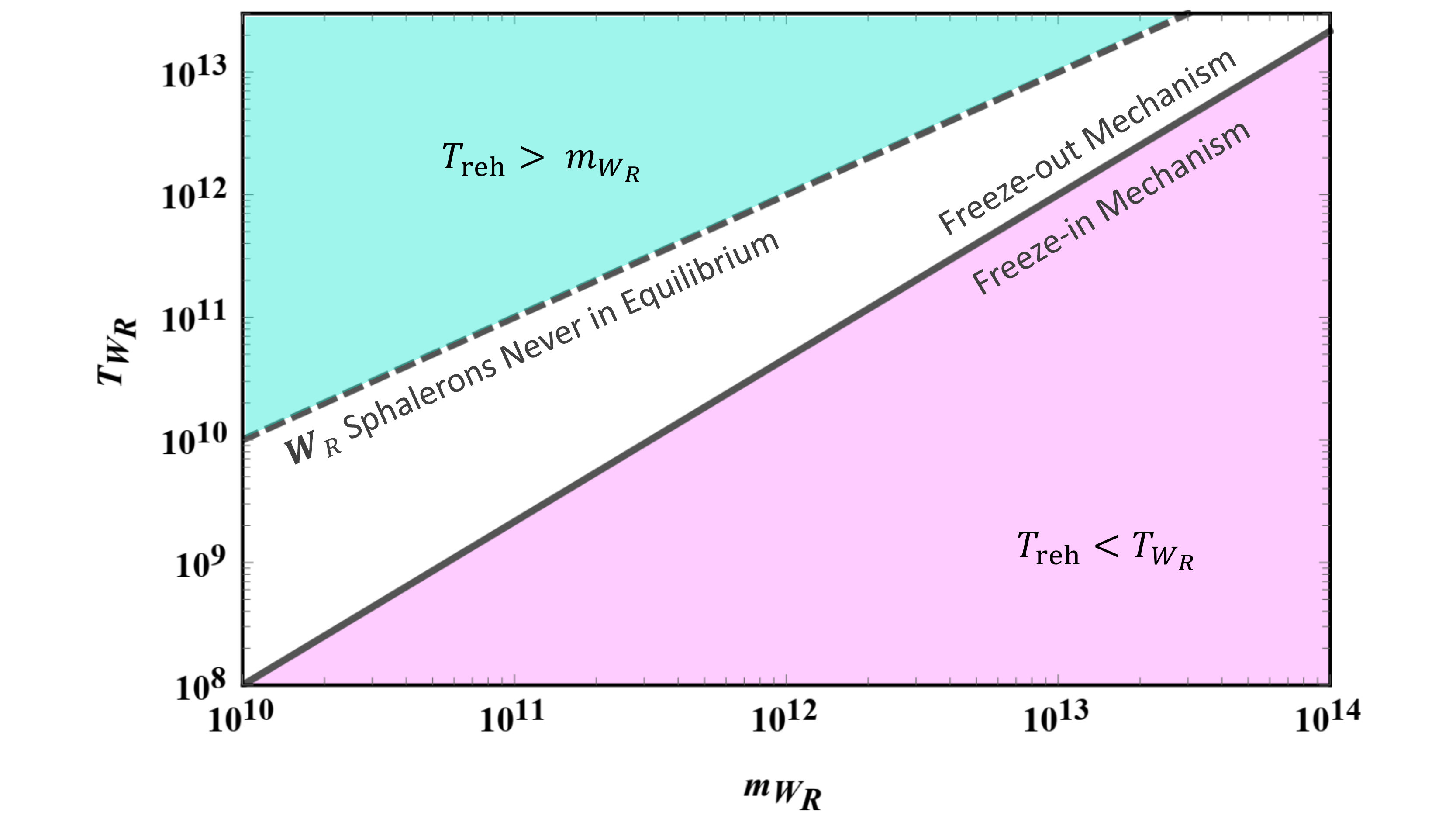}\\
   \caption{The freeze-out temperature of $\bW_R$ interactions in terms of $m_{W_R}$. The (pink) shaded area shows regions with $T_{{\rm reh}}< T_{W_R}$ (condition C2) and below the solid line, the secondary (post-inflationary) abundance of RHNs are not generated by freeze-out but instead by the freeze-in mechanism. In the (blue) shaded region we have $T_{\rm reh}>m_{W_R}$, so $\bF_R$ sphalerons are never in thermal equilibrium below the dashed line. }  \label{fig:TWR-Treh} 
\end{figure}

{\bf{$\rhd$~Condition C2:}} \newline
The $\bF_R$ gauge interaction has essential effects on thermal properties of our setup. After the 1st SSB, they keep sterile neutrinos in thermal equilibrium by scattering with the SM fermions. The temperature of the freeze-out of $\bF_R$ gauge field can be estimated as
\bea\label{TTWR}
T_{_{W_R}} \sim g_{*}^{\frac16} \bigg(\frac{m_{_{W_R}}}{10^{14}GeV}\bigg)^{\frac43}  \times 10^{13} ~GeV,
\eea
where $g_*$ is the number of relativistic degrees of freedom at $T_{_{W_R}}$ and $m_{_{W_{R}}}$ is the mass of $W^{\pm}_{R}$.
The particles which are only coupled through the $\bF_R$ interactions with the thermal bath, e.g. $\bN_1$, gets decoupled at this point. The thermal evolution after inflation depends on whether sterile neutrinos are in thermal equilibrium initially or not. If in thermal equilibrium, $\bF_R$ interactions generate thermal abundances of RHNs, i.e., freeze-out production.  The focus of this work, however, is the region in the parameter space in which $\bF_R$ interactions are never in thermal equilibrium, i.e. $T_{\rm reh}<T_{W_R}$ which demands (See Fig. \ref{fig:TWR-Treh})
\bea\label{HMpl}
\frac{H}{\mpl} \lesssim \frac32 \times 10^{-9} ~\exp[\frac{3(w_X+1)}{2}\Delta N]~\bigg(\frac{g_{\rm eff}}{10^2}\bigg)^{\frac12} \bigg(\frac{g_*}{10^2}\bigg)^{\frac13} \bigg( \frac{m_{_{W_R}}}{10^{14}GeV}\bigg)^{\frac83}.
\eea

The above condition guarantees that $\bN_i$ does not have thermal abundances by freeze-out mechanism. \footnote{
In case that $T_{_{W_R}}< T_{\rm reh}$, Fermi-type theory of $\bF_R$ field keeps sterile neutrinos in thermal equilibrium even at temperatures lower than $m_{_{W_R}}$. The freeze-out relic abundance of $\bN_i$ is $\frac{n_{N_i}}{s} = \frac{135\zeta(3)}{4\pi^4} \frac{1}{g_{*}(T_{_{W_R}})}$.} However, as we will see shortly, $\bF_R$ scatterings may still create a post inflationary abundance of RHNs via freeze-in mechanism.  Fig. \ref{fig:TWR-Treh} presents $T_{W_R}$ vs $m_{W_R}$ and the (pink) shaded area shows the region with $T_{\rm reh}<T_{_{W_R}}$ which is the focus of the current work. The (blue) shaded region marks where $T_{\rm reh}>m_{W_R}$.  Therefore, in the region of our interest the $SU(2)_R$ sphalerons are never in thermal equilibrium to cause any $\mB+\mL$ violating interaction (see Eq. \eqref{R-sph}). That is in contrast to the $SU(2)_L$ sphalerons which are in thermal equilibrium in the wide temperature interval of $m_{W_L}<T<10^{12}~GeV$. 
Another constraint on Hubble parameter in inflation comes from the current upper bound on the tensor to scalar ratio, $ r_{0.05} < 0.07$ at 95\% confidence \cite{Ade:2018gkx}, which implies $H\lesssim 10^{-5}~\mpl$.

\vskip 0.5cm

{\bf{$\rhd$~Restricted Condition C2:}}\newline
At reheating, the pre-existing RHNs, generated in inflation, is Eq. \eqref{alpha-inf}
\bea\label{npN}
n^{\rm p}_{\rm{N}_i} \approx  \frac13 \alpha_{\rm inf}(\xi)  \exp[-3\Delta N]~H^3.
\eea
One of the consequences of condition C2 is that the RHN does not have a thermal abundance, i.e. no freeze-out production.
However, post-inflationary $\bF_R$ scatterings produce RHNs via freeze-in as \cite{Dunsky:2020dhn}
\bea\label{nsN}
n^{s}_{\rm{N}_i} \approx 10^{13} \times \bigg(\frac{g_{\rm eff}}{10^2}\bigg) \bigg(\frac{ T_{\rm reh}}{\mpl}\bigg)^6  \bigg( ~\frac{10^{14}GeV}{m_{_{W_R}}}\bigg)^4  ~\mpl^3 ~.
\eea
The superscripts $p$ in Eq. \eqref{npN} and $s$ in Eq. \eqref{nsN} denote contributions of pre-existing (inflationary) and secondary (freeze-in) production respectively. One can restricted condition C2 such that this secondary RHN production is subleading comparing to the pre-existing one.
Using Eq. \eqref{Treh}, we can quantify restricted condition C2 as
\bea\label{ns-to-np}
\frac{n^{s}_{\rm{N}_i}}{n^{\rm p}_{\rm{N}_i}} \approx \frac{8 \times 10^{11}}{\alpha_{\rm inf}(\xi)}~\exp[-\frac32(1+3w_X)\Delta N] \bigg(\frac{10^2}{g_{\rm eff}}\bigg)^{\frac12}   \bigg( ~\frac{10^{14}GeV}{m_{_{W_R}}}\bigg)^4 < 1,
\eea
once inequality in Eq. \eqref{HMpl} holds.

\textit{Type-I Scenarios:} In this case, the $\bF_R$ gauge fields and $\rm{N}_i$s are massless in inflation. On the other hand, condition C2 demands $m_{_{W_R}}\gtrsim T_{\rm reh}$ which implies the first SSB must happen shortly after the end of inflation.

\begin{figure}[h]
  \centering\includegraphics[height=0.2\textheight]{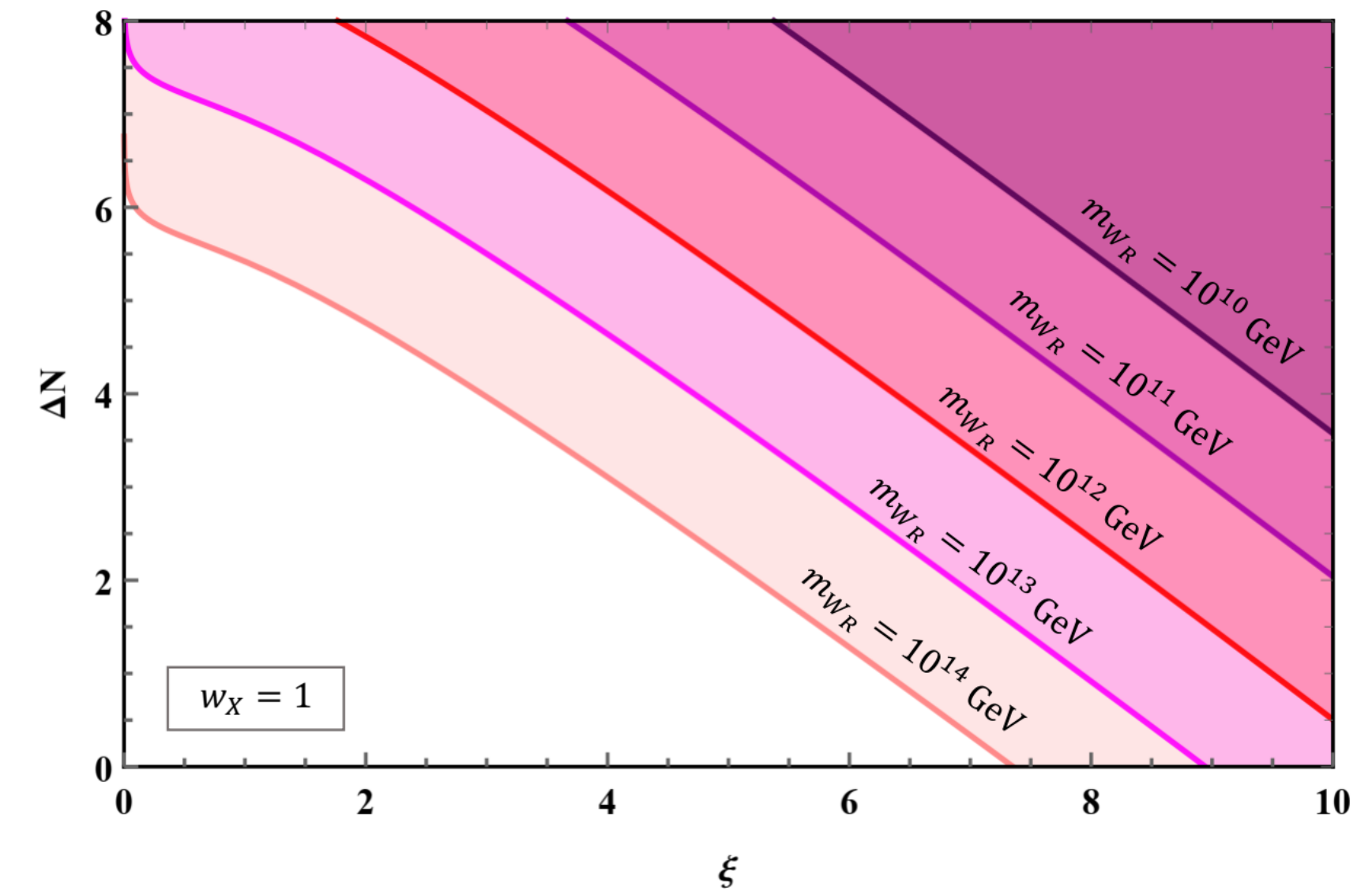} \includegraphics[height=0.2\textheight]{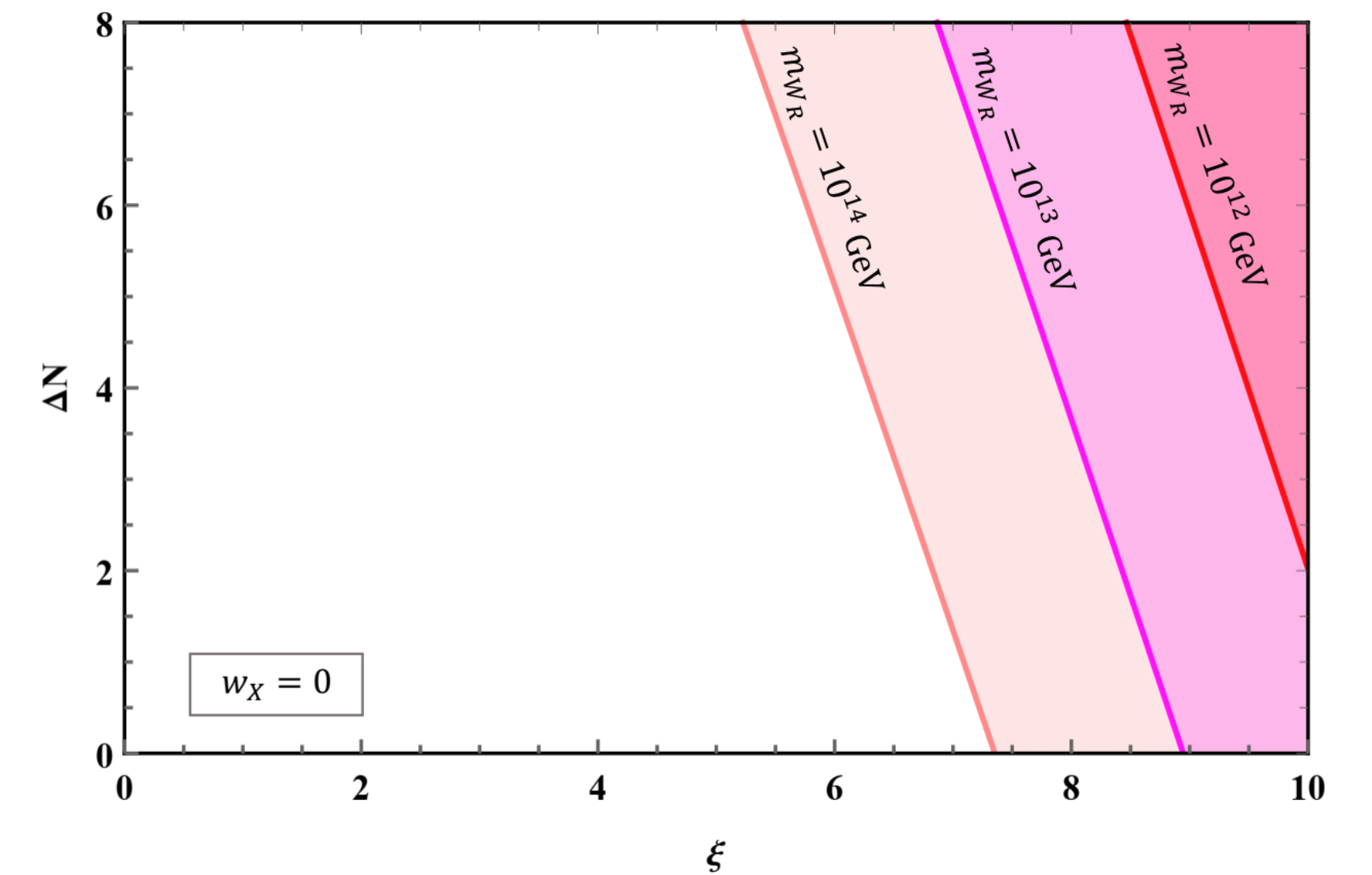}\\
   \caption{Condition restricted C2 in type-{\bf{I}} scenarios. The shaded area above each line corresponds to the accessible parameter space for a given $m_{W_R}$ that satisfies Eq. \eqref{ns-to-np}. The left and right panels show $w_X=1$ and $w_X=0$ respectively. The Max of $m_{_{W_R}}$ is set to be $10^{-2}$ GUT scale.}
  \label{fig:DeltaN-massless} 
\end{figure}

The shaded areas in Fig. \ref{fig:DeltaN-massless} show the parameter space in which restricted C2 is satisfied for a given $m_{W_R}$. As we see, most of the parameter space is accessible in case of $w_X=1$ while the case with $w_X=0$ requires larger values of $m_{_{W_R}}$ and $\Delta N$ and/or $\xi$.
The ratio of the pre-existing $\rm{N}_1$ to its freeze-in production in type-\textbf{I} can be analytically approximated as
\bea\label{ratio-freeze}
\frac{n^{s}_{\rm{N}_i}}{n^{\rm p}_{\rm{N}_i}} \sim 10^{11}~\exp[-\frac32(1+3w_X)\Delta N-2\pi \xi]\bigg(\frac{10^{14}GeV}{m_{_{W_R}}}\bigg)^4 .
\eea

\begin{figure}[h]
  \centering\includegraphics[height=0.2\textheight]{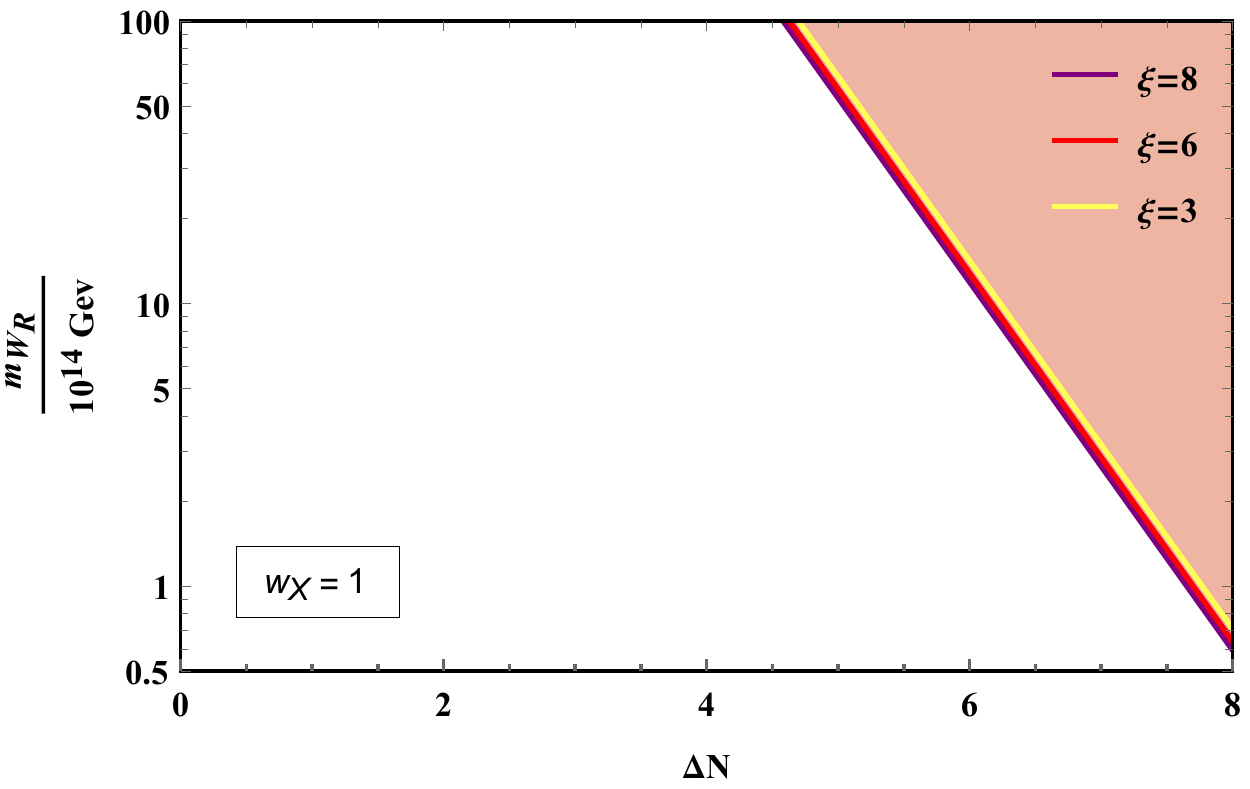}\\
   \caption{The parameter space $(\Delta N,m_{_{W_R}})$ for different values of $\xi$ that satisfies condition restricted C2 in type-{\textbf{II}} scenarios with $w_X=1$. The accessible region for $w_X=0$ case is in the region with $
   \Delta N>8$ and it is not shown here.}
  \label{fig:DeltaN-massive} 
\end{figure}

\textit{~Type-II Scenarios:} In this case the $SU(2)_R$ gauge fields and $\rm{N}_i$s are massive in inflation. However, $\rm{N}_1$ which is the dark matter candidate is very light compared to $H$. Therefore, the pre-existing $\rm{N}_1$ in Eq. \eqref{ns-to-np} is produced by the chiral anomaly in inflation. The mass of the $W_R$ can be roughly estimated as $m_{_{W_R}} \sim \sqrt{\frac{H}{\mpl}} \mpl$.
For $m_{_{W_R}} = 10^2 H$, Fig. \ref{fig:DeltaN-massive} shows the parameter space corresponding to each $\xi$ in which restricted C2 is satisfied. Comparing with the type-\textbf{I} scenario, the C2 is satisfied in a smaller part of the parameter space and only for $w_{X}=1$ case.

To summarize, condition restricted C2 prefers type-{\bf{I}} scenarios and $w_X=1$. In particular, it holds in a wide part of the parameter space when $\Lambda_F\lesssim \Lambda_{\rm inf}$, i.e., the first SSB coincides with the end of inflation. Interestingly, it relates left-right gauge symmetry breaking to a geometrical phase transition in cosmology, i.e., the end of exponential expansion of the Universe.
Moreover, as Fig.s. \ref{fig:DeltaN-massless}-\ref{fig:DeltaN-massive} show, it demands $m_{W_R}>10^{10}~GeV$ which is the scale suggested by the non-supersymmetric SO(10) GUT model with an intermediate left-right symmetry scale \cite{Rizzo:1981su, Bertolini:2012im, Deppisch:2017xhv}.

\subsection{Spectator Effects, RHN decay, and Matter Asymmetry}\label{spectator}
Throughout the Early Universe, particles experience a whole cascade of interactions that eventually equilibrate in the Early Universe. Many of them can potentially redistribute the initial asymmetries to the spectator degrees of freedom. These processes do not participate directly in the generation or washout of the asymmetries (hence the name spectator). Still, they have important effects in finial $\mB$ and $\mL$ by imposing certain relations between different species. In addition to the spectator effects, the CP asymmetric decay of $\bN_{2,3}$ produces SM leptons and simultaneously partially washes out the pre-existing lepton asymmetries. In this section, we consider washout effects, lepton flavor effects, and sphaleron processes. For the ease of notation, we denote the SM leptons as $\uL$, i.e.
\bea
\uL \equiv \mL_{\rm SM}.
\eea

 \subsection*{Spectator Effects}

The $SU(2)_{L}$ sphalerons ($SU(2)_{R}$ sphalerons) transmit the asymmetry from left-handed (right-handed) leptons to left-handed (right-handed) quarks and vice versa.  The $\bF_L$ gauge field is inactive and unimportant in inflation. Later on, however, they attain a thermal equilibrium, and together with $\bF_R$, they can have significant impacts on the final $\mB$ and $\mL$ asymmetries. The $\mB+\mL$ violating processes due to $W_{L,R}$ sphalerons shuffle the initial baryons and leptons coupled to them. In App. \ref{Sphaleron-App} we showed that $W_R$ sphalerons are never in thermal equilibrium in our setup (see also Fig. \ref{fig:TWR-Treh}). Hence they can not give rise to $\mB+\mL$ violating processes. After the $\bW_R$ and sterile neutrinos' freezeout, the SM particles remain in thermal equilibrium up to the electroweak scale. Quarks, SM leptons, and Higgs bosons interact via gauge and Yukawa interactions as well as non-perturbative sphaleron processes. All the SM gauge interactions and $W_L$ sphaleron processes are in equilibrium in the temperature range of $100~ GeV \lesssim T \lesssim 10^{12} ~GeV$. 
The thermal equilibrium of Yukawa interactions is flavor-dependent. Nevertheless, all of them are in equilibrium at $T<85 ~ TeV$ \cite{Bodeker:2019rvr}. Using the sphaleron effects and hypercharge constraint, we find that $\mB$, $\uL$, and $\mB-\uL$ are related as
\bea
n_{\mB} &=& c_{\rm sph} ~ n_{\mB-\uL} ,\\
n_{\uL} &=&  (c_{\rm sph}-1) ~ n_{\mB-\uL}.
\eea
where $c_{\rm sph}=\frac{28}{79}$ is the sphaleron conversion factor.

 \subsubsection*{Lepton Flavor Effects}

\begin{figure}[h!]
\centering
\includegraphics[height=0.22\textheight]{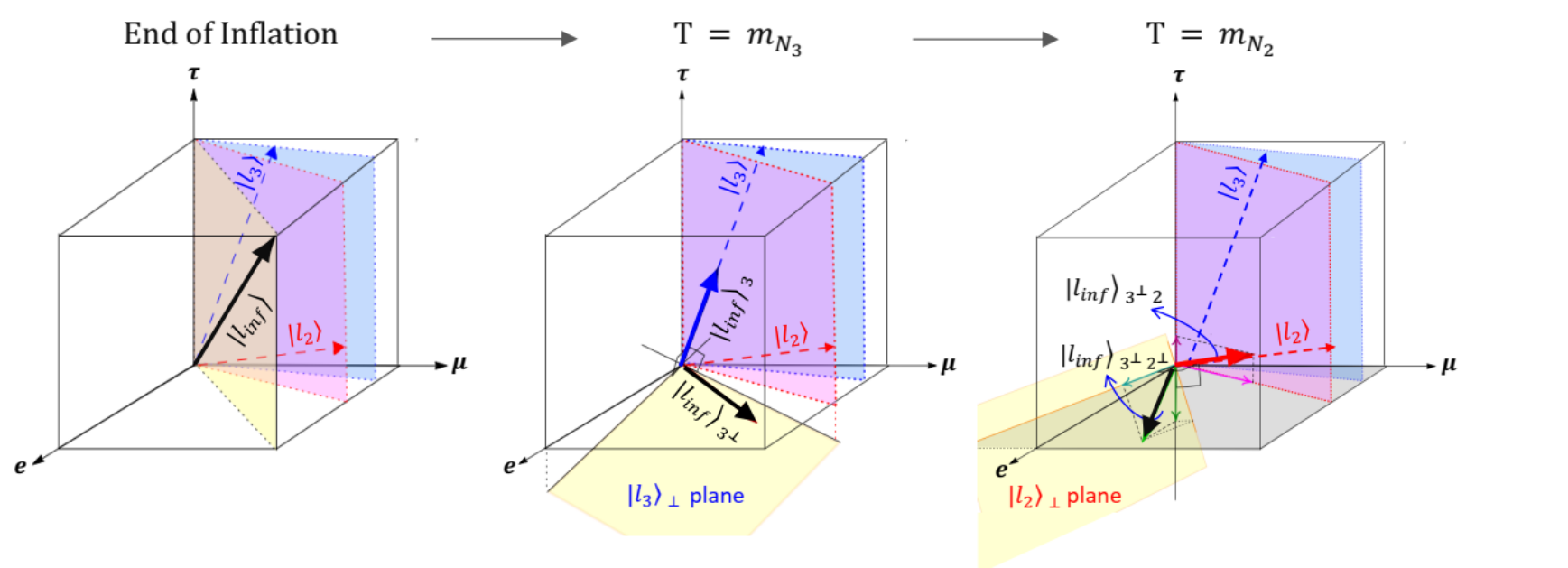}
\caption{The geometrical illustration of washout processes induced by the decay of $\bN_3$ and $\bN_2$. The left panel shows the SM leptonic states at the end of inflation $\lvert l_{inf} \rangle$ in black and $\lvert l_{3,2} \rangle$ in blue and red, respectively. The middle panel shows the SM lepton states at $T=m_{N_3}$ and the right panel presents the system at $T=m_{N_2}$. The black arrows in each panel show the pre-existing SM lepton asymmetry, which remains untouched by the washout effects. (Fig. adopted from ref. \cite{Maleknejad:2020yys})}
\label{fig:flavor} 
\end{figure}

One potentially very significant aspect of (post inflationary) leptogenesis is the flavor effect. The flavor-dependent washout and $\mL$ violating interactions can significantly change the value, and even sign of the final baryon asymmetry \cite{Abada:2006fw,Barbieri:1999ma,Blanchet:2006ch}. By the end of inflation, and due to our flavor blind CP violating source, we have a lepton quantum state $\lvert l_{inf} \rangle$  as
\bea
\lvert l_{inf} \rangle  \equiv  \sum_{\alpha=e,\mu,\tau} C^{inf}_{\alpha} \lvert \alpha \rangle  \where \quad C^{inf}_{\alpha} = \langle \alpha \vert l_{inf} \rangle.
\eea
The decays of the heavy sterile neutrinos modify these initial states. More precisely, the CP asymmetric decay of $\rm{N}_i$ produces leptons as  \footnote{The $ C_{i\alpha}$ coefficients are given by the Yukawa matrix. In terms of the active neutrino mass matrix we have $C_{i\alpha}= \frac{m_{\nu}^{\alpha i}}{\sqrt{(m_{\nu}^{\dag}m_{\nu})_{\alpha\alpha}}}$. Unlike $\lvert \alpha \rangle$s, $\lvert l_i \rangle$ does not form an orthonormal bases, i.e. in general $\langle l_i \lvert l_{j\neq i} \rangle \neq 0$.}
\bea
\ket{l_i} \equiv \sum_{\alpha=e,\mu,\tau} C_{i\alpha} \ket{\alpha}  \where C_{i\alpha} = \bra{\alpha}\ket{l_{i}},
\eea
and simultaneously washes out the pre-existing (inflationary) leptons in this direction, i.e.
\bea
\ket{l_{inf}}_{_{i}} \equiv  ~ \bra{l_{i}}\ket{l_{inf}} ~ \ket{l_{i}}.
\eea
However, the pre-existing leptons normal to $\lvert l_i \rangle$ direction, i.e.
\bea
\lvert l_{inf} \rangle_{_{i}^{\bot}} \equiv \lvert l_{inf} \rangle - \lvert l_{inf} \rangle_{_{i}},
\eea
elude the washout. As discussed earlier, we assume that $\bN_1$ has feeble Yukawa interactions with the SM and hence a DM candidate (condition C1). Therefore, only $\bN_2$ and $\bN_3$ contribute to the seesaw mechanism as well as decays and washouts. As a result, the component $\lvert l_{inf} \rangle_{{3}^{\bot}{2}^{\bot}}$ which is normal to both $\lvert l_{3} \rangle$ and $\lvert l_{2} \rangle$ remains as the remnant of the initial asymmetry. For the mass spectrum in Eq. \eqref{mass-h-RHN}, the corresponding Boltzmann equations and details are presented in App. \ref{Flavor-Sec} and here we report the final results.
The geometry of this process  in the SM flavor basis is schematically shown in Fig. \ref{fig:flavor}.

{\bf{$\rhd$~Condition C3:}}\newline
 The SM lepton asymmetry after decay of $\bN_2$ at $T=M_2\gtrsim 10^{9} ~GeV$ is \bea
n_{\mB-\uL} = n^{p,f}_{\mB-\uL} + n^{\rm{N}}_{\mB-\uL},
\eea
where $n^{p,f}_{\mB-\uL}$ is the remnant of the primordial asymmetry $n^{p,i}_{\mB-\uL}$, and $n^{\rm{N}}_{\mB-\uL}$ is the lepton number produced by the CP asymmetric decay of $\bN_2$ as
\bea
n^{\rm{N}}_{\mB-\uL} \approx \varepsilon_2 \kappa_2,
\eea
where $\varepsilon_2$ is the CP asymmetry and $\kappa_2$ is the associated efficiency factor. Interestingly, when flavour effects are considered, it is very difficult for the pre-existing asymmetry to be washed out by the RH neutrinos \cite{Bertuzzo:2010et,DiBari:2013qja}. The value of $n^{\rm{N}}_{\mB-\uL}$ depends on the leptonic Yukawa matrix and the unconstrained CP violating phases in the neutrino sector. In this work, we assume that the amount of this asymmetry is not sufficient to account for the observed matter asymmetry, i.e. condition C3:
\bea\label{CC3}  
  \frac{n^{\rm{N}}_{\mB-\uL}}{n^{p,f}_{\mB-\uL}} \ll 1.
\eea
Condition C3 is the opposite limit of what is assumed in leptogenesis scenarios \cite{Fukugita:1986hr}. 

\vskip 0.5cm

Finally the remnant of the primordial asymmetry is given as below in terms of the initial $\mB-\uL$
\bea
n_{\mB-\uL} \simeq n^{p,f}_{\mB-\uL} = \mathcal{C} ~ n^{p,i}_{\mB-\uL},
\eea
where $\mathcal{C}$ is a parameter less than one ( see Eq. \eqref{Eq-C} and Fig. \ref{fig:Af}). For most of the parameter space we have
\bea
\mathcal{C} \gtrsim \frac13.
\eea
Eliminating the effect of this pre-existing asymmetry is very hard and requires tightly fine-tuned relations between leptonic Yukawa couplings and the physics of inflation which is discussed in App \ref{Flavor-Sec}.

\section{Modern Era Baryon Asymmetry and Dark Matter}
\label{today}
In this section we work out the baryon to photon ratio and dark matter density today. Here we only consider type-\textbf{I} scenarios in which the 1st SSB happens after inflation. The remnants of the inflationary baryon and SM lepton asymmetries after the decay of the heavy RHNs and the getting redistributed by the spectator effects are respectively as 
\bea
n_{\mB}(a) &\simeq& 0.12 ~ \alpha_{\rm inf}(\xi) H^3  \exp[-3\Delta N] \bigg(\frac{a_{\rm reh}}{a}\bigg)^3,\\
n_{\mL_{\rm SM}}(a) &\simeq& - 0.18 ~n_{\mB}(a).
\eea
The final $\bN_1$ number density is 
\bea\label{nN1-number}
n_{{\rm N}_1}(a) \simeq  2.8 ~n_{\mB}(a) + n^s_{\rm N_1}(a),
\eea
where $n^s_{\rm N_1}$ is the secondary (freeze-in) production of $\bN_1$ given in Eq. \eqref{nsN}. As is assumed in \cite{Maleknejad:2020yys}, if the restricted version of condition C2 in Eq. \eqref{ns-to-np} holds, we have 
\bea\label{nN1-number}
n_{{\rm N}_1}(a) \simeq  2.8 ~n_{\mB}(a).
\eea
The particle production mechanism throughout cosmic evolution is then summarized in Fig. \ref{evolution}.

\subsection{Baryon to Photon Ratio}

\begin{figure}[h]
  \centering\includegraphics[height=0.22\textheight]{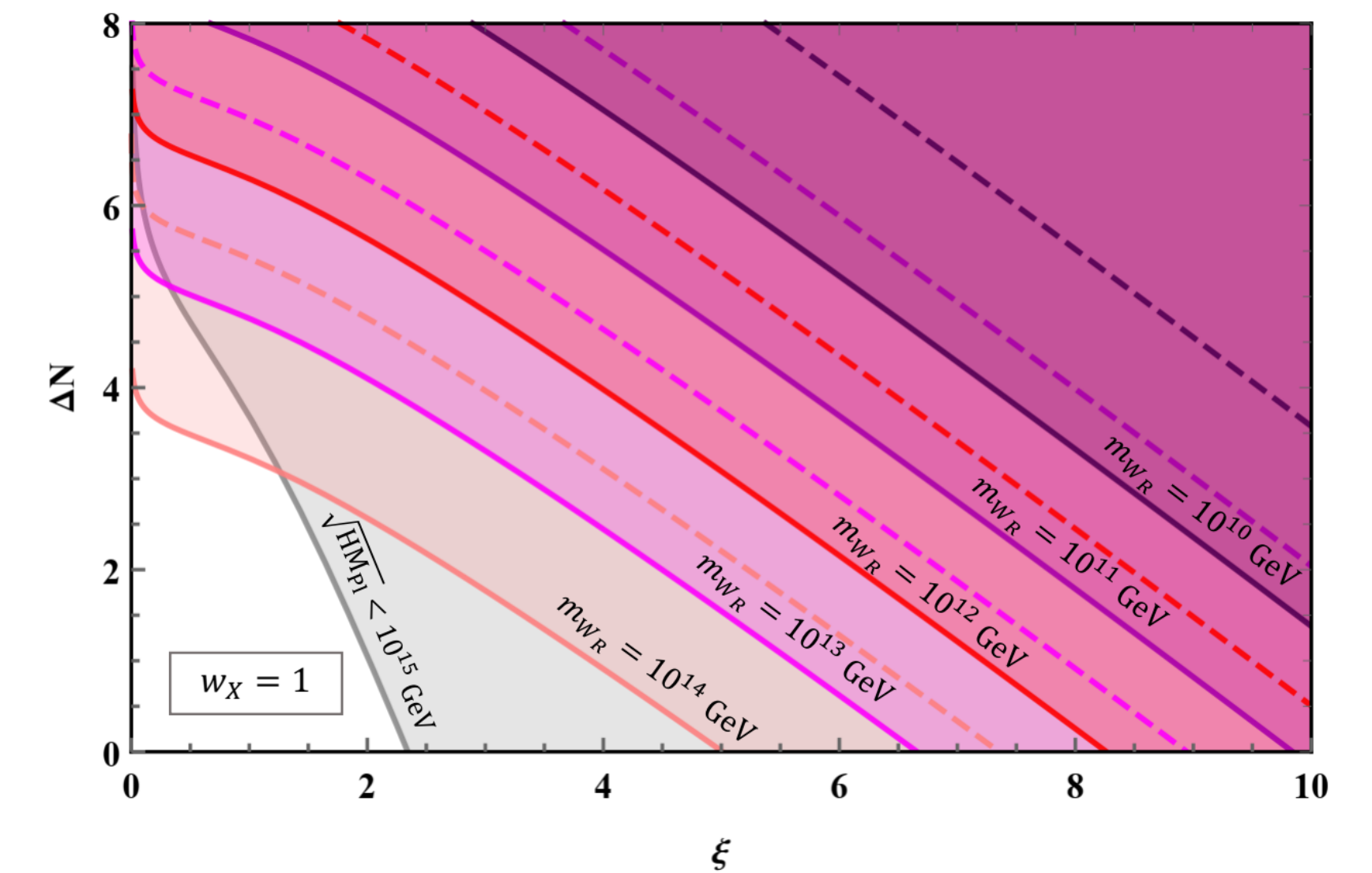}\includegraphics[height=0.22\textheight]{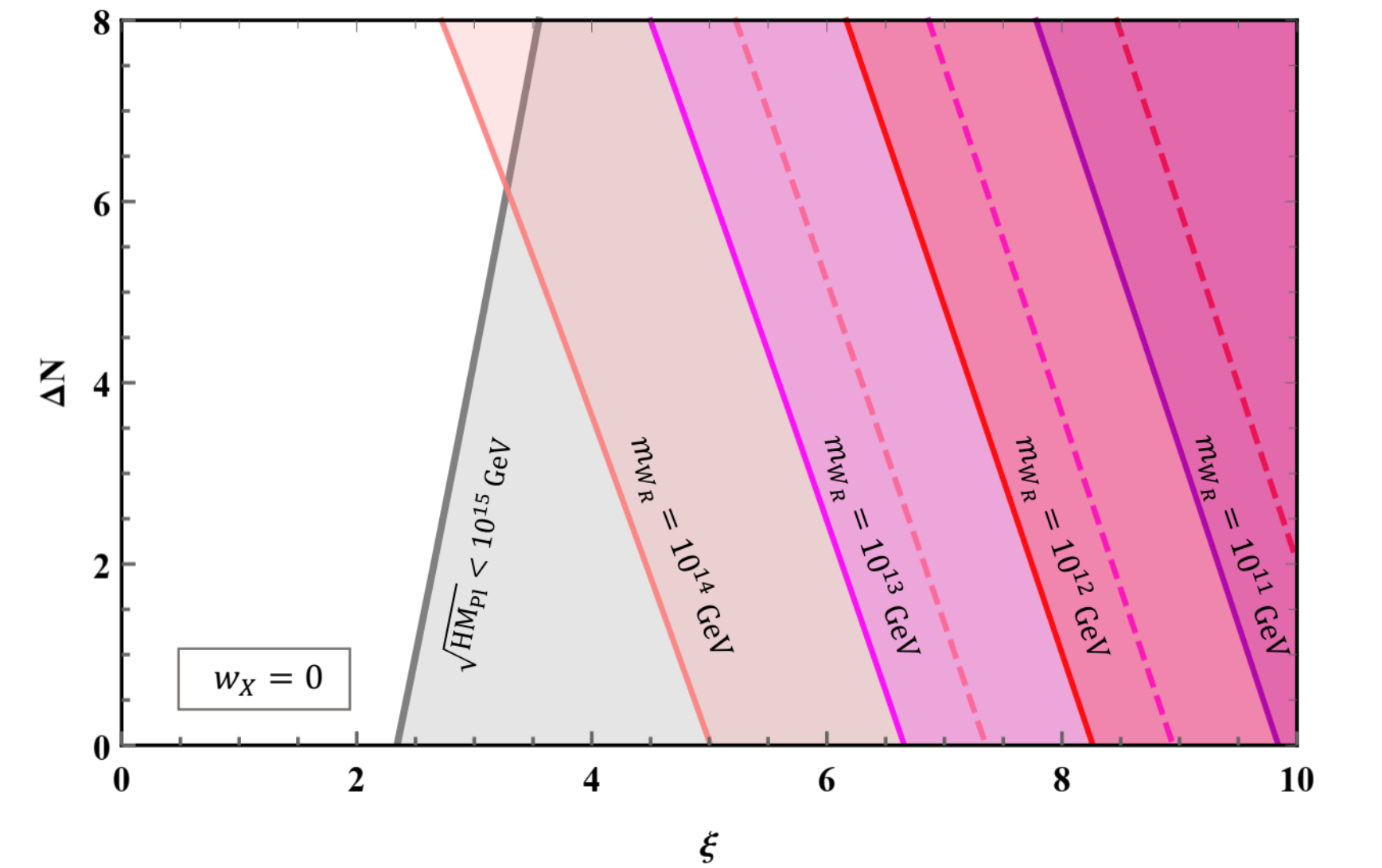}\\
    \caption{The accessible parameter space in terms of $\xi$, $\Delta N$, and $m_{W_R}$ for $w_X=1$ (Left Panel ) and $w_X=0$ (Right Panel). The color shaded areas (with solid line boundaries) present regions that Eq. \ref{condition-L} is satisfied while the gray shaded region shows areas associated with $\sqrt{H\mpl}<10^{15}~GeV$. The dashed lines present the boundaries corresponding to the same colors but with restricted C2 condition.}
  \label{fig:parameter-space}  
\end{figure}

To the best of our knowledge, the cosmos is highly matter-dominated. The baryon-antibaryon asymmetry can be quantified by the baryon to photon ratio at the present time as \cite{Ade:2015xua}
\bea
\eta^0_{\mB} = \frac{n^0_{\mB}}{n^0_{\gamma}} \simeq 6\times 10^{-10},
\eea
in which a $0$ superscript denotes the present time value. 
Our setup predicts the baryon to photon ratio as
\bea\label{eta-B-}  
\eta^0_{\mB} \approx 3 \bigg(\frac{g_{\rm eff}}{100}\bigg)^{\frac34}\frac{\alpha_{\rm inf}(\xi)}{\big(\delta_{\rm reh}\big)^{\frac34}} \bigg(\frac{H}{\mpl}\bigg)^{\frac32},
\eea
where $g_{\rm eff}=427/4$.
One can write $\eta^0_{\mB}$ in terms of the curvature power spectrum as
\bea
\eta^0_{\mB} \approx ~ 0.3 ~ \upbeta ~ P_{\zeta},
\eea
where $P_{\zeta}(k_0)= \frac{1}{2(2\pi)^2\epsilon} \big(\frac{H}{\mpl}\big)^{2}$ in which $\epsilon$ is the slow-roll parameter, and $\upbeta$ is 
\bea
\upbeta =   \frac{5~(4\pi)^2~\epsilon~ \alpha_{\rm inf}(\xi)}{\big(\delta_{\rm reh}\big)^{\frac34}} ~  \bigg(\frac{\mpl}{H}\bigg)^{\frac12}.  
\eea
To agree with the date, $\upbeta$ should be one and we have \bea\label{eta-H}
\frac{H}{\mpl} \approx 10^{-6} ~\alpha_{\rm inf}^{-\frac23}(\xi) ~ \delta_{\rm reh}^\frac12.
\eea
By this point, we have three constraints on $H$, i.e. Eq. \eqref{HMpl} imposed by C2, Eq. \eqref{eta-H} to explain the observed $\eta_{\mB}$, and the upper bound enforced by CMB data.
Combining Eq.s \eqref{HMpl} and \eqref{eta-H} gives
\bea \label{condition-L}
\alpha_{\rm inf}^{\frac23}(\xi) ~ \delta_{\rm reh}^{(\frac{1/3+w_X}{1/3-w_X})} ~ \bigg(\frac{m_{W_R}}{10^{14}~GeV}\bigg)^\frac83 \gtrsim 10^3,
\eea
which together with $\sqrt{H\mpl}<10^{15}~GeV$ specifies the accessible region of the parameter space. The color shaded areas (with solid line boundaries) in Fig. \ref{fig:parameter-space} show allowed parts
of the parameter space for different values of $m_{W_R}$ while the gray shaded area shows the region with $\sqrt{H\mpl}<10^{15}~GeV$. 
The boundaries of accessible parameters in the more restrictive case with condition restricted C2 in Eq. \eqref{ns-to-np} are shown with same color (dashed lines) in Fig. \ref{fig:parameter-space}. This setup can explain the observed $\eta_{\mB}$ for typical values of the parameters and in a wide range of the parameter space. Interestingly, it prefers left-right symmetry breaking scales above $10^{10}~GeV$, which is in the range  suggested by the non-supersymmetric SO(10) Grand Unified Theory with an intermediate left-right symmetry scale.

\subsection{Right-handed Neutrino as Cold Dark Matter}\label{NCDM}

 As discussed in Sec. \ref{Post-flation}, we assume that the lightest RHN, $\bN_1$, has feeble Yukawa couplings, hence decouples after freezeout of $W_R$ interactions at $T_{_{W_R}}$ with a relic density given as
 \bea\label{DM-relic}
 \Omega_{\rm N_1} \approx 2.8 ~ \frac{m_{\rm N_1}}{m_p} ~ \Omega_{\mB} (1+\frac{n^s_{\rm N_1}}{n^p_{\rm N_1}}),
 \eea
 where $\Omega_{\mB}$ is the baryon density parameter, $m_p$ is the proton mass.
  If $\bN_1$ makes all of the DM that we observer today, i.e. $\Omega_{\bN_1}\simeq 5 \Omega_{\mB}$, it specifies the mass of $\bN_1$ in terms of the proton mass as
 \bea\label{mN1-value}
 m_{\rm N_1} \approx \frac{1.8 ~ m_{p}}{(1+\frac{n^s_{\rm N_1}}{n^p_{\rm N_1}})}.
 \eea
 Condition C2 implies that $0\leq\frac{n^s_{\rm N_1}}{n^p_{\rm N_1}}<10^{6}$ which specifies the mass of $\bN_1$ in the wide range of a few $keV$ to a few $GeV$. That mass range is associated to different DM spectra from warm DM to cold DM.  On the other hand if following \cite{Maleknejad:2020yys} we consider  restricted condition C2 which grantees that the $\bN_1$ relic density is primordial, we have
 \bea
 \Omega_{\rm N_1} \approx 2.8 ~ \frac{m_{\rm N_1}}{m_p} ~ \Omega_{\mB},
 \eea
 which makes a specific prediction for the mass of $\bN_1$ as
 \bea\label{mN1-}
m_{\bN_1} \simeq 1.8 ~ m_p = 1.7~GeV.
\eea
That leads to a cold DM spectrum that is consistent with
structure formation. Next, we study the stability of $\bN_1$ as a DM particle.

 \subsubsection*{Decay of $\bN_1$}
 
 Given that $W_R$ is very heavy and freezes out early (see Sec. \ref{ThE}), the dominant decay channel of $\bN_1$ is $\bN_1\rightarrow 3\bn$ with the total decay width \cite{Pal:1981rm,Barger:1995ty}
\bea
\Gamma_{\rm N_1\rightarrow 3\nu}  =  \frac{G^2_{F}M_{\rm N_1}^5}{96~(2\pi)^3} \sum_{\alpha} \sin^2(2\theta_{\alpha,1}),
\eea
where $G_F$ is the Fermi constant, $\alpha=e, \nu, \tau$ and $\theta_{\alpha,1}$ are the mixing angles of left-handed neutrinos with $\bN_1$. Demanding that the lifetime of this process, $t_{\rm N_1}$, is larger than the age of the Universe, i.e. $t_{U}\approx 4.4 \times 10^{17}~s$, we arrive at
\bea
\frac{t_{\rm N_1}}{t_{U}} \approx \bigg(\frac{0.56~GeV}{M_{\rm N_1}}\bigg)^5 \bigg(\frac{10^{-26}}{\theta_1^2}\bigg),
\eea
where $\theta^2_1 \equiv \sum_{\alpha} \theta_{\alpha,1}^2$.
Demanding that $\bN_1$ is stable over the lifetime of the universe gives
\bea
\theta_1 < 10^{-13}.
\eea
In this framework, the generation mechanism of $\bN_1$ is independent of its Yukawa mixing with active neutrinos, and $\theta_1$ can be any number that satisfies the above upper bound. The next leading decay channel is the loop-mediated radiative decay of $\bN_1$ to active neutrinos and a gamma-ray photon with energy $E_{\gamma}\approx M_{\rm N_1}/2$ as \cite{Pal:1981rm}
\bea
\Gamma_{{\rm{N}}_1\rightarrow\gamma\nu} =  \frac{9\alpha_{em} G^2_{F}M_{N_1}^5}{64~(2\pi)^4} \sum_{\alpha} \sin^2(2\theta_{\alpha,1}) \sim 10^{-2} \Gamma_{{\rm{N}}_1\rightarrow3\nu}.
\eea
Although the radiative decay has a branching ratio of order $2\%$, it can provide upper bounds from not observing gamma-ray photons with energy
$E_{\gamma}$. The current strongest gamma-ray bounds in the GeV scale are on mass range 10-100~GeV \cite{Ackermann:2013uma} which is much heavier than our DM.

\section{Quick on Observational Constraints and Signatures}\label{Obser-sec}
In this section, we briefly discuss the cosmological, astrophysical, and collider constraints and signatures of our setup. The current work is based on embedding the minimal $SU(2)$-axion inflation model \cite{Maleknejad:2016qjz} in minimal LRSM \cite{Pati:1974y, Mohapatra:1974gc, Senjanovic:1975rk}. The cosmic perturbations of the minimal $SU(2)$-axion inflation in the presence of the gauge field VEV has been studied and compared with Planck data in \cite{Maleknejad:2016qjz}. In the current work, however, we sorely focus on scenarios with vanishing VEV. Hence it enjoys a wider accessible parameter space. As a cosmological smoking gun, all $SU(2)$-axion inflation models predict chiral \cite{Maleknejad:2012fw, Adshead:2013qp, Dimastrogiovanni:2012ew} and non-Gaussian \cite{Agrawal:2018mrg} gravitational wave background  which leads to parity odd CMB cross-spectra \cite{Thorne:2017jft}. This chiral GW Background (GWB) is blue tilted and can also be detected by future laser interferometer detectors. The parity odd features can be used as an observational marker to distinguish it from the standard GWB produced by the vacuum fluctuations \cite{Lue:1998mq,Saito:2007kt,Contaldi:2008yz}. This signal has been extensively studied in the literature. 
For an exhaustive discussion on the measurement of this effect, see \cite{Campeti:2020xwn}.

Direct production or virtual contributions at astrophysical and collider processes put several constraints on the charged and neutral $SU(2)_R$ gauge boson mass and mixing parameters. The $K_L - K_S$ kaon mass difference measurement \cite{Barenboim:1996nd} places a lower bound on the mass of $W_{R}$ as $m_{_{W_R}} > 1.6 ~TeV$ and the mixing angle between $Z_R$ and $Z_L$ is constrained to be less than $10^{-4}$. The possible low-energy $W_{R}$ has been the target of several LHC collaborations which puts the current bound as $m_{_{W_R}}> 3 ~ TeV$ \cite{Bertolini:2014sua}. For an exhaustive discussion of the phenomenological implications and constraints of LRSM, see \cite{Beringer:1900zz}. Our current setup with high scale $SU(2)_R$ SSB, i.e., $m_{W_R}> 10^{10}~GeV$, satisfies the above lower bounds. The most distinctive astrophysical signal of our DM candidate with GeV mass is the gamma-ray line at $E = m_{\rm N_1} /2$ produced in the one-loop decay $\bN_1\rightarrow \gamma \bn$. Gamma-ray lines have been probed by the Fermi-LAT \cite{Ackermann:2013uma}, H.E.S.S. \cite{Abramowski:2013ax}, and MAGIC telescopes \cite{Aleksic:2013xea}. However, the strongest current bounds are on DM masses above 10~GeV \cite{Ackermann:2013uma} which is heavier than our DM. 
We leave the further study of the observable signatures of this setup for future work.

\section{Conclusions}
\label{Sec/Sec-Concl}
Recently \cite{Maleknejad:2020yys} proposed a new particle physics model for inflation, based on embedding axion-inflation in gauge extensions of the SM. To unify cosmic inflation and BSM, it utilized the minimal Left-Right Symmetric Model (LRSM) \cite{Pati:1974y, Mohapatra:1974gc, Senjanovic:1975rk} with gauge group $SU(2)_L\times SU(2)_R \times U(1)_{\mB-\mL}$. As the name implies, the model includes $\bW_{R}$ gauge bosons and three right-handed neutrinos (RHN). As the inflaton field, an axion is added to the field content of LRSM, which is directly coupled to the $SU(2)_R$ gauge field. In this work, we presented the analytical and numerical details of this setup.

 LRSM in cosmology introduces a new fundamental cosmic scale, i.e., feeble scale $\Lambda_F$, where the extended gauge symmetry breaks down to the SM one, i.e., $SU(2)_R\times U(1)_{\mB-\mL} \rightarrow U(1)_Y$. At feeble scale, the $W_R^{\pm}$ and $Z_R^0$ become massive, and the RHNs acquire Majorana masses \cite{Mohapatra:1980yp}. Later, at the electroweak scale, the second spontaneous symmetry breaking happens, which gives mass to the SM particles. At this point, the left-handed neutrinos
acquire mass by seesaw mechanism \cite{Mohapatra:1980yp} (for cosmic evolution see Fig. \ref{cosmic-history}).
Based on the scale of inflation $\Lambda_{\rm inf}=\sqrt{H\mpl}$, feeble scale $\Lambda_F$, and the (possible) $SU(2)_R$ field’s VEV in inflation, one can separate four different types of scenarios (see table \ref{scenario}). Following \cite{Maleknejad:2020yys}, we solely focused on scenarios with vanishing $SU(2)_R$ VEV, i.e. type-{\bf{I}} ($\Lambda_{\rm inf}>\Lambda_F$) \& type-{\bf{II}} ($\Lambda_{\rm inf}<\Lambda_F$) scenarios.

The $SU(2)_R$ gauge field is produced by inflaton while other gauge fields, i.e., $SU(3)$, $SU(2)_L$ and $U(1)_{\mB-\mL}$, are diluted by the exponential expansion. The chiral anomaly of $\bW_R$ breaks $CP$ in physics of inflation and gives rise to simultaneous baryogenesis, leptogenesis, and RHN creation in inflation (see Eq.s \eqref{B-L-massless} \& \eqref{nNi} for type-{\bf{I}} and Eq.s \eqref{B-massive} \& \eqref{LN-massive-} for type-{\bf{II}} scenarios). Even in type-{\bf{I}} scenarios in which $\mB-\mL$ is a gauge symmetry in inflation, we have $\mB-\mL_{\rm SM}\neq 0$. For cosmic evolution after inflation, we future specified our parameters and imposed the three conditions which are used in \cite{Maleknejad:2020yys}. Condition C1 considered a hierarchical mass spectrum for RHNs with feeble Yukawa interactions for $\bN_1$ such that it is a DM candidate (Eq. \eqref{mass-h-RHN}). Condition C2 demands that $\bW_R$ is never in thermal equilibrium with the thermal bath. Consequently, it implies; 1) the $SU(2)_R$ sphalerons were never in equilibrium as well (Eq. \eqref{HMpl} and Fig. \ref{fig:TWR-Treh}), and 2) there is no secondary freeze-out production of RHNs. However, the post-inflationary scatterings of $\bW_R$ can generate RHNs via freeze-in mechanism (Eq. \eqref{nsN}). Following \cite{Maleknejad:2020yys}, one can also consider a restricted version of condition C2 which demands that this secondary RHN production is subleading comparing to the pre-existing one (Eq. \eqref{ns-to-np}). Finally, condition C3 assumed that the unconstrained CP-violating phases in the neutrino sector are not strong enough to make a sizable contribution to the matter asymmetry (Eq. \eqref{CC3}). C3 is the opposite limit of what is assumed in leptogenesis scenarios.

The lightest RHN gets decoupled after the freeze-out of $W_R$ field at $T_{W_R}$ (Eq. \eqref{TTWR}). The heavier RHNs decay after temperature gets below their masses, and the spectator effects reshuffle the primordial baryon and SM lepton numbers. The final baryon to photon ratio and DM relic density are presented in Eq.s \eqref{eta-B-} and \eqref{DM-relic} respectively. This setup can explain $\eta_{\mB}$ and $\Omega_{\rm DM}$ in a wide range of its parameter space (see Fig. \ref{fig:parameter-space}). 
If $\bN_1$ makes all the DM relic density, then its mass is in the range of $keV-GeV$. In case that restricted C2 condition holds, the mass is predicted to be $m_{\rm N_1}\approx 1.7~ GeV$, i.e. a cold DM spectra consistent with structure formation (Eq. \eqref{mN1-}). In that case, baryogenesis and DM today are the remnants of a pure quantum effect (chiral anomaly of $W_R$) in inflation.  Consequently, it can naturally explain the observed
coincidences among cosmological parameters, i.e., $\eta_{\mB}=0.3 P_{\zeta}$ and $\Omega_{\rm DM}=5\Omega_{\mB}$. Besides, this model is a complete
setup that can simultaneously provide plausible explanations for the phenomena (I-IV)  named in the introduction. The summary of this new mechanism is illustrated in Fig. \ref{evolution}. 

It is noteworthy to mention that we can couple the axion to both $SU(2)_R$ and $SU(2)_L$ gauge fields. However, the inflationary production of left-handed baryons and leptons by $SU(2)_L$ (i.e. $\mB_{\rm SM}=\mL_{\rm SM}$) will be completely washed out by the $SU(2)_L$ sphaleon effects which are in thermal equilibrium between $T_{\rm reh}$ and $m_{W_L}$. Since $SU(2)_L$-axion interaction leaves no fermionic remnants today, it is neglected in the minimal realization of this idea proposed in \cite{Maleknejad:2020yys}.

In this setup, P and CP are broken by the VEV of the axion and its interaction with the gauge field. It provides a deep connection between inflation, matter asymmetry, and DM relic density. This alternative mechanism, therefore, 
does not rely on the largeness of the unconstrained CP-violating phases in the neutrino sector
nor fine-tuned masses for the heaviest right-handed neutrinos. Interestingly, sufficient matter creation relates the feeble scale to a geometrical phase transition in cosmology, i.e., the end of exponential expansion of the Universe. Moreover, it demands $m_{W_R}>10^{10}~GeV$ (see Fig.s \ref{fig:DeltaN-massless}-\ref{fig:DeltaN-massive}) which is the scale suggested by the non-supersymmetric SO(10) GUT model with an intermediate left-right symmetry scale \cite{Rizzo:1981su, Bertolini:2012im, Deppisch:2017xhv}. The above relations between the energy scales may be hints of a fundamental connection that we leave for future work.  
As yet another added benefit, this setup comes with a cosmological smoking gun; chiral, non-Gaussian, and blue-tilted gravitational wave background, which can be probed by future CMB missions and laser interferometer detectors.  For an exhaustive discussion on the measurement of this effect, see \cite{Campeti:2020xwn}.

\acknowledgments

The author especially thanks Eiichiro Komatsu for insightful discussions and valuable input during previous collaborations. She also likes to thank G. Giudice, J. Kopp, and M. Shaposhnikov for valuable discussions. 


\appendix

\section{Overview of Minimal Left-Right Symmetric Theories}\label{LRSM-overview}
Here we review the aspects of minimal
left-right symmetric models (LRSM) \cite{Pati:1974y, Mohapatra:1974gc,Senjanovic:1975rk,Davidson:1978pm} that we need in this paper. 

\subsection*{Field and Matter Content:}

The model's field content is presented in Table \ref{Table1}, and in the following, we explain the gauge field, extended Higgs, and fermionic sectors, respectively. The baryon and lepton numbers are denoted by $\mB$ and $\mL$, respectively. Moreover, $L$ and $R$ subscribes represent left- and right-handed fields.

\vspace{0.4 cm}

$\blacktriangle$ {\it{Gauge Group}}
 of the minimal left right symmetric interaction (suppressing color) is 
\bea
\mathcal{G} = SU(2)_R\times SU(2)_L \times U(1)_{\mB-\mL},
\eea
where ($\bW_{R}, \gR$) and ($\bW_{L}, \gL$) are the $SU(2)_R$ and $SU(2)_L$ gauge fields respectively 
\bea
\bW_{R}= W^{a}_{R} ~ \bT_{aR} \an \bW_{L}= W^a_{L} ~ \bT_{aL},
\eea
and ($B_{\mu}, g_{_{\mB\mL}}$) is the $ U(1)_{\mB-\mL}$ gauge field which naturally identifies with the $\mB-\mL$ generator. 
Here $\bT^a_{L,R}$ are the generators of the $SU(2)_{L,R}$
\bea
\bT^a_{L,R} = \frac{\bt_a}{2},
\eea
where $\bt_a$ denotes the Pauli matrices which acts on the $SU(2)$-color indices.
The $\bW_{L,R}$ act on left- and right-handed fields respectively. The strength tensor of $\bW_{L,R}$ are given as
\bea
\bW_{\mu\nu} = \p_{\mu} \bW_{\nu} - \p_{\nu} \bW_{\mu} - i g [ \bW_{\mu},\bW_{\nu}].
\eea
The electric charge $Q$ is defined as
\bea
Q= \bT^3_{L} + \bT^3_{R} + \frac{\mB-\mL}{2} = \bT^{3}_{R} + Y,
\eea
where $Y$ is the hypercharge.

\vspace{0.4 cm}

$\blacktriangle$ {\it{Scalar Sector}} involves three Higgs fields, i.e. a Higgs bi-doublet to produce the Dirac masses, and two triplet Higgs to create Majorana masses for the neutrinos. The $SU(2)_L\times SU(2)_R$ bi-double with $\mB-\mL=0$ is 
\bea
\bPhi= \begin{pmatrix}
\Phi^{0}_1 & \Phi^{+}_2 \\
\Phi^{-}_1 & \Phi^{0}_2 
\end{pmatrix},
\eea
and the $SU(2)_{R,L}$ triplets with $\mB-\mL=2$ are given as
\bea
\bDelta_{R,L}= \begin{pmatrix}
\delta^{+} & ~\delta^{++} \\
\delta^{0} & -\delta^{+} 
\end{pmatrix}_{R,L}.
\eea
The gauge-covariant derivatives of $\bPhi$ and $\bD_{L,R}$ are given as
\bea
\mathcal{D}_{\mu} \bPhi &=& \p_{\mu}\bPhi -  i \gL \bW_{\mu L} \bPhi +  i \gR \bPhi \bW_{\mu R} ,\\
\mathcal{D}_{\mu} \bD_{L,R} &=& \p_{\mu} \bD_{L,R} - i g_{_{L,R}} [\bW_{\mu} , \bD]_{L,R} - ig_{_{\mB\mL}}  B_{\mu} \bD_{L,R}.
\eea
{\renewcommand{\arraystretch}{1.2}
\setlength{\extrarowheight}{1pt}
\begin{table}
\begin{center}
\begin{tabular}
{|c|c|c|c|}
\cline{1-4}\rule{0pt}{20pt}
& LRSM Sector & Left-handed & Right-handed \\[2ex]
\hline
\rule{0pt}{15pt}
\multirow{2}{*}{\begin{sideways} Gauge~ \end{sideways}~\begin{sideways} Fields~ \end{sideways}} & $SU(2)_{R} \times SU(2)_L$ & $\bW_{L}$ & $\bW_{R}$ \\[1ex]
\cline{2-4} \rule{0pt}{16pt}
& $U(1)_{\mB-\mL}$ & \multicolumn{2}{c|}{$B_{\mu}$}  \\[1ex]
\hline \rule{0pt}{30pt}
\multirow{2}{*}{\begin{sideways} Fermions~ \end{sideways}} & quarks &   $~q_{iL} =  \begin{pmatrix}
{\boldsymbol{u}}_{iL} \\ {\boldsymbol{d}}_{iL}
\end{pmatrix} ~ : ~ ({\bf{1}}, {\bf{2}}, ~\frac13)~$ & $~q_{iR} = \begin{pmatrix}
{\boldsymbol{u}}_{iR} \\ {\boldsymbol{d}}_{iR}
\end{pmatrix} ~ : ~ ({\bf{2}}, {\bf{1}}, ~\frac13 )~$   \\[1pt] 
\rule{0pt}{30pt}
& leptons  & $l_{iL} = ~\begin{pmatrix}
{\bn}_{iL} \\ {\boldsymbol{l}}_{iL}
\end{pmatrix}~ : ~ ({\bf{1}}, {\bf{2}}, -1)~$ & $l_{iR} = ~\begin{pmatrix}
{\bn}_{iR} \\ {\boldsymbol{l}}_{iR}
\end{pmatrix}~ : ~ ({\bf{2}}, {\bf{1}}, -1)~$ \\ [3ex]
\hline
\rule{0pt}{25pt}
\multirow{2}{*}{\begin{sideways} ~Scalars~~~~~~~ \end{sideways}} & $\begin{matrix} \textmd{Higgs SU(2)} \\
\textmd{bi-doublet} \end{matrix}$ & \multicolumn{2}{c|}{ ~~~~~~$~{\boldsymbol{\Phi}} = \begin{pmatrix}
\Phi^0_1 ~~ \Phi^+_2 \\ \Phi^-_1 ~ \Phi^0_2
\end{pmatrix} ~ : ~ ({\bf{2}}, {\bf{2}}, ~0)~$}  \\ [3ex]
\cline{2-4}
\rule{0pt}{25pt}
&  $ \begin{matrix} \textmd{Higgs SU(2)} \\
\textmd{triplets} \end{matrix}$ & $~{\boldsymbol{\Delta}}_L = \begin{pmatrix}
\delta^+ ~~ \delta^{++} \\ \delta^{0} ~ -\delta^{+}
\end{pmatrix}_{\!_{L}}  :  ({\bf{1}}, {\bf{3}},2)$   &  $~{\boldsymbol{\Delta}}_R = \begin{pmatrix}
\delta^{+} ~~ \delta^{++} \\ \delta^{0} ~ -\delta^{+}
\end{pmatrix}_{\!_{R}}  :  ({\bf{3}}, {\bf{1}},2)$ \\ [4ex]
\hline
\end{tabular}
\end{center}
\caption{The field content of the minimal Left-Right Symmetric Model (LRSM) extension of the SM. }
\label{Table1}
\end{table}}

The theory of the Higgs sector is given as
\bea
\mathcal{L}_{_{Higgs}} &=& -{\rm{Tr}}\big[(\mathcal{D}_{\mu} \bD_{R})^{\dag}\mathcal{D}^{\mu} \bD_{R}\big] -{\rm{Tr}}\big[(\mathcal{D}_{\mu} \bD_{L})^{\dag}\mathcal{D}^{\mu} \bD_{L}\big] - {\rm{Tr}}\big[(\mathcal{D}_{\mu} \bPhi)^{\dag}\mathcal{D}^{\mu} \bPhi\big] \nonumber\\
&& - V_{_{Higgs}}(\bPhi,\bD_L,\bD_R),
\eea
where the Higgs potential $V_{_{Higgs}}(\bPhi,\bD_{R},\bD_{L})$ is the most general renormalizable, gauge and parity invariant  potential for $\bPhi$ and $\bD_{L,R}$ \cite{Mohapatra:1980yp,Deshpande:1990ip}. The  Higgs mass spectrum and the scale of each spontaneous symmetry breaking are given by minimizing the Higgs potential. Here we are interested in the cosmological consequences of such potential. For an exhaustive discussion we refer the interested reader to \cite{Maiezza:2016ybz, Dev:2016dja, Dev:2018foq}.

$\blacktriangle$ {\it{Fermionic Sector}}
 consists of three generations of quarks and leptons as
\bea
q_{iL,R} = ~\begin{pmatrix}
{\boldsymbol{u}}_{i} \\ {\boldsymbol{d}}_{i}
\end{pmatrix}_{\!L,R} \an 
l_{iL,R} = ~\begin{pmatrix}
{\bn}_{i} \\ {\boldsymbol{l}}_{i}
\end{pmatrix}_{\!L,R},
\eea
where $\bn_{iR}$ are three RHNs interacting via the $SU(2)_R$ and $U(1)_{\mB-\mL}$. Given that we assume neutrinos are Majorana, and define two Majorana fields associated to the left- and right-handed neutrinos as
\bea
\bn_i \equiv \bn_{iL}+\bn_{iL}^c \an \bN_i \equiv \bn_{iR}+\bn_{iR}^c,
\eea
where the $c$ superscript denotes the charge conjugated field.
For simplicity, we present the left- and right-handed fermions  collectively as 
\bea
\Psi_{JL,R}=(q_1,q_2,q_3,l_1,l_2,l_3)_{L,R},
\eea
which are specified by the Lagrangian
\bea\label{L-Psi}
\mathcal{L}_{\Psi} = i \sum_{J=1}^6 \bar{\Psi}_{JR}\bs^{\mu} \mathcal{D}_{\mu} \Psi_{JR} + i \bar{\Psi}_{JL}\bar{\bs}^{\mu} \mathcal{D}_{\mu} \Psi_{JL},
\eea
where the spinor gauge-covariant derivatives are
\bea
\mathcal{D}_{\mu} \Psi_{L,R} &=& (D_{\mu} -  i g_{_{L,R}} \bW_{\mu L,R} - \frac{i g_{_{\mB\mL}}(\mB-\mL)}{2} B_{\mu}) \Psi_{L,R},\\
D_{\mu} & \equiv & \p_{\mu} + \omega_{\mu},
\eea
where $\omega_{\mu} $ is the spin connection. \footnote{ The spin connection is defined as $\omega_{\mu} \equiv \frac{i}{2}\omega^{\alpha\beta}_{~~\mu}\Sigma_{\alpha\beta}$ where
$\Sigma_{\alpha\beta} = \frac{i}{4}[\gamma_{\alpha},\gamma_{\beta}]$ and $
{\bf{\omega}}^{\alpha\beta}_{~~\mu} \equiv {\bf{e}}^{\alpha\nu}  \nabla_{\mu} {\bf{e}}^{\beta}_{~\nu}$.}
For the cosmological background, we have $\bs^{\mu}\omega_{\mu}=\frac32 H \I$ and 
\bea
\bs^{\mu}=(\I, \frac1a\bs_i) \an \bar{\bs}^{\mu}=(\I, -\frac1a\bs_i),
\eea
where $\bs_i$ are the Pauli matrices which carries spatial index. \footnote{Note that $\bs_{\mu}$ is the curved space form of the flat space $\bs_{\alpha}=(\I, \bs_{i})$ as $\bs_{\mu}= e^{\alpha}_{\mu} \bs_{\alpha}$ where $ e^{\alpha}_{\mu}$ are the tetrads.} The fermions pick up their mass by the Yukawa interactions 
\bea\label{LY}
\mathcal{L}_{Y} &=&  -  \bar{q}_{iL}\big( y^{q}_{ij}\bPhi  + \tilde{y}^{q}_{ij} \tilde{\bPhi} \big) {q}_{jR} - \bar{l}_{iL} \big(  y^{l}_{ij} \bPhi + \tilde{y}^{l}_{ij} \tilde{\bPhi} \big) {l}_{jR}  - \frac12 Y^{R}_{ij} ~ \bar{l}^c_{iR} \tilde{\bD}_R~ {l}_{jR} - \frac12 Y^{L}_{ij} ~ \bar{l}^c_{iL} \tilde{\bD}_L~ {l}_{jL} \nonumber\\
&& + h.c.,
\eea
where $l^{c}_{Ri}= C ~l^{*}_{Ri}$ is the charged conjugated $l_{Ri}$, and
\bea
\tilde{\bPhi} \equiv \bt_2 \bPhi^* \bt_2 \an \tilde\bD \equiv i\bt_2 \bD.
\eea

\subsection*{Symmetry Breaking Structure, New Fundamental Scale, and Mass:}

Once the neutral component of $\bD_{R}$ acquires a VEV as
\bea
\langle \bD_R \rangle =  \begin{pmatrix} 
0 ~ & ~0 \\
\kappa_{R} ~& ~ 0
\end{pmatrix},
\eea
both of the $\mB-\mL$ and left-right symmetries are spontaneously broken. 
That introduces a new fundamental scale, i.e. $\Lambda_F = \kappa_{R}$, which is much higher than the EW scale, $\Lambda_{W} \simeq 246~ GeV$. The 1st SSB breaks the gauge symmetry down to the SM electroweak symmetry as
\bea
SU(2)_L \times SU(2)_R \times U(1)_{\mB-\mL} \xrightarrow[\Lambda_{F}]{1^{st}~ \rm{SSB}} SU(2)_L\times U(1)_Y.
\eea
All non-Standard Model heavy particle masses are related to the VEV of $\bD_R$. The charged and neutral $SU(2)_R$ gauge bosons pick up the following masses
\bea
m_{_{W_{R}}} = \gR \kappa_{R} \an  m_{_{Z_R}} =  \frac{\gBL}{\gY}m_{W_{R}},
\eea
where $\gY$ is given as
\bea
\gY = \frac{\gBL \gR}{\sqrt{\gBL^2+\gR^2}}.
\eea 
The right-handed neutrinos get Majorana mass terms as
\bea\label{Majorana-mass-N}
\mathcal{L}_{Y}^{SSB1} = \frac{\kappa_{R}}{2} Y^R_{ij} \bn_{jR}^{T}C  \bn_{iR}  + h.c.  ,
\eea
which leads to the Majorana mass matrix $
M_{Rij} = \kappa_{R}~Y^R_{ij} $. Finally, when the temperature drops below the EW phase transition, i.e. $T=\Lambda_{ W}$, the 2nd SSB happens and the neutral components of the bi-doublet receive its VEVs as
\bea
\langle \bPhi \rangle =  \frac{1}{\sqrt{2}}\begin{pmatrix} 
\kappa_1 ~ & ~ 0 \\
0 ~& ~ \kappa_2
\end{pmatrix}.
\eea
That breaks the gauge symmetry to $U(1)_{\rm em}$, i.e.
\bea
 SU(2)_L \times U(1)_Y \xrightarrow[\Lambda_{ W}]{2^{nd}~ \rm{SSB}}  U(1)_{\rm em},
\eea
which provides Dirac masses for the SM particles, SM neutrinos included. After 2nd SSB, therefore, all the SM massive particles pick a Dirac mass similar to SM. The interaction between $\bD_{R}$ and $\bPhi$ with $\bD_L$ in Higgs potential imposes a VEV for the latter once the former fields acquired their VEVs. The VEV of $\bD_{L}$ is of the order of $\mathcal{O}(\frac{\langle \bPhi\rangle^2}{\kappa_R}) \ll \langle \bPhi\rangle$ \cite{Mohapatra:1980yp}. The value of the $\kappa_{1,2}$ is related to the EW scale $\kappa$ as
\bea
\kappa^2_1+\kappa^2_2 = \kappa^2 = (246 ~ GeV)^2.
\eea
In the limit of our interest, $\kappa_{R}\gg \kappa_{1},\kappa_{2},\kappa_L$ in which the left and right charged and neutral gauge bosons are decoupled. Thus, we can consider $W_{L,R}^{\pm}$ and $Z_{L,R}$ as physical states. Here for simplicity we also assume $$\kappa_1\ll \kappa_2 \an \kappa_2\simeq  \kappa.$$
 We summarize the symmetry-breaking structure of the setup in Table \ref{Table2}. Below we will discuss its consequences on active neutrinos.

\setlength{\extrarowheight}{1pt}
\begin{table}[H]
\begin{center}
\begin{tabular}
{|c|c|c|}
\hline
Symmetry Group & After 1st SSB  & After 2nd SSB \\ 
$SU(2)_R \times SU(2)_L \times U(1)_{\mB-\mL}$ & $SU(2)_L \times U(1)_Y$ & $U(1)_{\rm em}$ \\
\hline
\rule{0pt}{20pt}
Higgs VEV & $\langle \bD_R \rangle =  \begin{pmatrix}
0~~ & 0 \\
\kappa_{_{R}} & 0
\end{pmatrix}$ & $\langle \bPhi \rangle = \frac{1}{\sqrt{2}}\begin{pmatrix}
\kappa_1 & ~0 \\
0 & ~\kappa_2
\end{pmatrix} $ \& $\langle \bD_L \rangle = \begin{pmatrix}
0~~ & 0 \\
\kappa_{_{L}} & 0
\end{pmatrix}$ \\ [3ex]
\hline
\rule{0pt}{15pt}
Massive Particles & $W^{\pm}_R$, $Z_R$  ~ and ~ $\bN_i$ 
& SM ~ and ~ $\bn_i$ \textit{(Seesaw type-I \& II)}
\\ [1ex]
\hline
\end{tabular}
\end{center}
\caption{The spontaneous symmetry breaking structure of the minimal LRSM.}
\label{Table2}
\end{table}

\subsection*{Neutrino masses; Natural Seesaw Mechanism:}

The first SSB provides Majorana masses for the RHNs and the second SSB gives Dirac masses to neutrinos as well as an induced Majorana mass to left-handed neutrinos. The neutrino mass matrix as
\bea
M_{\nu} = \begin{pmatrix} M_{L} & m_{D}\\
m_{D}^T & M_{R}
\end{pmatrix},
\eea
where the Majorana mass matrices $M_{R,L}$ are
\bea
M_{R ij} =  \kappa_R  ~ Y_{ij}^{R} \an M_{L ij} = \kappa_L  ~ Y_{ij}^{L}  \sim \mathcal{O}(\frac{\kappa^2}{\kappa_R}),
\eea
and the Dirac mass matrix is
\bea
m_{D ij} = \frac{\kappa}{2} \tilde{y}^l_{ij}.
\eea
Given the fact that $m_D\ll M_{R}$, we can diagonalize the mass matrix and find the masses of the active neutrinos as
\bea
m_{\nu} \approx - m_{D}^{T} M^{-1}_{R} m_{D} + M_{L}  =   \frac14 \frac{(\kappa^{*}\tilde{y}^l )^2}{ \kappa_R~Y^R} + \kappa_L~ Y^L.
\eea
Note that $\kappa_L$ is a (small) induced VEV and the contribution of both the first term (seesaw type-I) and the second term (seesaw type-II) are of the same order. Thus, in minimal LRSM, the neutrino mass is a hybrid seesaw type-I, and II \cite{Mohapatra:1980yp}.

\subsubsection*{Experimental Constraints on Parameters:}~

Various Experimental limits can be placed on the mass scales and mixing parameters of the LRSM. First, considering the charged lepton Yukawa couplings as a guide to the neutrino ones suggests $10^{-10} \lesssim \frac{y^2}{Y} \lesssim 1$ which implies a successful seesaw requires 
\bea
10~ TeV \lesssim \kappa_R \lesssim 10^{15} ~GeV.
\eea
Next, regardless of the details of the SSB, there is a theoretical lower bound on the ratio of $\gR$ to $\gL$ \cite{Dev:2016dja} 
\bea\label{lower-gR}
\frac{\gR}{\gL} \geq \tan \theta_w \simeq 0.55.
\eea
That gives $m_{_{Z_R}}\approx 1.7 m_{_{W_R}}$.
Finally, there are several constraints for the right-handed charged and neutral gauge boson mass and mixing parameters. These arise due to their direct production or virtual contributions at colliders or astrophysical processes. The $K_L - K_S$ kaon mass difference measurement \cite{Barenboim:1996nd} places a lower bound on the mass of $W_{R}$ as $m_{_{W_R}} > 1.6 ~TeV$ and the mixing angle between $Z_R$ and $Z_L$ is constrained to be less than $10^{-4}$. The possible low-energy $W_{R}$ has been the target of several LHC collaborations which puts the current bound as $m_{_{W_R}}> 3 ~ TeV$ \cite{Bertolini:2014sua}. For an exhaustive discussion of the phenomenological implications and constraints of LRSM, see \cite{Beringer:1900zz}.

\subsection*{Gauge coupling evolution}\label{RG}

There is a significant difference between a high scale inflation and electroweak scale. Thus the running of the gauge couplings might be sizable. In the one-loop approximation, the RGE for the $SU(\mN_c)$ gauge coupling with $\mN_f$ Weyl or Majorana fermions in the fundamental representation and $\mN_s$ Higgs fields in the $R_s$ representation is given as 
\bea\label{running}
\frac{dg_i}{d\ln(\frac{k}{\mu})} =  b_i \frac{g_i^3}{(4\pi)^2},
\eea
where $k$ is the momentum, $\mu$ is a given scale associated with our renormalization and $b_i$ is 
\bea
b_i = - \bigg[ \frac{11}{3}\mN_c - \frac13 \mN_{f} - \frac13 \mN_{s} T(R_s) \bigg].
\eea
Here $T(R)$ is the index of the irreducible representation
$T(R) \delta_{ab} \equiv {\rm{Tr}}(\bT_a\bT_b)$, 
where for fields in the fundamental representation of $SU(\mN_c)$ it is $T(R_{\rm{fund}})=\frac12$ and for the adjoint representation $T(R_{\rm{adj}})=\mN_c$.
The $SU(2)_L$ and $SU(2)_R$ gauge fields with the Higgs bi-doublet and triplet have $b_{_{R,L}}=-\frac{7}{3}$.
Given $\gL(m_{_{Z_L}}) \simeq 0.65$, the RGE determines the $L$ gauge coupling at the GUT scale (assuming inflation happens around GUT, i.e. $k=H\simeq 10^{13}$ GeV) as 
\bea\label{gL-RG}
\gL(H) \simeq 0.56.
\eea
The gauge coupling of $SU(2)_R$ at the scale of inflation is 
\bea\label{gR-RG}
\gR(H) \simeq 0.56 ~\frac{\gR(m_{_{Z_L}})}{\gL(m_{_{Z_L}})}.
\eea
Using the theoretical lower bound on $\gR$ in Eq. \eqref{lower-gR}, we arrive at
\bea\label{gR-RG--}
0.3 \leq \gR(H) \leq 0.56.
\eea

\section{Massive Sterile Neutrino production in Inflation}\label{massive-N}

This appendix presents the analytical calculations of massive RHN production by the axion in inflation. In this work, we restrict ourselves to the cases with $\langle \bW_R \rangle =0$. \footnote{The fermion production by the Schwinger effect with $\langle \bW_R \rangle \neq 0$ in $SU(2)$-axion inflation is studied in \cite{Maleknejad:2019hdr}.} From Eq. \eqref{EOM-Lepton}, we find the linearized field equation of $\bn_{jR}$ as
\bea\label{neutrino-EOM-App}
(i\bs^{\mu}\p_{\mu} +\frac{3i}{2} H - 2\tilde\xi H) \bn_{jR} -   m_{N_j} \bn^c_{jR} \simeq 0.
\eea
As a Majorana fermion, $\bN_{j}\equiv \bn_{jR}+\bn_{jR}^c$ can be decomposed as
\bea\label{Psi-Majorana}
\bN_{j}= \sum_{s=\pm} \frac{1}{a^{\frac32}}\int d^3k \left[X^s_{j{\bf{k}}}(\tau)c^s_{j\bk}  e^{i\bk.\bx} +Y^s_{j{\bf{k}}}(\tau)c^{s\dagger}_{j\bk} e^{-i\bk.\bx}\right]~{\bf{E}}^s_{{\bf{k}}}\,,
\eea
where $c^s_{j\bk}$ and $c^{s\dag}_{j\bk}$ are the annihilation and creation operators of the RHNs as
\bea
\{c^{s}_{i\bk},c^{s'\dag}_{j\bk'}\} = \delta^{ss'} \delta_{ij} \delta^{(3)}(\bk-\bk'),
\eea
and ${\bf{E}}^{\pm}_{{\bf{k}}}$ are the $\pm\frac12$ helicity polarization states 
\bea
{\bf{E}}^+_{{\bf{k}}}= \frac{k_{\alpha}{\boldsymbol{\sigma}}^{\alpha}}{\sqrt{2k(k+k_3)}} \begin{pmatrix}
1 \\ 0
\end{pmatrix} \an 
{\bf{E}}^-_{{\bf{k}}}= \frac{k_{\alpha}\bar{{\boldsymbol{\sigma}}}^{\alpha}}{\sqrt{2k(k+k_3)}} \begin{pmatrix}
0 \\ 1
\end{pmatrix}.
\eea
These helicity 2-spinors are the eigenstates of the helicity operator and satisfy ${\bf{E}}^{-s}_{{\bf{k}}}=-is\bs_2 {\bf{E}}^{s*}_{{\bf{k}}}$. The Majorana condition then requires
\bea
Y^s_{j\bk}=sX^{-s*}_{j\bk}.
\eea
The pair of coupled first order differential equations for $\bn_{iR}$ and $\bn^{c}_{iR}$ coming from Eq. \eqref{neutrino-EOM-App} can be decoupled into two second order differential equations for the mode functions $X^{\pm}_{j\bk}(\tau)$. Upon field redefinition
\bea
\tilde{X}^{s}_{j\bk} \equiv \sqrt{2\x} X^{s}_{j\bk},
\eea
we have 
\bea
\p_{\x}^2 \tilde{X}^s_{j\bk} + \big[ 1 - \frac{2i\kappa_s}{\x} + \frac{\frac14-\mu^2_j}{\x^2} \big] \tilde{X}^s_{j\bk}  =0,
\eea
where $\kappa_{s}$ and $\mu_j$ are
\bea
\kappa_{s} = s(\frac{1}{2}+2i\tilde\xi) \an \mu^2_j=- \big(\frac{m_{\rm{N}_j}}{H}\big)^2 -(2\tilde\xi)^2.
\eea
Setting the initial conditions with Bunch-Davies vacuum, the solutions are
\bea
\tilde{X}^+_{j\bk} &=& \frac{1}{(2\pi)^{\frac32}} e^{-\xi\pi} W_{\kappa_{+},\mu_j}(-2i\x),\\
\tilde{X}^-_{j\bk} &=& -\frac{i}{(2\pi)^{\frac32}}(\frac{m_{\rm{N}_j}}{H}) e^{\xi\pi} W_{\kappa_{-},\mu_j}(-2i\x).
\eea
Notice that the $-\frac12$ helicity mode of the right-handed neutrinos is proportional to their mass. Thus, as we expected, the massless $\bn_{jR}$ has only the $+\frac12$ helicity state. 

Working out the mode functions of the massive sterile neutrinos, we are ready to compute its contribution to the lepton number in Eq. \eqref{Int-n-m} as 
\bea\label{Int-n-m-app}
\bar{n}_{\rm{N}_j} \equiv \int d^3k \langle \bn_{jR}^{\dag} \bn_{jR}\rangle = - \frac{\tilde\xi}{\pi}~\bigg(\frac{m_{\rm{N}_i}}{H}\bigg)^2~ H^3  \mathcal{D}(\tilde\xi,m_{\rm{N}_j}).
\eea
Using point splitting regularization, we analytically calculated the above momentum integral in \cite{Maleknejad:2019hdr}. \footnote{The details of the calculation and point splitting regularization that is used in computing the integral \eqref{Int-n-m-app} can be found in Appendix D of \cite{Maleknejad:2019hdr}. Notice that $\kappa_{+}$ and $\kappa_{-}$ parameters in the current work are denoted as $\kappa_{+}$ and $\tilde{\kappa}_{-}$ in \cite{Maleknejad:2019hdr}. The dimensionless parameter $\xi_A$ in the latter is associated with the VEV of the $SU(2)$ gauge field, which is set to zero in the current work.} We here use the final result which is
\bea\label{fin-J+---}
&& \bigg(\frac{m_{\rm{N}_i}}{H}\bigg)^2~\mathcal{D}(\tilde\xi,m_{\rm{N}_j}) =   \nonumber\\
&& \frac{1}{2\pi} \bigg\{ \frac{2}{3} (1-2\kappa_{I}^2) \bigg(1 - \frac{\lvert \mu_j\rvert}{\kappa_{I}} \frac{\sinh(2\kappa_{I}\pi)}{\sinh(2\lvert\mu_j\rvert\pi)}\bigg) + \frac{m_{\rm{N}_j}^2}{H^2} \bigg( 2
-4  \psi^{(0)}(1)  - \frac{8\lvert \mu_j\rvert}{3\kappa_{I} }\frac{\sinh(2\kappa_{I}\pi)}{\sinh(2\lvert\mu_j\rvert\pi)}\bigg)  \nonumber\\
&& + \frac{m_{\rm{N}_j}^2}{H^2}  \sum_{s=\pm}  {\rm{Re}}\bigg[  
   \frac{e^{2\lvert\mu_j\rvert\pi}-e^{-2s\kappa_{I}\pi}}{\sinh(2\lvert\mu_j\rvert\pi)} \psi^{(0)}(-is\kappa_{I} -i\lvert\mu_j\rvert)  - \frac{e^{-2\lvert\mu_j\rvert\pi}-e^{-2s\kappa_{I}\pi}}{\sinh(2\lvert\mu_j\rvert\pi)} \psi^{(0)}(-is\kappa_{I}+i\lvert\mu_j\rvert)
 \bigg] \bigg\},\nonumber\\
\eea
 in which $\kappa_I \equiv 2\tilde\xi$ and $\psi^{(0)}(z) \equiv \frac{d\Gamma(z)}{z}$ is the digamma function.

\section{Phenomenological Model of Reheating}

\begin{figure}
\centering
\includegraphics[height=0.205\textheight]{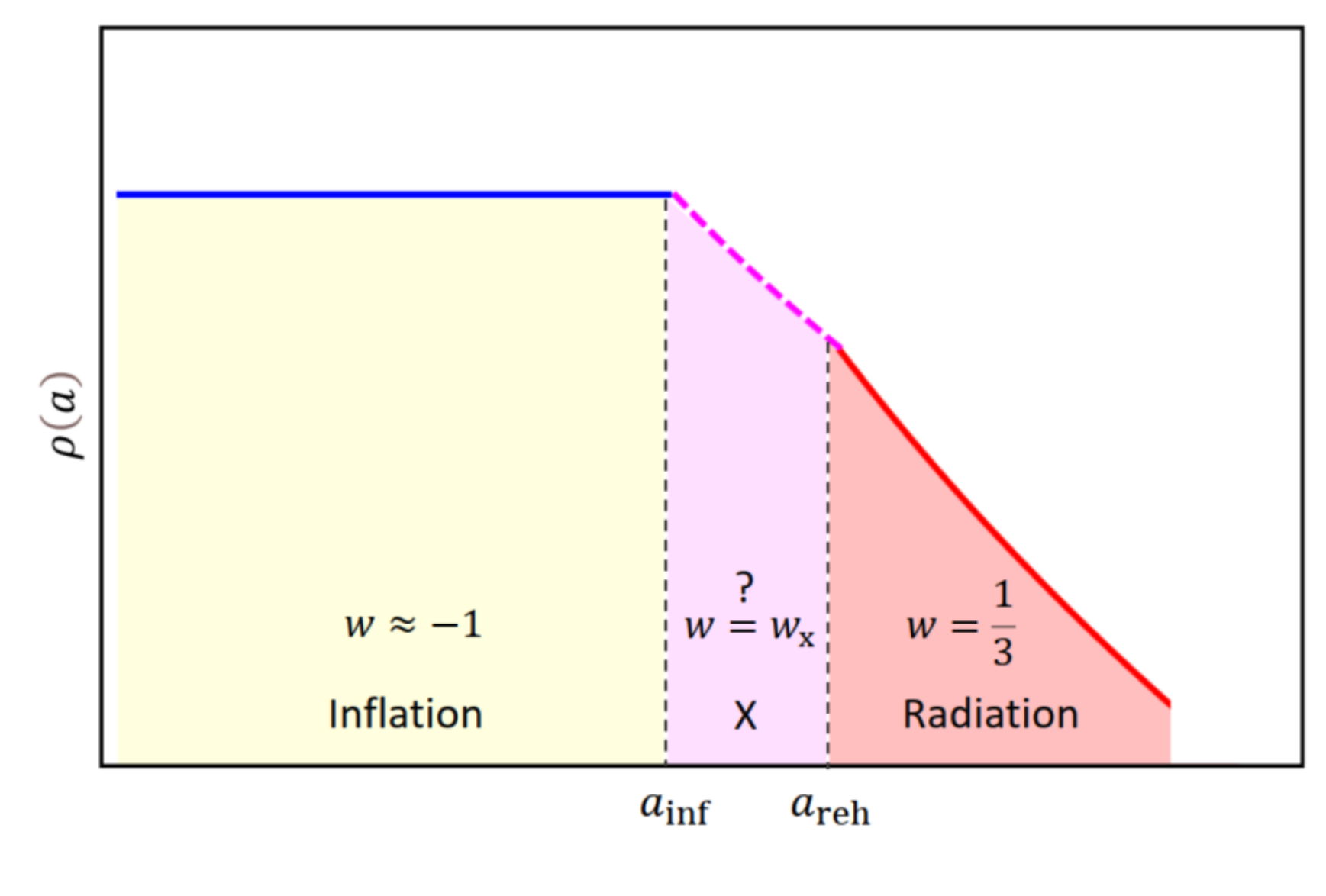} \includegraphics[height=0.205\textheight]{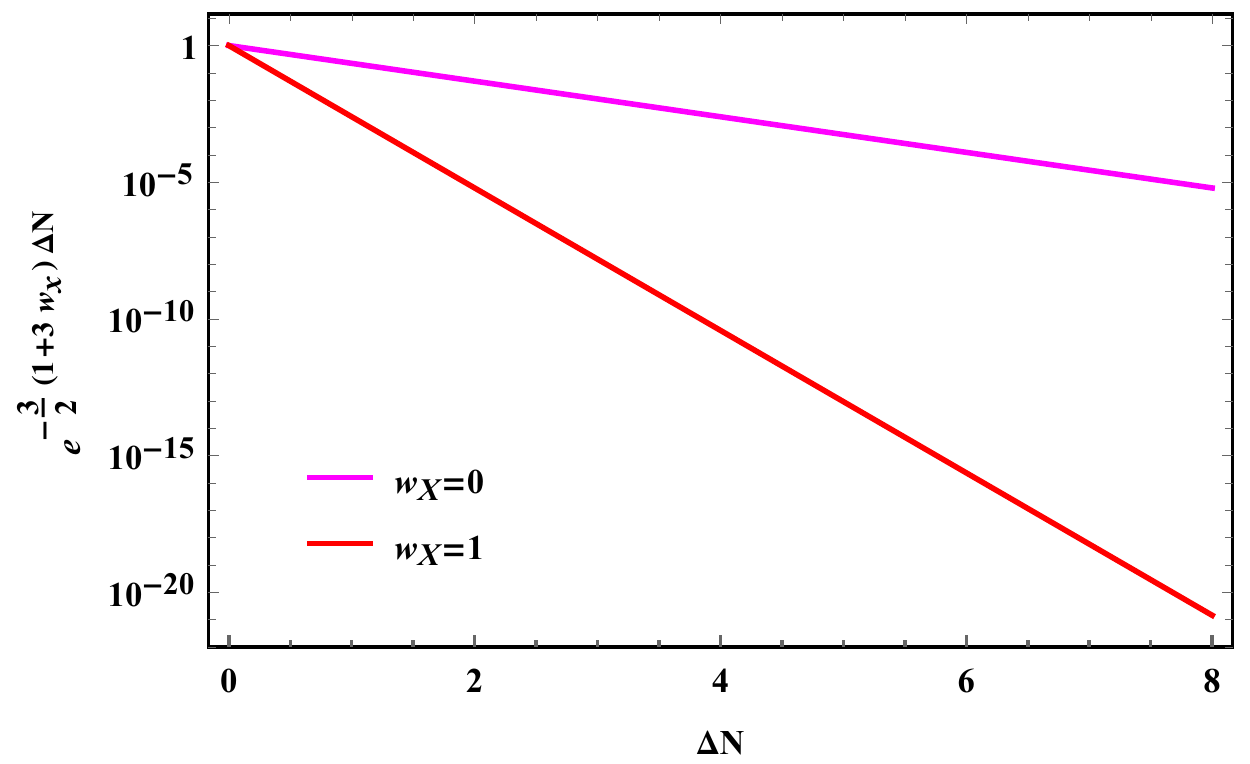}
\caption{Left Panel: The energy density of Universe vs scale factor. The dashed (pink) line which connects inflation to radiation era is a possible unknown intermediate phase with an average equation of state $w=w_X$. Right Panel: The prefactor  $e^{-\frac32(3w_X+1)\Delta N}$ in Eq. \eqref{ns-to-np} vs $\Delta N=\ln\big(\frac{a_{\rm reh}}{a_{\rm inf}}\big)$. (Fig. adopted from ref. \cite{Maleknejad:2020yys})}
\label{omega-x} 
\end{figure}

Reheating starts at some point after the end of inflation and ends at $a_{\rm reh}$ with the formation of a dominant thermal bath with temperature $\treh$. Yet, the precise physics of reheating is not well understood. Depends on the details of the post-inflation physics, there may be an intermediate phase $X$ with the average equation of state $w_X$, which connects inflation to the final thermal bath (See Fig. \ref{omega-x}). To quantify our analysis and capture these ambiguities, in this appendix, we introduce a phenomenological model for reheating. Next, we compute the entropy injection by the decay of RHNs in our setup.

In that case, the energy density at the end of reheating is related to $\rho_{\rm inf}$ as
\bea
\rho_{\rm reh}=\delta_{\rm reh} ~\rho_{\rm inf}~\big(\frac{a_{\rm inf}}{a_{\rm reh}}\big)^{4}.
\eea
The parameter $\delta_{\rm reh}$ is the efficiency of the reheating process 
\bea\label{varepsilon}
\delta_{\rm reh} \approx {\rm{exp}}\big(-(3w_X-1)\Delta N\big),
\eea
which models our ignorance about physics of reheating in terms of two parameters; $w_X$ and $\Delta N$ given as
\bea
\Delta N = \ln(\frac{a_{\rm reh}}{a_{\rm inf}}),
\eea
which is the number of e-folds between end of inflation  until the formation of the thermal bath. 
The ratio $\frac{n^{s}_{\rm{N}}}{n^{p}_{\rm{N}}}$ in Eq. \eqref{ns-to-np} is related to $\Delta N$ as 
\bea\label{alpha-Delta-N}
\frac{n^{s}_{\rm{N}}}{n^{p}_{\rm{N}}} \propto e^{-\frac32(1+3w_X)\Delta N}.
\eea

Two possible scenarios for the intermediate phase, i.e. $X$-era in Fig. \ref{omega-x}, are: 1) inflation ends in a short period of matter domination with $w_X=0$ with reheating efficiency parameter as
\bea
\delta_{\rm reh} \simeq \big(\frac{a_{\rm reh}}{a_{\rm inf}}\big) = e^{\Delta N} >1,
\eea
which gives
\bea
\frac{n^{s}_{\rm{N}}}{n^{p}_{\rm{N}}} \propto e^{-\frac32\Delta N},
\eea
or 2) inflation ends with domination of the kinetic term such that $w_X=1$ and $\delta_{\rm reh}$ is
\bea
\delta_{\rm reh} \simeq \big(\frac{a_{\rm inf}}{a_{\rm reh}}\big)^2 =e^{-2\Delta N} <1,
\eea
which gives
\bea
\frac{n^{s}_{\rm{N}}}{n^{p}_{\rm{N}}} \propto  ~e^{-6\Delta N} .
\eea
The factor $e^{-\frac32(3w_X+1)\Delta N}$ for these two preheating scenarios are presented in the right panel of Fig. \ref{omega-x}.

\subsection{Entropy Injection}\label{entropy-sub}

 The out of thermal equilibrium decay of heavy RHNs, i.e. $\bN_{2,3}$, at $T=m_{\rm N_{2,3}}$ injects entropy to the hot plasma and increase the energy of radiation as
\bea
\rho_{\rm rad} (a_{_{{\rm N}_i}}) = S^{\frac43}  ~ \bigg(\frac{a_{\rm reh}}{a_{_{{\rm N}_i}}}\bigg)^4 ~\rho_{\rm reh},
\eea
where $S$ is the entropy injection factor given as
\bea
S = 1 +  \frac{1}{3} ~ \bigg(\frac{m_{\rm{N}_i}}{\mpl}\bigg) \bigg(\frac{H}{\mpl}\bigg) \bigg(\frac{a_{_{{\rm N}_i}}}{a_{\rm reh}}\bigg) \bigg(\frac{n_{\rm N_i}(a_{\rm reh})}{\delta_{\rm reh} H^3}\bigg),
\eea
where $n_{\rm N_i}(a_{\rm reh})$ is the total number density of $\bN_i$, i.e. 
\bea
n_{\rm N_i} \equiv n^p_{\rm N_i} + n^s_{\rm N_i}.
\eea
From Eq. \eqref{nsN}, the freeze-in part of the density is 
\bea\label{Eq-1-n}
\frac{n^s_{\rm N_i}(a_{\rm reh})}{\delta_{\rm reh} H^3} \approx 3 \times 10^{11} ~ \exp{[-\frac{(13-3w_X)}{4}\Delta N]},
\eea
while the contribution of the primordial density in Eq. \eqref{npN} gives 
\bea\label{Eq-2-n}
\frac{n^p_{\rm N_i}(a_{\rm reh})}{\delta_{\rm reh} H^3} \approx \frac{\alpha_{\rm inf}(\xi)}{3} ~ \exp{[-(4-3w_X)\Delta N]}.
\eea
Given that $\frac{H}{\mpl}<10^{-5}~GeV$ and the mass of the heaviest RHN is around $10^{12}~GeV$, Eq.s \eqref{Eq-1-n} implies that contribution of $n^s_{\rm N_i}$ (freeze-in mechanism) to the entropy injection is negligible in our setup.  Therefore, we have
\bea\label{Entropy}
S \approx 1 +  10^{-7} ~ \alpha_{\rm inf}(\xi)   \exp{[-(4-3w_X)\Delta N]} \bigg(\frac{H}{\mpl}\bigg).
\eea
In case that the entropy injection is sizable in our setup, the baryon to photon ratio is
\bea\label{eta-H---}
\eta^0_{\mB} \approx 3 \bigg(\frac{g_{\rm eff}}{100}\bigg)^{\frac34}\frac{\alpha_{\rm inf}(\xi)}{\big(\delta_{\rm reh}\big)^{\frac34}~S} \bigg(\frac{H}{\mpl}\bigg)^{\frac32},
\eea
To agree with the date, we need
\bea\label{eta-H--}
\frac{H}{\mpl} \approx 10^{-6} ~\alpha_{\rm inf}^{-\frac23}(\xi) ~ \delta_{\rm reh}^\frac12 S^{\frac23}.
\eea
Combining \eqref{Entropy} and \eqref{eta-H--}, we find a cubic algebraic equation for $S^{\frac12}$, i.e. 
\bea
S-A(\xi,\Delta N) ~ S^{\frac23}-1=0,
\eea
where $A(\xi,\Delta N)$ is
\bea
A(\xi,\Delta N) = 10^{-13} \alpha^{\frac13}_{\rm inf}(\xi) \exp[-\frac{(7-3w_X)}{2}\Delta N].
\eea
The quantity $A(\xi,\Delta N)$ is negligible in the region of our interest (see Fig. \ref{fig:alpha-inf-}). Therefore, our setup has negligible entropy injection
\bea
S\simeq 1.
\eea\label{reheat}

\section{Spectator Effects}\label{spectator-App}
This Appendix is devoted to the spectator effects on matter asymmetry. First, we work out the temperature windows in which each of the $W_{L,R}$ sphalerons are in thermal equilibrium and hence can violate the left-/right-handed $\mB+\mL$. Next, we discuss the lepton flavor effects in our setup.

\subsection{${\bf{W}}_{L,R}$ Sphalerons}
\label{Sphaleron-App}

The $SU(2)_{L,R}$ sphaleron transitions start getting in thermal equilibrium once 
\bea
\frac{\Gamma_{\rm sph}^{L,R}}{T^3}\gtrsim H(T),
\eea
where $\Gamma_{\rm sph}^{L,R}$ is the transition rate per unit time per unit volume so dimensional estimate gives $\Gamma_{\rm sph}^{L,R}\sim (\alpha_{_{L,R}} T)^4$ where $\alpha_{_{L,R}}=\frac{g_{_{L,R}}^2}{4\pi}$. Using lattice simulations the transition rate for the $SU(2)_L$ weak sphalerons has been found in \cite{Bodeker:1999zt} as
\bea
\Gamma_{\rm sph}^{L} = \chi' \alpha_{_{L}}^5 T^4,
\eea
where $\chi' \approx 18$ and the extra $\alpha_{_L}$ factor is due to specific plasma effects \cite{Arnold:1996dy}. The $W_{R,L}$ switch off after their corresponding scale of SSB. 
 Therefore, the $SU(2)_L$ weak lepton and baryon violating processes are in thermal equilibrium in the wide temperature interval 
\bea
100~GeV < T_{\rm sph}^L < 10^{12} ~GeV.
\eea
As a rough estimate, we assume that the same relation holds for the $SU(2)_R$ sphalerons, i.e. 
\bea
\Gamma_{\rm sph}^{R} \sim \bigg(\frac{\alpha_{_R}}{\alpha_{_L}}\bigg)^5 \Gamma_{\rm sph}^{L}.
\eea
Thus, $W_R$ sphalerons are in thermal equilibrium in the following interval
\bea\label{R-sph}
m_{W_R} \leq T^R_{\rm sph} \leq \bigg(\frac{\gR}{\gL}\bigg)^{10} \times 10^{12}~ GeV.
\eea
Given that in our setup $T_{{\rm reh}}< T_{W_R}<m_{W_R}$,  the $W_R$ sphalerons are never in equilibrium to cause any $\mB+\mL$ violating interaction.

 \subsection{Lepton Flavor Effects} \label{Flavor-Sec}

One potentially very significant aspect of leptogenesis is the flavor effects. The flavor-dependent washout and $\mL$ violating interactions can significantly change the value, and even sign of the final baryon asymmetry \cite{Abada:2006fw,Barbieri:1999ma,Blanchet:2006ch}. By the end of inflation, we have a lepton (anti-lepton) quantum state $\lvert l_{inf} \rangle$ ($\lvert \bar{l}_{inf} \rangle$) as
\bea
\lvert l_{inf} \rangle  \equiv  \sum_{\alpha=e,\mu,\tau} C^{\rm inf}_{\alpha} \lvert \alpha \rangle  \an
\lvert \bar{l}_{inf} \rangle  \equiv  \sum_{\alpha=e,\mu,\tau} \bar{C}^{\rm inf}_{\alpha} \lvert \alpha \rangle,
\eea
where $C^{inf}_{\alpha}$ and $\bar{C}^{inf}_{\alpha}$ are specified by the physics of inflation as
\bea
C^{inf}_{\alpha} = \langle \alpha \vert l_{inf} \rangle \an \bar{C}^{inf}_{\alpha} = \langle \bar{\alpha} \vert \bar{l}_{inf} \rangle.
\eea
The composition of this primordial initial leptons and their CP conjugated anit-leptons are different. The CP violating decays of the heavy sterile neutrinos can modify these initial states. At very high temperatures $T\gg 10^{12}~GeV$, however, the interactions are still flavor blind and we can describe leptons as a coherent superposition of charged leptons as
\bea
\lvert l_i \rangle \equiv \sum_{\alpha=e,\mu,\tau} C_{i\alpha} \lvert \alpha \rangle  \quad \textmd{with} \quad  C_{i\alpha} = \langle \alpha \vert l_{i} \rangle,
\eea
where $ C_{i\alpha}$ are coefficients given by the Yukawa matrix which in terms of the active neutrino mass matrix we have $C_{i\alpha}= \frac{m_{\nu}^{\alpha i}}{\sqrt{(m_{\nu}^{\dag}m_{\nu})_{\alpha\alpha}}}$. The flavored decay parameters of $\bN_i$ to $\boldsymbol{l}_{\alpha}$ are defined as
 \bea
 K_{i\alpha} \equiv \frac{\Gamma(\bN_i \rightarrow \Phi^{\dag} \boldsymbol{l}_{\alpha}) + \bar{\Gamma}(\bN_i \rightarrow \Phi^{\dag} \boldsymbol{l}_{\alpha})}{H(T=M_i)} \where \Gamma(\bN_i \rightarrow \Phi^{\dag} \boldsymbol{l}_{\alpha}) = \frac{M_i~Y^{\dag}_{i\alpha}Y_{\alpha i}}{8\pi},
 \eea
and $K_{i} = \sum_{\alpha} K_{i\alpha}$. The Yukawa couplings of neutrinos contain several CP-violating phases which remain unconstrained by the current data. Therefore, the decay of sterile neutrinos can be a CP asymmetric process quantified as
\bea
 \varepsilon_{i\alpha} \equiv \frac{\Gamma(\bN_i \rightarrow \Phi^{\dag} \boldsymbol{l}_{\alpha}) - \bar{\Gamma}(\bN_i \rightarrow \Phi^{\dag} \boldsymbol{l}_{\alpha})}{\Gamma(\bN_i \rightarrow \Phi^{\dag} \boldsymbol{l}_{\alpha}) + \bar{\Gamma}(\bN_i \rightarrow \Phi^{\dag} \boldsymbol{l}_{\alpha})},
\eea
where $ \varepsilon_{i} = \sum_{\alpha}  \varepsilon_{i\alpha} $ is the CP-asymmetry. 

In the light of the current neutrino oscillations data, the RH neutrino mass spectrum turns out to be typically highly hierarchical. For the sake of concreteness, in this work, we consider
\bea\label{Apx-mN}
m_{{\rm N}_3}\gtrsim 10^{12}~GeV \gg m_{{\rm N}_2} \gtrsim 10^{9}~GeV \gg m_{{\rm N}_1},
\eea
where $m_{{\rm N}_1}$ is assumed to be lower than the EW scale. Furthermore, we assume that the lightest sterile neutrino has feeble Yukawa interactions with the SM and hence a DM candidate, i.e. 
\bea
K_{1\alpha}\ll 1.
\eea
Therefore, only the two heavy sterile neutrinos, $\bN_2$ and $\bN_3$ contribute to the seesaw mechanism as well as decays and washouts. Moreover, due to the hierarchical neutrino mass spectrum, the decays and washout of $\bN_2$ and $\bN_3$ occur in separate stages with no overlaps. As a result, the decay processes can be studied by the following semi-classical Boltzmann equations for 
$\eta_X \equiv \frac{n_{X}}{n_{\gamma}}$ ($\eta_X^{\rm eq}= \frac{n^{\rm eq}_{X}}{n_{\gamma}}$)
\bea
\frac{d\eta_{{\rm N}_i}}{dz_i} &=& -D_i ~ (\eta_{{\rm N}_i}-\eta_{{\rm N}_i}^{\rm eq}),\\ \label{Bolzmann-2}
\frac{d\eta_{_{\updelta_{i}}}}{dz_i} &=& \varepsilon_i~D_i ~ (\eta_{{\rm N}_i}-n_{{\rm N}_i}^{\rm eq}) -  W_i ~ \eta_{_{\updelta_i}}, \\ \label{Bolzmann-3}
\frac{d\eta_{_{\updelta_{i^{\bot}}}}}{dz_i} &=& 0.
\eea
where $i=2,3$, $z_i=\frac{m_{N_i}}{T}$, and $D_i,W_i $ are the decay, and washout terms, and $\updelta$ is
\bea
\updelta \equiv \mB - \uL.
\eea
The decay terms and the related washout terms are given as
\bea
D_i(z_i) \equiv \frac{\Gamma_i(z_i)}{Hz_i} \an W_i(z_i) \equiv \frac12 D_i(z_i) \frac{n^{\rm{eq}}_{{\rm{N}}_i}(z_i)}{n^{\rm{eq}}_{\uL}}.
\eea
At very high temperatures $T\gg 10^{12}~GeV$, the interactions are flavor blind and we can describe leptons as a coherent superposition of charged leptons.  At temperatures $10^9~GeV < T <10^{12}~GeV$, $\tau$ lepton-Higgs interactions are fast and destroy the coherence of the lepton states produced by $\bN_i$ decay. Therefore the Boltzmann Eq.s \eqref{Bolzmann-2} and \eqref{Bolzmann-3} are effectively described by two incoherent SM flavors, i.e., $\tau$ and $\tau^{\bot}=e+\mu$. The SM lepton asymmetry after decay of $\bN_2$ at $T=M_2\gtrsim 10^{9} ~GeV$ is 
\bea
n_{\updelta}(z_2) = n^{p,f}_{\updelta}(z_2) + n^{\rm{N}}_{\updelta}(z_2),
\eea
where $n^{p,f}_{\updelta}(z_2)$ is the contribution of the primordial asymmetry $n^{p,i}_{\updelta}$, as
\bea
n^{p,f}_{\updelta}(z_2) = \mathcal{C} ~ n^{p,i}_{\updelta} + e^{-\int^{z_2}_{1} \frac{dW_i(z')}{dz'} dz'} n^{p,i}_{\updelta_{3^{\bot}}},
\eea
and $n^{\rm{N}}_{\updelta}(z_2)$ is the lepton number produced by the CP asymmetric decay of $\bN_2$, i.e.
\bea
n^{\rm{N}}_{\updelta}(z_2) \approx \varepsilon_2 \kappa_2(z_2),
\eea
in which $\kappa_2(z_2)$ is the efficiency factor of the CP asymmetric decay. In this work we are interested in the limit 
\bea
\frac{n^{\rm{N}}_{\updelta}(z_2) }{n^{p,f}_{\updelta}} \ll 1 \quad \textmd{(condition C3)}.
\eea
As a result, the SM lepton asymmetry after the washout effects is
\bea
n^{p,f}_{\updelta} = \mathcal{C} ~ n^{p,i}_{\updelta}.
\eea
For our $\bN_i$ mass spectrum given in Eq. \eqref{Apx-mN}, the decay process consists of two separate stages, which we will study in the following to find the desired  $\mathcal{C}$.

\vspace{0.7 cm}

{\bf{First Stage~-}} decay of $\bN_3$ ($m_{{\rm N}_3}\gtrsim 10^{12} ~ GeV$):  \newline
The decay of $\bN_3$ washes out the pre-existing asymmetry in the direction of heavy neutrino lepton flavor $\vert l_3 \rangle$ while leaves the component normal to it unchanged. The pre-existing asymmetry can be decomposed as
\bea
n_{\updelta}^p = n_{\updelta_{3}}^p +   n_{\updelta_{3^{\bot}}}^p,
\eea
where $ n_{\updelta_{3}}^p$ ($ n_{\updelta_{3^{\bot}}}^p$) is the asymmetry parallel (perpendicular) to $\vert l_3 \rangle$. The above superposition sum is due to the linearity of the Boltzmann equations. The residual values of the primordial asymmetries are 
\bea\label{Eq-N3-d}
n^p_{\updelta_3} = A^0_{3} e^{-\frac{3\pi}{8}K_{3}}~ n_{\updelta}^{p,i} \an
n^p_{\updelta_{3^{\bot}}} = (1-A^0_{3}) ~n_{\updelta}^{p,i},
\eea
where $A^0_{3}$ is the tree-level probability of the primordial asymmetry to be in the direction of $\vert l_3 \rangle$.

{\bf{Second Stage~-}} decay of $\bN_2$  ($ 10^{12}~GeV \gg m_{{\rm N}_2}\gtrsim 10^{9} ~ GeV$): \newline
This stage of our post inflationary evolution can be effectively described by two SM flavors, i.e., $(\tau,\tau^{\bot}=e+\mu)$ and two relevant flavors of the sterile neutrinos $(\rm{N}_2,\rm{N}_3)$. At temperatures $10^{9}~GeV< T <10^{12} ~GeV$, $\tau$ lepton-Higgs interactions are fast and destroy the coherence of the lepton states produced by $\bN_i$ decay. Thus we need to consider separate Boltzmann equations for the components parallel and orthogonal to $\tau$, i.e. 
\bea
n^p_{\updelta_{3}} = n^p_{\updelta_{3\tau}} + n^p_{\updelta_{3\tau^{\bot}}},
\eea
in which
\bea
n^p_{\updelta_{3\tau}} = A^0_{3\tau} n^p_{\updelta_{3}} \an n^p_{\updelta_{3\tau^{\bot}}} = (1-A^0_{3\tau}) n^p_{\updelta_{3}},
\eea
where the probabilities $A^0_{i\tau}$ ($i=1,2,3$) are given in terms of the flavored decay parameters as
\bea
A^0_{i\tau}=\frac{K_{i\tau}}{\sum_{\alpha} K_{i\alpha}}.
\eea
 From that we can define 
\bea
n^p_{\updelta_{\tau}} \equiv n^p_{\updelta_{3\tau}} + n^p_{\updelta_{3^{\bot}\tau}} \an n^p_{\updelta_{\tau^{\bot}}} \equiv n^p_{\updelta_{3\tau^{\bot}}} + n^p_{\updelta_{3^{\bot}\tau^{\bot}}}.
\eea
Using Eq. \eqref{Eq-N3-d}, we find the explicit form of $n^p_{\updelta_{\tau}}$ and $n^p_{\updelta_{\tau}^{\bot}}$ as
\bea
n^p_{\updelta_{\tau}} &=& \big[ A^0_{3\tau} A^0_{3} e^{-\frac{3\pi}{8}K_3} +(1-A^0_{3\tau}) (1-A^0_{3}) \big] n_{\updelta}^{p,i} ,\\
n^p_{\updelta_{\tau}^{\bot}} &=& \big[ (1-A^0_{3\tau}) A^0_{3} e^{-\frac{3\pi}{8}K_{3}} + A^0_{3\tau} (1-A^0_3)\big] n_{\updelta}^{p,i}.
\eea
At temperatures $T\sim m_{{\rm N}_2}$, the $\bN_2$ wash-out processes act on the flavored asymmetries. The final residual asymmetries in the end of its decay process is
\bea
n^{p,f}_{\updelta_{\tau2}} = A^0_{2\tau} e^{-\frac{3\pi}{8}K_2} n_{\updelta_{\tau}}^p \an n^{p,f}_{\updelta_{\tau2^{\bot}}} = (1-A^0_{2\tau}) n^p_{\updelta_{\tau}}.
\eea
Similar relations hold for $\tau^{\bot}=e+\mu$. Fig. \ref{fig:flavor} shows the geometrical structure of the flavor effects in the flavor space.

\begin{figure}[h!]
\centering
\includegraphics[height=0.3\textheight]{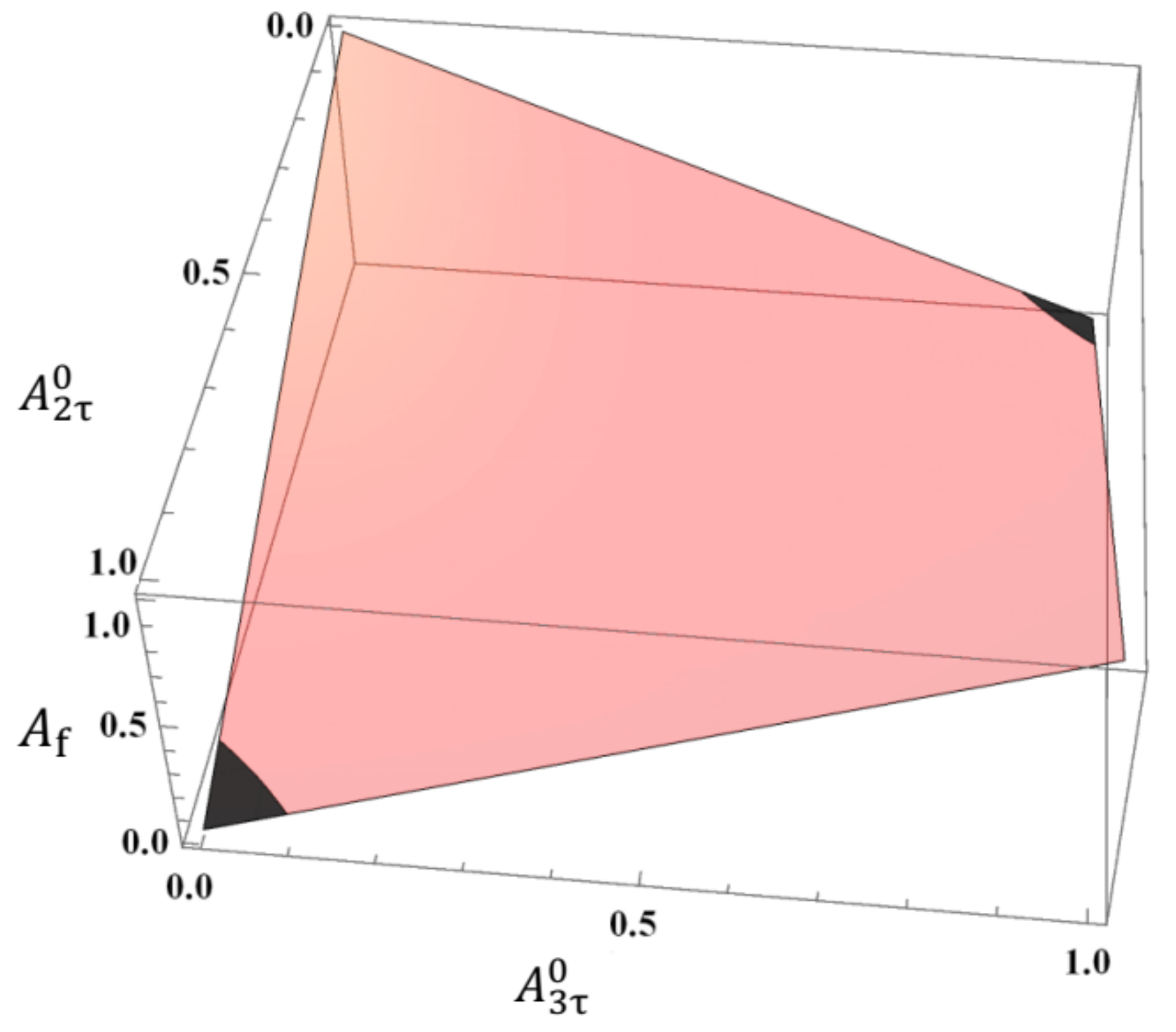} 
\caption{The flavor parameter $\mathcal{A}_f$ in terms of $A^0_{2\tau}$ and $A^0_{3\tau}$. The dark shaded area shows regions with $\mathcal{A}_f<0.1$.}
\label{fig:Af} 
\end{figure}

 The final residual asymmetry in the SM lepton frame is 
\bea
n^{p,f}_{\updelta_{\tau}} &=& A^0_{2\tau} e^{-\frac{3\pi}{8}K_2} n_{\updelta_{\tau}}^p + (1-A^0_{2\tau}) n^p_{\updelta_{\tau}} ,\\
n^{p,f}_{\updelta_{\tau^{\bot}}} &=& (1-A^0_{2\tau}) e^{-\frac{3\pi}{8}K_2} n_{\updelta_{\tau}^{\bot}}^p + A^0_{2\tau} n^p_{\updelta_{\tau}^{\bot}}.
\eea
Considering the most conservative assumption that the decaying terms experience strong washout effects and are negligible, the final remnant of the primordial (inflationary) asymmetry is 
\bea
n^{p,f}_{\updelta} = n^{p,f}_{\updelta_{\tau}} + n^{p,f}_{\updelta_{\tau^{\bot}}} = \mathcal{C}~ n^{p,i}_{\updelta},
\eea
where $\mathcal{C}$ is 
\bea\label{Eq-C}
\mathcal{C} \simeq (1-A^0_{3})    \big(1 - A^0_{2\tau} - A^0_{3\tau} + 2A^0_{2\tau}A^0_{3\tau}\big).
\eea
Fig. \ref{fig:Af} presents
\bea
\mathcal{A}_f \equiv  \big(1 - A^0_{2\tau} - A^0_{3\tau} + 2A^0_{2\tau}A^0_{3\tau}\big).
\eea
vs $A^0_{2\tau}$ and $A^0_{3\tau}$ where the dark shaded area denotes regions with $\mathcal{A}_f<0.1$. As we see, in most of its parameter space, $\mathcal{A}_f$ is close to one with an average value as
\bea
\bar{\mathcal{A}}_f = \frac12.
\eea
Given that our inflationary primordial asymmetry is flavor blind, it is a plausible assumption to consider $A^0_3=\frac13$. For typical values of flavored decay rates, the remnant of the primordial asymmetry is significant which is related to the inflationary asymmetry as
\bea\label{pre-l}
\frac13 \lesssim \mathcal{C} = \frac{n^{p,f}_{\updelta} }{n^{p,i}_{\updelta}} <1.
\eea

Interestingly, eliminating the effect of this pre-existing asymmetry requires tightly fine-tuned relations between the flavored decay rates, hence on leptonic Yukawa couplings, and the flavor-space direction of the inflationary asymmetry. More precisely, one needs either i) $\vert l_{inf} \rangle$ coincides with one of $\vert l_{2} \rangle$ and $\vert l_{3} \rangle$, or ii)  $\vert l_{2} \rangle$ and $\vert l_{3} \rangle$ are perpendicular to each other which $\vert l_{inf} \rangle$ is in the plane of $\vert l_{2} \rangle-\vert l_{3} \rangle$. As a result, the relation presented in Eq. \eqref{pre-l} is a good estimate for most of the possible flavor parameter space.

\bibliographystyle{JHEP}
\bibliography{mybib}

\providecommand{\href}[2]{#2}\begingroup\raggedright\begin{thebibliography}{10}

\bibitem{Maleknejad:2020yys}
A.~Maleknejad, \emph{{SU(2)$_R$ and its Axion in Cosmology: A common Origin for
  Inflation, Cold Sterile Neutrinos, and Baryogenesis}},
  \href{https://arxiv.org/abs/2012.11516}{{\ttfamily 2012.11516}}.

\bibitem{Pati:1974y}
J.~C. Pati and A.~Salam, \emph{{Lepton Number as the Fourth Color}},
  \href{https://doi.org/10.1103/PhysRevD.10.275}{\emph{Phys. Rev. D} {\bfseries
  10} (1974) 275}.

\bibitem{Mohapatra:1974gc}
R.~Mohapatra and J.~C. Pati, \emph{{A Natural Left-Right Symmetry}},
  \href{https://doi.org/10.1103/PhysRevD.11.2558}{\emph{Phys. Rev. D}
  {\bfseries 11} (1975) 2558}.

\bibitem{Senjanovic:1975rk}
G.~Senjanovic and R.~N. Mohapatra, \emph{{Exact Left-Right Symmetry and
  Spontaneous Violation of Parity}},
  \href{https://doi.org/10.1103/PhysRevD.12.1502}{\emph{Phys. Rev. D}
  {\bfseries 12} (1975) 1502}.

\bibitem{Davidson:1978pm}
A.~Davidson, \emph{{}},
  \href{https://doi.org/10.1103/PhysRevD.20.776}{\emph{Phys. Rev. D} {\bfseries
  20} (1979) 776}.

\bibitem{Mohapatra:1980qe}
R.~N. Mohapatra and R.~Marshak, \emph{{Local B-L Symmetry of Electroweak
  Interactions, Majorana Neutrinos and Neutron Oscillations}},
  \href{https://doi.org/10.1103/PhysRevLett.44.1316}{\emph{Phys. Rev. Lett.}
  {\bfseries 44} (1980) 1316}.

\bibitem{Mohapatra:1980yp}
R.~N. Mohapatra and G.~Senjanovic, \emph{{Neutrino Masses and Mixings in Gauge
  Models with Spontaneous Parity Violation}},
  \href{https://doi.org/10.1103/PhysRevD.23.165}{\emph{Phys. Rev. D} {\bfseries
  23} (1981) 165}.

\bibitem{Maiezza:2016ybz}
A.~Maiezza, G.~Senjanovi\'c and J.~C. Vasquez, \emph{{Higgs sector of the
  minimal left-right symmetric theory}},
  \href{https://doi.org/10.1103/PhysRevD.95.095004}{\emph{Phys. Rev. D}
  {\bfseries 95} (2017) 095004}
  [\href{https://arxiv.org/abs/1612.09146}{{\ttfamily 1612.09146}}].

\bibitem{Freese:1990rb}
K.~Freese, J.~A. Frieman and A.~V. Olinto, \emph{{Natural inflation with pseudo
  - Nambu-Goldstone bosons}},
  \href{https://doi.org/10.1103/PhysRevLett.65.3233}{\emph{Phys. Rev. Lett.}
  {\bfseries 65} (1990) 3233}.

\bibitem{Pajer:2013fsa}
E.~Pajer and M.~Peloso, \emph{{A review of Axion Inflation in the era of
  Planck}}, \href{https://doi.org/10.1088/0264-9381/30/21/214002}{\emph{Class.
  Quant. Grav.} {\bfseries 30} (2013) 214002}
  [\href{https://arxiv.org/abs/1305.3557}{{\ttfamily 1305.3557}}].

\bibitem{McAllister:2014mpa}
L.~McAllister, E.~Silverstein, A.~Westphal and T.~Wrase, \emph{{The Powers of
  Monodrom}}, \href{https://doi.org/10.1007/JHEP09(2014)123}{\emph{JHEP}
  {\bfseries 09} (2014) 123} [\href{https://arxiv.org/abs/1405.3652}{{\ttfamily
  1405.3652}}].

\bibitem{Maleknejad:2011sq}
A.~Maleknejad and M.~Sheikh-Jabbari, \emph{{Non-Abelian Gauge Field
  Inflation}}, \href{https://doi.org/10.1103/PhysRevD.84.043515}{\emph{Phys.
  Rev. D} {\bfseries 84} (2011) 043515}
  [\href{https://arxiv.org/abs/1102.1932}{{\ttfamily 1102.1932}}].

\bibitem{Maleknejad:2011jw}
A.~Maleknejad and M.~Sheikh-Jabbari, \emph{{Gauge-flation: Inflation From
  Non-Abelian Gauge Fields}},
  \href{https://doi.org/10.1016/j.physletb.2013.05.001}{\emph{Phys. Lett. B}
  {\bfseries 723} (2013) 224}
  [\href{https://arxiv.org/abs/1102.1513}{{\ttfamily 1102.1513}}].

\bibitem{Adshead:2012kp}
P.~Adshead and M.~Wyman, \emph{{Chromo-Natural Inflation: Natural inflation on
  a steep potential with classical non-Abelian gauge fields}},
  \href{https://doi.org/10.1103/PhysRevLett.108.261302}{\emph{Phys. Rev. Lett.}
  {\bfseries 108} (2012) 261302}
  [\href{https://arxiv.org/abs/1202.2366}{{\ttfamily 1202.2366}}].

\bibitem{SheikhJabbari:2012qf}
M.~Sheikh-Jabbari, \emph{{Gauge-flation Vs Chromo-Natural Inflation}},
  \href{https://doi.org/10.1016/j.physletb.2012.09.014}{\emph{Phys. Lett. B}
  {\bfseries 717} (2012) 6} [\href{https://arxiv.org/abs/1203.2265}{{\ttfamily
  1203.2265}}].

\bibitem{Baumann:2014nda}
D.~Baumann and L.~McAllister, \emph{{Inflation and String Theory}}, Cambridge
  Monographs on Mathematical Physics. Cambridge University Press, 5, 2015,
  \href{https://doi.org/10.1017/CBO9781316105733}{10.1017/CBO9781316105733},
  [\href{https://arxiv.org/abs/1404.2601}{{\ttfamily 1404.2601}}].

\bibitem{Maleknejad:2016qjz}
A.~Maleknejad, \emph{{Axion Inflation with an SU(2) Gauge Field: Detectable
  Chiral Gravity Waves}},
  \href{https://doi.org/10.1007/JHEP07(2016)104}{\emph{JHEP} {\bfseries 07}
  (2016) 104} [\href{https://arxiv.org/abs/1604.03327}{{\ttfamily
  1604.03327}}].

\bibitem{Maleknejad:2012fw}
A.~Maleknejad, M.~Sheikh-Jabbari and J.~Soda, \emph{{Gauge Fields and
  Inflation}}, \href{https://doi.org/10.1016/j.physrep.2013.03.003}{\emph{Phys.
  Rept.} {\bfseries 528} (2013) 161}
  [\href{https://arxiv.org/abs/1212.2921}{{\ttfamily 1212.2921}}].

\bibitem{Maleknejad:2018nxz}
A.~Maleknejad and E.~Komatsu, \emph{{Production and Backreaction of Spin-2
  Particles of $SU(2)$ Gauge Field during Inflation}},
  \href{https://doi.org/10.1007/JHEP05(2019)174}{\emph{JHEP} {\bfseries 05}
  (2019) 174} [\href{https://arxiv.org/abs/1808.09076}{{\ttfamily
  1808.09076}}].

\bibitem{Bastero-Gil:2016qru}
M.~Bastero-Gil, A.~Berera, R.~O. Ramos and J.~G. Rosa, \emph{{Warm Little
  Inflaton}}, \href{https://doi.org/10.1103/PhysRevLett.117.151301}{\emph{Phys.
  Rev. Lett.} {\bfseries 117} (2016) 151301}
  [\href{https://arxiv.org/abs/1604.08838}{{\ttfamily 1604.08838}}].

\bibitem{Kamali:2019ppi}
V.~Kamali, \emph{{Warm pseudoscalar inflation}},
  \href{https://doi.org/10.1103/PhysRevD.100.043520}{\emph{Phys. Rev. D}
  {\bfseries 100} (2019) 043520}
  [\href{https://arxiv.org/abs/1901.01897}{{\ttfamily 1901.01897}}].

\bibitem{Berghaus:2019whh}
K.~V. Berghaus, P.~W. Graham and D.~E. Kaplan, \emph{{Minimal Warm Inflation}},
  \href{https://doi.org/10.1088/1475-7516/2020/03/034}{\emph{JCAP} {\bfseries
  03} (2020) 034} [\href{https://arxiv.org/abs/1910.07525}{{\ttfamily
  1910.07525}}].

\bibitem{Lozanov:2018kpk}
K.~D. Lozanov, A.~Maleknejad and E.~Komatsu, \emph{{Schwinger Effect by an
  $SU(2)$ Gauge Field during Inflation}},
  \href{https://doi.org/10.1007/JHEP02(2019)041}{\emph{JHEP} {\bfseries 02}
  (2019) 041} [\href{https://arxiv.org/abs/1805.09318}{{\ttfamily
  1805.09318}}].

\bibitem{Maleknejad:2019hdr}
A.~Maleknejad, \emph{{Dark Fermions and Spontaneous $CP$ violation in
  $SU(2)$-axion Inflation}},
  \href{https://doi.org/10.1007/JHEP07(2020)154}{\emph{JHEP} {\bfseries 07}
  (2020) 154} [\href{https://arxiv.org/abs/1909.11545}{{\ttfamily
  1909.11545}}].

\bibitem{Mirzagholi:2019jeb}
L.~Mirzagholi, A.~Maleknejad and K.~D. Lozanov, \emph{{Production and
  backreaction of fermions from axion-$SU(2)$ gauge fields during inflation}},
  \href{https://doi.org/10.1103/PhysRevD.101.083528}{\emph{Phys. Rev. D}
  {\bfseries 101} (2020) 083528}
  [\href{https://arxiv.org/abs/1905.09258}{{\ttfamily 1905.09258}}].

\bibitem{Sakharov:1967dj}
A.~Sakharov, \emph{{Violation of CP Invariance, C asymmetry, and baryon
  asymmetry of the universe}},
  \href{https://doi.org/10.1070/PU1991v034n05ABEH002497}{\emph{Sov. Phys. Usp.}
  {\bfseries 34} (1991) 392}.

\bibitem{Maleknejad:2014wsa}
A.~Maleknejad, \emph{{Chiral Gravity Waves and Leptogenesis in Inflationary
  Models with non-Abelian Gauge Fields}},
  \href{https://doi.org/10.1103/PhysRevD.90.023542}{\emph{Phys. Rev. D}
  {\bfseries 90} (2014) 023542}
  [\href{https://arxiv.org/abs/1401.7628}{{\ttfamily 1401.7628}}].

\bibitem{Maleknejad:2016dci}
A.~Maleknejad, \emph{{Gravitational leptogenesis in axion inflation with SU(2)
  gauge field}},
  \href{https://doi.org/10.1088/1475-7516/2016/12/027}{\emph{JCAP} {\bfseries
  12} (2016) 027} [\href{https://arxiv.org/abs/1604.06520}{{\ttfamily
  1604.06520}}].

\bibitem{Caldwell:2017chz}
R.~Caldwell and C.~Devulder, \emph{{Axion Gauge Field Inflation and
  Gravitational Leptogenesis: A Lower Bound on B Modes from the
  Matter-Antimatter Asymmetry of the Universe}},
  \href{https://doi.org/10.1103/PhysRevD.97.023532}{\emph{Phys. Rev. D}
  {\bfseries 97} (2018) 023532}
  [\href{https://arxiv.org/abs/1706.03765}{{\ttfamily 1706.03765}}].

\bibitem{Alexander:2018fjp}
S.~Alexander, E.~McDonough and D.~N. Spergel, \emph{{Chiral Gravitational Waves
  and Baryon Superfluid Dark Matter}},
  \href{https://doi.org/10.1088/1475-7516/2018/05/003}{\emph{JCAP} {\bfseries
  05} (2018) 003} [\href{https://arxiv.org/abs/1801.07255}{{\ttfamily
  1801.07255}}].

\bibitem{Adshead:2013qp}
P.~Adshead, E.~Martinec and M.~Wyman, \emph{{Gauge fields and inflation: Chiral
  gravitational waves, fluctuations, and the Lyth bound}},
  \href{https://doi.org/10.1103/PhysRevD.88.021302}{\emph{Phys. Rev. D}
  {\bfseries 88} (2013) 021302}
  [\href{https://arxiv.org/abs/1301.2598}{{\ttfamily 1301.2598}}].

\bibitem{Dimastrogiovanni:2012ew}
E.~Dimastrogiovanni and M.~Peloso, \emph{{Stability analysis of chromo-natural
  inflation and possible evasion of Lyth\textquoteright{}s bound}},
  \href{https://doi.org/10.1103/PhysRevD.87.103501}{\emph{Phys. Rev. D}
  {\bfseries 87} (2013) 103501}
  [\href{https://arxiv.org/abs/1212.5184}{{\ttfamily 1212.5184}}].

\bibitem{Agrawal:2018mrg}
A.~Agrawal, T.~Fujita and E.~Komatsu, \emph{{Tensor Non-Gaussianity from
  Axion-Gauge-Fields Dynamics : Parameter Search}},
  \href{https://doi.org/10.1088/1475-7516/2018/06/027}{\emph{JCAP} {\bfseries
  06} (2018) 027} [\href{https://arxiv.org/abs/1802.09284}{{\ttfamily
  1802.09284}}].

\bibitem{Campeti:2020xwn}
P.~Campeti, E.~Komatsu, D.~Poletti and C.~Baccigalupi, \emph{{Measuring the
  spectrum of primordial gravitational waves with CMB, PTA and Laser
  Interferometers}},  \href{https://arxiv.org/abs/2007.04241}{{\ttfamily
  2007.04241}}.

\bibitem{Thorne:2017jft}
B.~Thorne, T.~Fujita, M.~Hazumi, N.~Katayama, E.~Komatsu and M.~Shiraishi,
  \emph{{Finding the chiral gravitational wave background of an axion-SU(2)
  inflationary model using CMB observations and laser interferometers}},
  \href{https://doi.org/10.1103/PhysRevD.97.043506}{\emph{Phys. Rev. D}
  {\bfseries 97} (2018) 043506}
  [\href{https://arxiv.org/abs/1707.03240}{{\ttfamily 1707.03240}}].

\bibitem{Silverstein:2008sg}
E.~Silverstein and A.~Westphal, \emph{{Monodromy in the CMB: Gravity Waves and
  String Inflation}},
  \href{https://doi.org/10.1103/PhysRevD.78.106003}{\emph{Phys. Rev. D}
  {\bfseries 78} (2008) 106003}
  [\href{https://arxiv.org/abs/0803.3085}{{\ttfamily 0803.3085}}].

\bibitem{Flauger:2009ab}
R.~Flauger, L.~McAllister, E.~Pajer, A.~Westphal and G.~Xu, \emph{{Oscillations
  in the CMB from Axion Monodromy Inflation}},
  \href{https://doi.org/10.1088/1475-7516/2010/06/009}{\emph{JCAP} {\bfseries
  06} (2010) 009} [\href{https://arxiv.org/abs/0907.2916}{{\ttfamily
  0907.2916}}].

\bibitem{Adshead:2016omu}
P.~Adshead, E.~Martinec, E.~I. Sfakianakis and M.~Wyman, \emph{{Higgsed
  Chromo-Natural Inflation}},
  \href{https://doi.org/10.1007/JHEP12(2016)137}{\emph{JHEP} {\bfseries 12}
  (2016) 137} [\href{https://arxiv.org/abs/1609.04025}{{\ttfamily
  1609.04025}}].

\bibitem{Adshead:2017hnc}
P.~Adshead and E.~I. Sfakianakis, \emph{{Higgsed Gauge-flation}},
  \href{https://doi.org/10.1007/JHEP08(2017)130}{\emph{JHEP} {\bfseries 08}
  (2017) 130} [\href{https://arxiv.org/abs/1705.03024}{{\ttfamily
  1705.03024}}].

\bibitem{Peccei:1977hh}
R.~Peccei and H.~R. Quinn, \emph{{CP Conservation in the Presence of
  Instantons}}, \href{https://doi.org/10.1103/PhysRevLett.38.1440}{\emph{Phys.
  Rev. Lett.} {\bfseries 38} (1977) 1440}.

\bibitem{Weinberg:1996kr}
S.~Weinberg, \emph{{The quantum theory of fields. Vol. 2: Modern
  applications}}. Cambridge University Press, 8, 2013.

\bibitem{Adler:1969gk}
S.~L. Adler, \emph{{Axial vector vertex in spinor electrodynamics}},
  \href{https://doi.org/10.1103/PhysRev.177.2426}{\emph{Phys. Rev.} {\bfseries
  177} (1969) 2426}.

\bibitem{Bell:1969ts}
J.~Bell and R.~Jackiw, \emph{{A PCAC puzzle: $\pi^0 \to \gamma \gamma$ in the
  $\sigma$ model}}, \href{https://doi.org/10.1007/BF02823296}{\emph{Nuovo Cim.
  A} {\bfseries 60} (1969) 47}.

\bibitem{Nemevsek:2012cd}
M.~Nemevsek, G.~Senjanovic and Y.~Zhang, \emph{{Warm Dark Matter in Low Scale
  Left-Right Theory}},
  \href{https://doi.org/10.1088/1475-7516/2012/07/006}{\emph{JCAP} {\bfseries
  07} (2012) 006} [\href{https://arxiv.org/abs/1205.0844}{{\ttfamily
  1205.0844}}].

\bibitem{Ade:2018gkx}
{\scshape BICEP2, Keck Array} collaboration, P.~A.~R. Ade et~al., \emph{{BICEP2
  / Keck Array x: Constraints on Primordial Gravitational Waves using Planck,
  WMAP, and New BICEP2/Keck Observations through the 2015 Season}},
  \href{https://doi.org/10.1103/PhysRevLett.121.221301}{\emph{Phys. Rev. Lett.}
  {\bfseries 121} (2018) 221301}
  [\href{https://arxiv.org/abs/1810.05216}{{\ttfamily 1810.05216}}].

\bibitem{Dunsky:2020dhn}
D.~Dunsky, L.~J. Hall and K.~Harigaya, \emph{{Sterile Neutrino Dark Matter and
  Leptogenesis in Left-Right Higgs Parity}},
  \href{https://arxiv.org/abs/2007.12711}{{\ttfamily 2007.12711}}.

\bibitem{Rizzo:1981su}
T.~G. Rizzo and G.~Senjanovic, \emph{{Can There Be Low Intermediate Mass Scales
  in Grand Unified Theories?}},
  \href{https://doi.org/10.1103/PhysRevLett.46.1315}{\emph{Phys. Rev. Lett.}
  {\bfseries 46} (1981) 1315}.

\bibitem{Bertolini:2012im}
S.~Bertolini, L.~Di~Luzio and M.~Malinsky, \emph{{Seesaw Scale in the Minimal
  Renormalizable SO(10) Grand Unification}},
  \href{https://doi.org/10.1103/PhysRevD.85.095014}{\emph{Phys. Rev. D}
  {\bfseries 85} (2012) 095014}
  [\href{https://arxiv.org/abs/1202.0807}{{\ttfamily 1202.0807}}].

\bibitem{Deppisch:2017xhv}
F.~F. Deppisch, T.~E. Gonzalo and L.~Graf, \emph{{Surveying the SO(10) Model
  Landscape: The Left-Right Symmetric Case}},
  \href{https://doi.org/10.1103/PhysRevD.96.055003}{\emph{Phys. Rev. D}
  {\bfseries 96} (2017) 055003}
  [\href{https://arxiv.org/abs/1705.05416}{{\ttfamily 1705.05416}}].

\bibitem{Bodeker:2019rvr}
D.~B\"odeker and D.~Schr\"oder, \emph{{Kinetic equations for sterile neutrinos
  from thermal fluctuations}},
  \href{https://doi.org/10.1088/1475-7516/2020/02/033}{\emph{JCAP} {\bfseries
  02} (2020) 033} [\href{https://arxiv.org/abs/1911.05092}{{\ttfamily
  1911.05092}}].

\bibitem{Abada:2006fw}
A.~Abada, S.~Davidson, F.-X. Josse-Michaux, M.~Losada and A.~Riotto,
  \emph{{Flavor issues in leptogenesis}},
  \href{https://doi.org/10.1088/1475-7516/2006/04/004}{\emph{JCAP} {\bfseries
  04} (2006) 004} [\href{https://arxiv.org/abs/hep-ph/0601083}{{\ttfamily
  hep-ph/0601083}}].

\bibitem{Barbieri:1999ma}
R.~Barbieri, P.~Creminelli, A.~Strumia and N.~Tetradis, \emph{{Baryogenesis
  through leptogenesis}},
  \href{https://doi.org/10.1016/S0550-3213(00)00011-0}{\emph{Nucl. Phys. B}
  {\bfseries 575} (2000) 61}
  [\href{https://arxiv.org/abs/hep-ph/9911315}{{\ttfamily hep-ph/9911315}}].

\bibitem{Blanchet:2006ch}
S.~Blanchet, P.~Di~Bari and G.~Raffelt, \emph{{Quantum Zeno effect and the
  impact of flavor in leptogenesis}},
  \href{https://doi.org/10.1088/1475-7516/2007/03/012}{\emph{JCAP} {\bfseries
  03} (2007) 012} [\href{https://arxiv.org/abs/hep-ph/0611337}{{\ttfamily
  hep-ph/0611337}}].

\bibitem{Bertuzzo:2010et}
E.~Bertuzzo, P.~Di~Bari and L.~Marzola, \emph{{The problem of the initial
  conditions in flavoured leptogenesis and the tauon $N_2$-dominated
  scenario}},
  \href{https://doi.org/10.1016/j.nuclphysb.2011.03.027}{\emph{Nucl. Phys. B}
  {\bfseries 849} (2011) 521}
  [\href{https://arxiv.org/abs/1007.1641}{{\ttfamily 1007.1641}}].

\bibitem{DiBari:2013qja}
P.~Di~Bari and L.~Marzola, \emph{{SO(10)-inspired solution to the problem of
  the initial conditions in leptogenesis}},
  \href{https://doi.org/10.1016/j.nuclphysb.2013.10.027}{\emph{Nucl. Phys. B}
  {\bfseries 877} (2013) 719}
  [\href{https://arxiv.org/abs/1308.1107}{{\ttfamily 1308.1107}}].

\bibitem{Fukugita:1986hr}
M.~Fukugita and T.~Yanagida, \emph{{Baryogenesis Without Grand Unification}},
  \href{https://doi.org/10.1016/0370-2693(86)91126-3}{\emph{Phys. Lett. B}
  {\bfseries 174} (1986) 45}.

\bibitem{Ade:2015xua}
{\scshape Planck} collaboration, P.~Ade et~al., \emph{{Planck 2015 results.
  XIII. Cosmological parameters}},
  \href{https://doi.org/10.1051/0004-6361/201525830}{\emph{Astron. Astrophys.}
  {\bfseries 594} (2016) A13}
  [\href{https://arxiv.org/abs/1502.01589}{{\ttfamily 1502.01589}}].

\bibitem{Pal:1981rm}
P.~B. Pal and L.~Wolfenstein, \emph{{Radiative Decays of Massive Neutrinos}},
  \href{https://doi.org/10.1103/PhysRevD.25.766}{\emph{Phys. Rev. D} {\bfseries
  25} (1982) 766}.

\bibitem{Barger:1995ty}
V.~D. Barger, R.~Phillips and S.~Sarkar, \emph{{Remarks on the KARMEN
  anomaly}}, \href{https://doi.org/10.1016/0370-2693(95)00486-5}{\emph{Phys.
  Lett. B} {\bfseries 352} (1995) 365}
  [\href{https://arxiv.org/abs/hep-ph/9503295}{{\ttfamily hep-ph/9503295}}].

\bibitem{Ackermann:2013uma}
{\scshape Fermi-LAT} collaboration, M.~Ackermann et~al., \emph{{Search for
  Gamma-ray Spectral Lines with the Fermi Large Area Telescope and Dark Matter
  Implications}}, \href{https://doi.org/10.1103/PhysRevD.88.082002}{\emph{Phys.
  Rev. D} {\bfseries 88} (2013) 082002}
  [\href{https://arxiv.org/abs/1305.5597}{{\ttfamily 1305.5597}}].

\bibitem{Lue:1998mq}
A.~Lue, L.-M. Wang and M.~Kamionkowski, \emph{{Cosmological signature of new
  parity violating interactions}},
  \href{https://doi.org/10.1103/PhysRevLett.83.1506}{\emph{Phys. Rev. Lett.}
  {\bfseries 83} (1999) 1506}
  [\href{https://arxiv.org/abs/astro-ph/9812088}{{\ttfamily
  astro-ph/9812088}}].

\bibitem{Saito:2007kt}
S.~Saito, K.~Ichiki and A.~Taruya, \emph{{Probing polarization states of
  primordial gravitational waves with CMB anisotropies}},
  \href{https://doi.org/10.1088/1475-7516/2007/09/002}{\emph{JCAP} {\bfseries
  09} (2007) 002} [\href{https://arxiv.org/abs/0705.3701}{{\ttfamily
  0705.3701}}].

\bibitem{Contaldi:2008yz}
C.~R. Contaldi, J.~Magueijo and L.~Smolin, \emph{{Anomalous CMB polarization
  and gravitational chirality}},
  \href{https://doi.org/10.1103/PhysRevLett.101.141101}{\emph{Phys. Rev. Lett.}
  {\bfseries 101} (2008) 141101}
  [\href{https://arxiv.org/abs/0806.3082}{{\ttfamily 0806.3082}}].

\bibitem{Barenboim:1996nd}
G.~Barenboim, J.~Bernabeu, J.~Prades and M.~Raidal, \emph{{Constraints on the
  $W_{R}$ mass and CP violation in left-right models}},
  \href{https://doi.org/10.1103/PhysRevD.55.4213}{\emph{Phys. Rev. D}
  {\bfseries 55} (1997) 4213}
  [\href{https://arxiv.org/abs/hep-ph/9611347}{{\ttfamily hep-ph/9611347}}].

\bibitem{Bertolini:2014sua}
S.~Bertolini, A.~Maiezza and F.~Nesti, \emph{{Present and Future K and B Meson
  Mixing Constraints on TeV Scale Left-Right Symmetry}},
  \href{https://doi.org/10.1103/PhysRevD.89.095028}{\emph{Phys. Rev. D}
  {\bfseries 89} (2014) 095028}
  [\href{https://arxiv.org/abs/1403.7112}{{\ttfamily 1403.7112}}].

\bibitem{Beringer:1900zz}
{\scshape Particle Data Group} collaboration, J.~Beringer et~al., \emph{{Review
  of Particle Physics (RPP)}},
  \href{https://doi.org/10.1103/PhysRevD.86.010001}{\emph{Phys. Rev. D}
  {\bfseries 86} (2012) 010001}.

\bibitem{Abramowski:2013ax}
{\scshape H.E.S.S.} collaboration, A.~Abramowski et~al., \emph{{Search for
  Photon-Linelike Signatures from Dark Matter Annihilations with H.E.S.S.}},
  \href{https://doi.org/10.1103/PhysRevLett.110.041301}{\emph{Phys. Rev. Lett.}
  {\bfseries 110} (2013) 041301}
  [\href{https://arxiv.org/abs/1301.1173}{{\ttfamily 1301.1173}}].

\bibitem{Aleksic:2013xea}
J.~Aleksi\'c et~al., \emph{{Optimized dark matter searches in deep observations
  of Segue 1 with MAGIC}},
  \href{https://doi.org/10.1088/1475-7516/2014/02/008}{\emph{JCAP} {\bfseries
  02} (2014) 008} [\href{https://arxiv.org/abs/1312.1535}{{\ttfamily
  1312.1535}}].

\bibitem{Deshpande:1990ip}
N.~Deshpande, J.~Gunion, B.~Kayser and F.~I. Olness, \emph{{Left-right
  symmetric electroweak models with triplet Higgs}},
  \href{https://doi.org/10.1103/PhysRevD.44.837}{\emph{Phys. Rev. D} {\bfseries
  44} (1991) 837}.

\bibitem{Dev:2016dja}
P.~S.~B. Dev, R.~N. Mohapatra and Y.~Zhang, \emph{{Probing the Higgs Sector of
  the Minimal Left-Right Symmetric Model at Future Hadron Colliders}},
  \href{https://doi.org/10.1007/JHEP05(2016)174}{\emph{JHEP} {\bfseries 05}
  (2016) 174} [\href{https://arxiv.org/abs/1602.05947}{{\ttfamily
  1602.05947}}].

\bibitem{Dev:2018foq}
P.~Bhupal~Dev, R.~N. Mohapatra, W.~Rodejohann and X.-J. Xu, \emph{{Vacuum
  structure of the left-right symmetric model}},
  \href{https://doi.org/10.1007/JHEP02(2019)154}{\emph{JHEP} {\bfseries 02}
  (2019) 154} [\href{https://arxiv.org/abs/1811.06869}{{\ttfamily
  1811.06869}}].

\bibitem{Bodeker:1999zt}
D.~Bodeker, G.~D. Moore and K.~Rummukainen, \emph{{Hard thermal loops and the
  sphaleron rate on the lattice}},
  \href{https://doi.org/10.1016/S0920-5632(00)91745-6}{\emph{Nucl. Phys. B
  Proc. Suppl.} {\bfseries 83} (2000) 583}
  [\href{https://arxiv.org/abs/hep-lat/9909054}{{\ttfamily hep-lat/9909054}}].

\bibitem{Arnold:1996dy}
P.~B. Arnold, D.~Son and L.~G. Yaffe, \emph{{The Hot baryon violation rate is O
  (alpha-w**5 T**4)}},
  \href{https://doi.org/10.1103/PhysRevD.55.6264}{\emph{Phys. Rev. D}
  {\bfseries 55} (1997) 6264}
  [\href{https://arxiv.org/abs/hep-ph/9609481}{{\ttfamily hep-ph/9609481}}].

\end{thebibliography}\endgroup

\end{document}